**White Paper on**

# The *Majorana* Zero-Neutrino Double-Beta Decay Experiment

The *Majorana Collaboration*:

Brown University, Providence, RI

Institute for Theoretical and Experimental Physics, Moscow, Russia

Joint Institute for Nuclear Research, Dubna, Russia

Lawrence Berkeley National Laboratory, Berkeley, CA

Lawrence Livermore National Laboratory, Livermore, CA

Los Alamos National Laboratory, Los Alamos, NM

New Mexico State University, Carlsbad, NM

Oak Ridge National Laboratory, Oak Ridge, TN

Osaka University, Osaka, Japan

Pacific Northwest National Laboratory, Richland, WA

Queen's University, Kingston, Canada

Triangle Universities Nuclear Laboratory, Durham, NC

University of Chicago, Chicago, IL

University of South Carolina, Columbia, SC

University of Tennessee, Knoxville, TN

University of Washington, Seattle, WA

# Abstract

The objective of the Majorana Experiment is to study neutrinoless double beta decay (0νββ) with an effective Majorana-neutrino mass sensitivity below 50 meV in order to characterize the Majorana nature of the neutrino, the Majorana mass spectrum, and the absolute mass scale. An experimental study of the neutrino mass scale implied by neutrino oscillation results is now technically within our grasp. This exciting physics goal is best pursued using the well-established technique of searching for 0νββ of $^{76}$Ge, augmented with recent advances in signal processing and detector design. The Majorana Experiment will consist of a large mass of $^{76}$Ge in the form of high-resolution intrinsic germanium detectors located deep underground within a low-background shielding environment. Observation of a sharp peak at the ββ endpoint will quantify the 0νββ half-life and thus the effective Majorana mass of the electron neutrino. In addition to the modest R&D program, we present here an overview of the entire project in order to help put in perspective the scope, the low level of technical risk, and the readiness of the Collaboration to immediately begin the undertaking.

## *The Majorana Collaboration*

*Brown University, Providence, RI*
    Richard Gaitskell
*Institute for Theoretical and Experimental Physics, Moscow, Russia*
    Alexander Barabash, Sergey Konovalov, Vladimir Umatov, Igor Vanushin
*Joint Institute for Nuclear Research, Dubna, Russia*
    Viktor Brudanin, Viatcheslav Egorov, Oleg Kochetov, Viatcheslav Sandukovsky
*Lawrence Berkeley National Laboratory, Berkeley, CA*
    Yuen-Dat Chan, Paul Fallon, Reyco Henning, Kevin Lesko, Augusto Macchiavelli, Alan Poon
*Lawrence Livermore National Laboratory, Livermore, CA*
    Kai Vetter
*Los Alamos National Laboratory, Los Alamos, NM*
    Thedore Ball, Steven Elliott, Victor Gehman, Andrew Hime, Dongming Mei, Geoffrey Mills, Richard Van de Water, Jan Wouters
*New Mexico State University, Carlsbad NM*
    Joel Webb
*Oak Ridge National Laboratory, Oak Ridge, TN*
    Thomas V. Cianciolo, Krzysztof P. Rykaczewski, Robert Grzywacz, David Radford, Cyrus Baktash
*Osaka University, Osaka Japan*
    Hiroyasu Ejiri, Ryuta Hazama, Masaharu Nomachi
*Pacific Northwest National Laboratory, Richland, WA*
    Craig Aalseth, Dale Anderson, Richard Arthur, Ronald Brodzinski, Glen Dunham, James Ely, Shelece Easterday, Todd Hossbach, David Jordan, Richard Kouzes, Harry Miley, William Pitts, Robert Thompson, Ray Warner
*Queen's University, Kingston, Canada*
    Aksel Hallin, Art McDonald
*Triangle Universities Nuclear Laboratory, Durham, NC*
    Art Champagne, Jeremy Kephart, Ryan Rohm, Konstantin Saburov, Werner Tornow, Albert Young
*University of Chicago, Chicago, IL*
    Juan Collar
*University of South Carolina, Columbia, SC*
    Frank Avignone, Richard Creswick, Horatio A. Farach, George King, John M. Palms
*University of Tennessee, Knoxville TN*
    William Bugg, Yuri Efremenko
*University of Washington, Seattle, WA*
    Tom Burritt, Peter Doe, Mark Howe, Kareem Kazkaz, Hamish Robertson, John Wilkerson

## Table of Contents

# 1.0 Executive Summary

### 1.1 Purpose of Experiment

The objective of the Majorana Experiment is to study neutrinoless double beta decay (0νββ) with an effective Majorana-neutrino mass sensitivity below 50 meV in order to characterize the Majorana nature of the neutrino, the Majorana mass spectrum, and the absolute mass scale. An experimental study of the neutrino mass scale implied by neutrino oscillation results is now technically within our grasp. This exciting physics goal is best pursued using the well-established technique of searching for 0νββ of $^{76}$Ge, augmented with recent advances in signal processing and detector design. The Majorana Experiment will consist of a large mass of $^{76}$Ge in the form of high-resolution intrinsic germanium detectors located deep underground within a low-background shielding environment. Observation of a sharp peak at the ββ endpoint will quantify the 0νββ half-life and thus the effective Majorana mass of the electron neutrino. In addition to the modest R&D program, we present here an overview of the entire project in order to help put in perspective the scope, the low level of technical risk, and the readiness of the Collaboration to immediately begin the undertaking.

### 1.2 Research and Development Requirements

The Majorana proposal is based on well-established technology that does not require proof-of-principle research and development. However, there are two R&D projects that are currently underway to optimize the engineering design of the Majorana Experiment. These projects are called SEGA (Segmented Enriched Germanium Assembly) and MEGA (Multiple Element Germanium Array). Not only do these efforts help to optimize the Majorana design but also they will achieve physics goals themselves. The three projects address the main technical goals of the overall Majorana Experiment.

- SEGA: The goal is to optimize the previously successful, signal processing techniques for crystals whose charge collection is segmented.
- MEGA: The goal is to optimize the arrangement and packaging for multiple crystals sharing a single cooling system.
- Majorana: Implement the optimum configuration determined by the SEGA and MEGA activities, but in addition operate with a large quantity of enriched Ge material to reach a significant sensitivity for the 0νββ half-life.

In SEGA, the low background and specialized signal processing is expected to produce an interesting dark matter result in only a few months of counting. After this initial goal, results such as the precise measurement of the two-neutrino half-life of $^{76}$Ge can be achieved (values from previous experiments vary considerably).

MEGA consists of an array of 18 detectors. All detectors will make use of pulse-shape analysis and some fraction of the detectors will be segmented. This arrangement should provide excellent sensitivity for study of inclusive ββ-decays to excited states in $^{76}$Ge, $^{82}$Se, $^{96}$Zr, $^{100}$Mo, $^{130}$Te, and $^{150}$Nd. Further improvement in dark matter sensitivity is also expected. Once the goals of MEGA are complete, the apparatus may be used for screening materials for Majorana construction, other underground experiments, or ultra-trace environmental radiological measurements. The combination of low background and special signal processing would make this arrangement among the most sensitive and selective available for sample counting anywhere in the world.

The Majorana Experiment will consist of a few hundred crystals enriched in $^{76}$Ge grouped into a collection of modules. All crystals will be segmented and instrumented for pulse-shape analysis. This modular arrangement results in a small footprint and allows easy access to modules. The low-risk Reference Plan is to cool the germanium using well-understood conventional techniques. More than 20 years of double-beta decay experience by the Majorana Collaboration members and the lessons learned from SEGA and MEGA ensure that the instrumentation, analysis techniques, and packaging will be proven and the engineering risk will be minimal.

### 1.3 Anticipated Sensitivity

The Majorana Collaboration estimate of the ultimate sensitivity of the experiment is based on a background model described in Section 3. The background model is motivated in part from early IGEX[1] data and predicts an achievable 0νββ half-life limit of over $10^{27}$ y within 5 years of initial receipt of the enriched material and an "asymptotic" limit of $4 \times 10^{27}$ y: a factor of ~200 improvement over current limits. Depending on the nuclear matrix elements chosen, the effective neutrino mass sensitivity reaches below |<m$_\nu$>| ~40 meV which is within the range implied by recent neutrino oscillation experiments.

### 1.4 Major Requirements

This sensitivity can be reached with about 2500 kg-y of data from enriched Ge (86% in $^{76}$Ge) detectors operated with backgrounds lower than previously obtained. Thus a deep underground location, an active veto system, and carefully designed shielding are required. In addition, signal processing techniques, detector segmentation, and underground material preparation will be needed. These have been developed.

### 1.5 Basic Timeline

Beyond the first year when the laboratory is being prepared, the rate of germanium enrichment is a significant variable. Assuming that the rate is rapidly ramped to 200 kg/y, about 4 years of production will be required. Since we will be fabricating detectors from the onset, we will achieve ~2000 kg-y of data within 5 years of the start of enrichment.

### 1.6 Current Status

Through Collaboration pooling of resources, our first (isotopically enriched) segmented detector for SEGA has been delivered. For MEGA, the 16 outer crystals are in hand and the special cryostat is under construction. We are developing space underground for SEGA and MEGA, as well as designing the data sharing and data hosting needed for a geographically dispersed collaboration. The national enthusiasm for developing underground science facilities and performing underground science in the United States has lent additional momentum to our collaborative effort. This proposal is requesting funds to implement the Majorana Experiment.

---

[1] The International Germanium Experiment (IGEX) amassed 8.9 kg-yr of data using $^{76}$Ge detectors.

## 2.0   Majorana Science Motivation

It is now possible for next-generation 0νββ experiments to access the neutrino mass range of interest suggested by recent studies of neutrino oscillations. Actually, a well-designed germanium detector array can find the effective mass, if the massive neutrinos are Majorana particles and the neutrino mass spectrum is quasi-degenerate or inverted hierarchy. In fact, many theories of the fundamental particle interactions predict that massive neutrinos are Majorana in nature. Hence the experiment may establish the Majorana nature, the mass spectrum and the absolute mass scale of the neutrino. This well-established technique has been augmented by the availability of isotopic enrichment facilities, dramatic improvements in germanium spectroscopy, and new US underground laboratory initiatives. The realization that the technology is available to achieve such a fundamental physics goal provides the basic motivation for the Majorana Experiment.

To convey the importance of neutrino mass, and of the Majorana Experiment, we present a physics motivation and a brief recapitulation of past double-beta decay experiments, drawing heavily on our own completed work. We present our new technological capabilities to show how the Majorana Collaboration is positioned to make rapid strides toward the measurement of the effective Majorana neutrino mass, as well as impacting other science areas.

### 2.1 Motivation of $^{76}$Ge 0ν Double-Beta Decay

Ordinary beta decay of many even-even nuclei is energetically forbidden. However, a process in which a nucleus changes its atomic number (Z) by two while simultaneously emitting two beta particles is energetically possible for some of these nuclei. Such a process is called double-beta decay. (Most of our discussions will consider only double β$^-$ decays because of the larger phase space.) Two-neutrino double-beta decay (2νββ), defined by

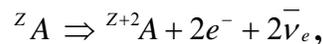

$$^{z}A \Rightarrow {}^{z+2}A + 2e^- + 2\bar{\nu}_e,$$

is an allowed second-order weak process that occurs in nature, although its rate is extremely low. Half-lives for this decay mode have been measured at ~10$^{19}$ years or longer.

A more interesting process is zero-neutrino double-beta decay (0νββ),

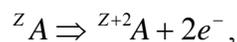

$$^{z}A \Rightarrow {}^{z+2}A + 2e^-,$$

where no neutrinos are emitted in the decay process. It is clear from this equation that unlike 2νββ, 0νββ violates lepton number conservation and hence requires physics beyond the standard model. One can visualize 0νββ as an exchange of a virtual neutrino between two neutrons within the nucleus. In the framework of the standard model of SU$_L$(2)×U(1) of weak interactions, the first neutron emits a right-handed anti-neutrino.

However, the second neutron requires the absorption of a left-handed neutrino. In order for this to happen, the neutrino would have to be massive so that it is not in a pure helicity state, and the neutrino and anti-neutrino would have to be indistinguishable. That is, the neutrino would have to be a massive Majorana particle. The Dirac or Majorana nature of the neutrino is an important open question. Neutrinoless double-beta decay is the only known practical way to determine if neutrinos are Majorana particles.

A complete understanding of the neutrino mass matrix depends on three types of data, each analogous to one leg of a three-legged stool. They are: neutrino oscillations, tritium beta-decay measurement, and neutrinoless double-beta decay. Each is necessary for a complete picture. The results of atmospheric and solar neutrino oscillation experiments and reactor neutrino experiments indicate that neutrinos have mass. This in itself excites interest in $0\nu\beta\beta$ experiments. However, neutrino oscillation experiments only measure the difference in the squares of the masses of the mass eigenstates along with the mixing angles. Therefore, they indicate only the relative mass scale of the neutrinos. Even so, these experiments show that at least one neutrino has a mass greater than ~50 meV. As a result, measurements of the absolute mass scale on this order are extremely exciting. The absolute scale can only be obtained from direct mass measurements, $^3$H end point measurements [Osi01], Cosmology [Spe03], or in the case of Majorana neutrinos, more sensitively by neutrinoless double-beta decay. Neutrinoless double-beta decay experiments are the only proposed method of measuring neutrino mass that have the potential to reach this interesting level of sensitivity and the only practical method to discern the Dirac or Majorana nature of the neutrino.

Neutrinoless double-beta decay is by now an old subject [Ell02]. What is new is the fact that positive observation of neutrino oscillations in atmospheric neutrinos [Tos01], in solar neutrinos [Ahm01, Ahm02, Fak02] and in reactor neutrinos [Eug03] gives new motivation for more sensitive searches. In fact, published constraints on the mixing angles of the neutrino-mixing matrix make a strong case that if neutrinos are Majorana particles, there are many scenarios in which next generation double-beta decay experiments should be able to observe the phenomenon and measure the effective Majorana mass of the electron neutrino, $\left|\langle m_\nu \rangle\right|$; this would provide a measure of the neutrino mass scale, $m$. The time for large, next generation $0\nu\beta\beta$ experiments has arrived, for if the mass scale is below ~0.35 $eV$, $0\nu\beta\beta$ may be the only hope for measuring it in the near future.

The most sensitive experiments carried out so far have probed the decay $^{76}$Ge→$^{76}$Se + 2e$^-$ with specially built Ge detectors fabricated from germanium isotopically enriched from 7.8% to 86% in $^{76}$Ge. The Heidelberg-Moscow Experiment [Kla01] and the International Germanium Experiment (IGEX) [Aal02] have placed lower bounds on the half-life for this process of $1.9\times10^{25}$ $y$ (90% $CL$) and $1.6\times10^{25}$ $y$ (90% $CL$) respectively. A recent claim for the observation of $0\nu\beta\beta$ [Kla01c, Kla02] ($T_{1/2} = 1.5\times10^{25}$ $y$) has been very controversial [Aal02a, Har02, Fer02, Zde02]. The Majorana experiment will be more than sensitive enough to verify or repudiate this claim.

### 2.1.1   Neutrinoless Double-Beta Decay

Many processes have been proposed to drive neutrinoless double-beta decay: for example intrinsic right-handed currents and the exchange of supersymmetric particles. Regardless of the process however, the existence of $0\nu\beta\beta$ implies the existence of a non-zero Majorana mass term for the neutrino [Sch82]. Only the process involving the exchange of a light Majorana neutrino will be discussed here.

The decay rate for this process is expressed as follows:

$$\left[ T_{1/2}^{0\nu} \right]^{1} = G^{0\nu}(E_0, Z) \left| \langle m_\nu \rangle \right|^2 \left| M_f^{0\nu} - (g_A / g_V)^2 M_{GT}^{0\nu} \right|^2. \tag{1}$$

In equation (1), $G^{0\nu}$ is the two-body phase-space factor including coupling constant, $M_f^{0\nu}$ and $M_{GT}^{0\nu}$ are the Fermi and Gamow-Teller nuclear matrix elements, respectively. The constants $g_A$ and $g_V$ are the axial-vector and vector relative weak coupling constants, respectively. The quantity $\left| \langle m_\nu \rangle \right|$ is the effective Majorana electron neutrino mass given by:

$$\left| \langle m_\nu \rangle \right| \equiv \left| \left| U_{e1}^L \right|^2 m_1 + \left| U_{e2}^L \right|^2 m_2\, e^{i\phi_2} + \left| U_{e3}^L \right|^2 m_3\, e^{i\phi_3} \right|, \tag{2}$$

where the $U$'s are the elements of the Weak Mixing Matrix, $e^{i\phi_2}$ and $e^{i\phi_3}$ are the relative CP phases ($\pm 1$ for CP conservation) and $m_{1,2,3}$ are the neutrino mass eigenvalues. $\phi_2$ is a pure Majorana phase ($\phi_2 = 2(\alpha_2 - \alpha_1)$) and $\phi_3 = -2(\delta + \alpha)$ is a mixture of Dirac and Majorana types, where $\alpha$ and $\delta$ are defined in the next section.

### 2.1.2   The Neutrino Mixing Matrix

The conventional form of the neutrino Mixing Matrix follows that suggested by the Particle Data Book [PDG02] for the Weak Mixing Matrix:

$$\begin{bmatrix} |\nu_e\rangle \\ |\nu_\mu\rangle \\ |\nu_\tau\rangle \end{bmatrix} = \begin{bmatrix} c_{12}c_{13} & s_{12}c_{13} & s_{13}e^{-i\delta} \\ -s_{12}c_{23} - c_{12}s_{23}s_{13}e^{i\delta} & c_{12}c_{23} - s_{12}s_{23}s_{13}e^{i\delta} & s_{23}c_{13} \\ s_{12}s_{23} - c_{12}c_{23}s_{13}e^{i\delta} & -c_{12}s_{23} - s_{12}c_{23}s_{13}e^{i\delta} & c_{23}c_{13} \end{bmatrix} \begin{bmatrix} e^{i\alpha_1} & 0 & 0 \\ 0 & e^{i\alpha_2} & 0 \\ 0 & 0 & 1 \end{bmatrix} \begin{bmatrix} |\nu_1\rangle \\ |\nu_2\rangle \\ |\nu_3\rangle \end{bmatrix}, \tag{3}$$

where $c_{ij} = \cos\theta_{ij}$ and, $s_{ij} = \sin\theta_{ij}$ and we multiply by an additional diagonal matrix that contains Majorana CP phases that do not appear in neutrino oscillations. While this looks very complicated and populated with many unknowns, neutrino oscillation data have constrained all three of the angles $\theta_{12}$, $\theta_{23}$, and $\theta_{13}$, while $\alpha_1$, $\alpha_2$ and $\delta$ and hence $\phi_2$ and $\phi_3$ are unknown.

The atmospheric neutrino oscillation data [Tos01, Hai03] indicates that $\theta_{23}$ is near $45^{\rm o}$. The solar and reactor neutrino oscillation data [Ahm01, Bah03, Fak02, Egu03 for example] indicates that $\theta_{12}$ is about $30^{\rm o}$ and the CHOOZ and Palo Verde experiments [Apo99, Boe01] indicate that $\theta_{13}$ is near $0^{\rm o}$. In addition, these experiments have indicated values of $\delta m_S^2 \approx 7 \times 10^{-5} \text{eV}^2$ for solar/reactor neutrino oscillations and $\delta m_{AT}^2 \approx 2 \times 10^{-3} \text{eV}^2$ for atmospheric neutrino oscillations.

Although the values of these angles are still fairly uncertain, one can use them to write the Mixing Matrix as:

$$U \cong \begin{bmatrix} c_{12} & s_{12} & 0 \\ -s_{12}c_{23} & c_{12}c_{23} & s_{23} \\ s_{12}s_{23} & -c_{12}s_{12} & c_{23} \end{bmatrix} \cong \begin{vmatrix} \frac{\sqrt{3}}{2} & \frac{1}{2} & 0 \\ -\frac{1}{2\sqrt{2}} & \frac{\sqrt{3}}{2\sqrt{2}} & \frac{1}{\sqrt{2}} \\ \frac{1}{2\sqrt{2}} & -\frac{\sqrt{3}}{2\sqrt{2}} & \frac{1}{\sqrt{2}} \end{vmatrix}. \tag{4}$$

Where we have suppressed the diagonal phase matrix for the time being, and $c_{12} \cong \sqrt{3}/2$, $s_{12} = 1/2$, and $c_{23} = 1/\sqrt{2} = s_{23}$ were used for the numerical values for the second matrix.

### 2.1.3 Neutrino Mass Patterns

The measured values of $\delta m_S^2$ (solar) and $\delta m_{AT}^2$ (atmospheric) given earlier motivate the pattern of masses in two possible hierarchy schemes shown in Fig. 2-1.

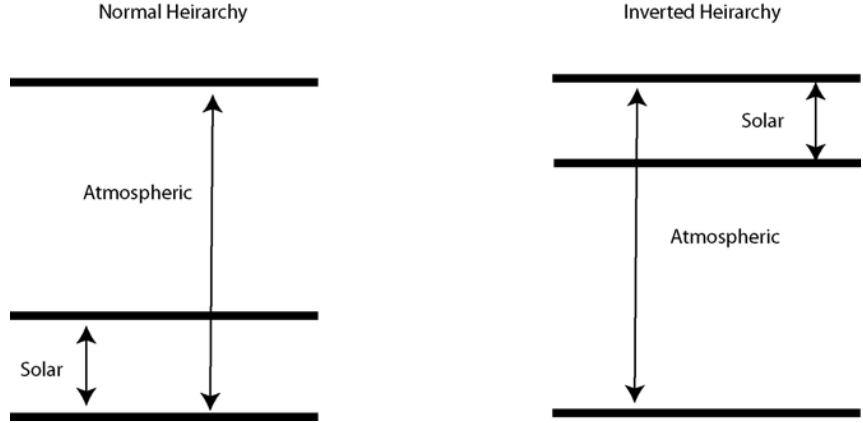

**Figure 2-1 Normal and inverted mass hierarchies. In both cases, our notation defines the lightest mass as m₁ and the heaviest as m₃.**

Defining the lightest mass as $m_1$ and the heaviest mass as $m_3$, we can write $m_2 = \sqrt{\sqrt{\delta m_S^2 + m_1^2}}$ and $m_3 = \sqrt{\sqrt{\delta m_{AT}^2 + m_1^2}}$ in the case of normal hierarchy and $m_2 = \sqrt{\sqrt{\delta m_{AT}^2 - \delta m_S^2 + m_1^2}}$ and $m_3 = \sqrt{\sqrt{\delta m_{AT}^2 + m_1^2}}$ in the case of inverted hierarchy. From these we can write $|\langle m_\nu \rangle|$, for normal and inverted hierarchy respectively, in terms of mixing angles, $\delta m_S^2$, $\delta m_{AT}^2$, and CP phases as [Bar02, Pas02]:

$$\left| \langle m_\nu \rangle \right| = \left| c_{13}^2 c_{12}^2 m_1 + c_{13}^2 s_{12}^2 e^{i\phi_2} \sqrt{\delta m_S^2 + m_1^2} + s_{13}^2 e^{i\phi_3} \sqrt{\delta m_{AT}^2 + m_1^2} \right| \tag{5}$$

$$\left| \left\langle m_\nu \right\rangle \right| = \left| \; s_{13}^2 m_1 + c_{13}^2 c_{12}^2 e^{i\phi_2} \sqrt{\delta m_{AT}^2 - \delta m_s^2 + m_1^2} + c_{13}^2 s_{12}^2 e^{i\phi_3} \sqrt{\delta m_{AT}^2 + m_1^2} \; \right| \tag{6}$$

With the approximation $\theta_{13} \equiv 0$ and the further approximation of $\delta m_S^2 << \delta m_{AT}^2$, equations (5) and (6) can be written as follows in equations (7) and (8) respectively:

$$\left| \left\langle m_\nu \right\rangle \right| = m_1 \left| \; c_{12}^2 + s_{12}^2 e^{i\phi_2} \left( 1 + \frac{\delta m_s^2}{2m_1^2} \right) \; \right| \tag{7}$$

$$\left| \left\langle m_\nu \right\rangle \right| = \sqrt{m_1^2 + \delta m_{AT}^2} \; \left| \; c_{12}^2 e^{i\phi_2} + s_{12}^2 e^{i\phi_3} \; \right| \tag{8}$$

These approximate expressions are accurate to a few percent. It should be noted here that the observed value of $\theta_{12} \sim 30$ deg, together with the values for $\delta m_s$ and $\delta m_{AT}$ are crucial for making the effective mass observable by realistic $0\nu\beta\beta$ detectors even in the small $m_1$ region. In case of the normal hierarchy, the effective mass can be an order of 10 meV even with $m_1 << 10$ meV, and in case of an inverted hierarchy, it can be an order of 45 meV. Numerical values for $\left| \left\langle m_\nu \right\rangle \right|$ are obtained from the equations (7) and (8) by using the observed value of $\theta_{12} = 30$ deg and the central values for the $\delta m^2$ as summarized above, and are given in Table 1. In Fig. 2-2, the range of possible values of $\left| \left\langle m_\nu \right\rangle \right|$ are shown for the general case that includes CP violation. Qualitatively, one should consider $\sqrt{\delta m_{At}^2} \approx 45$ meV as the physics driver for the next generation of experiments.

**Table 2-2-1 Approximate numerical predictions of $\left| \left\langle m_\nu \right\rangle \right|$ in milli-electron volts for both hierarchies and CP phase relations.**

| Normal Hierarchy | | | | Inverted Hierarchy | | | |
|---|---|---|---|---|---|---|---|
| $e^{i\phi_2} = -1$ | | $e^{i\phi_2} = +1$ | | $e^{i\phi_2} = -e^{i\phi_3}$ | | $e^{i\phi_2} = +e^{i\phi_3}$ | |
| $m_1$ meV | $\left| \left\langle m_\nu \right\rangle \right|$ | $m_1$ meV | $\left| \left\langle m_\nu \right\rangle \right|$ | $m_1$ meV | $\left| \left\langle m_\nu \right\rangle \right|$ | $m_1$ meV | $\left| \left\langle m_\nu \right\rangle \right|$ |
| 20 | 10 | 20 | 20 | 0 | 22 | 0 | 45 |
| 60 | 30 | 60 | 60 | 30 | 27 | 30 | 54 |
| 100 | 50 | 100 | 100 | 100 | 55 | 100 | 110 |
| 200 | 100 | 200 | 200 | 200 | 103 | 200 | 205 |
| 400 | 200 | 400 | 400 | 400 | 201 | 400 | 403 |

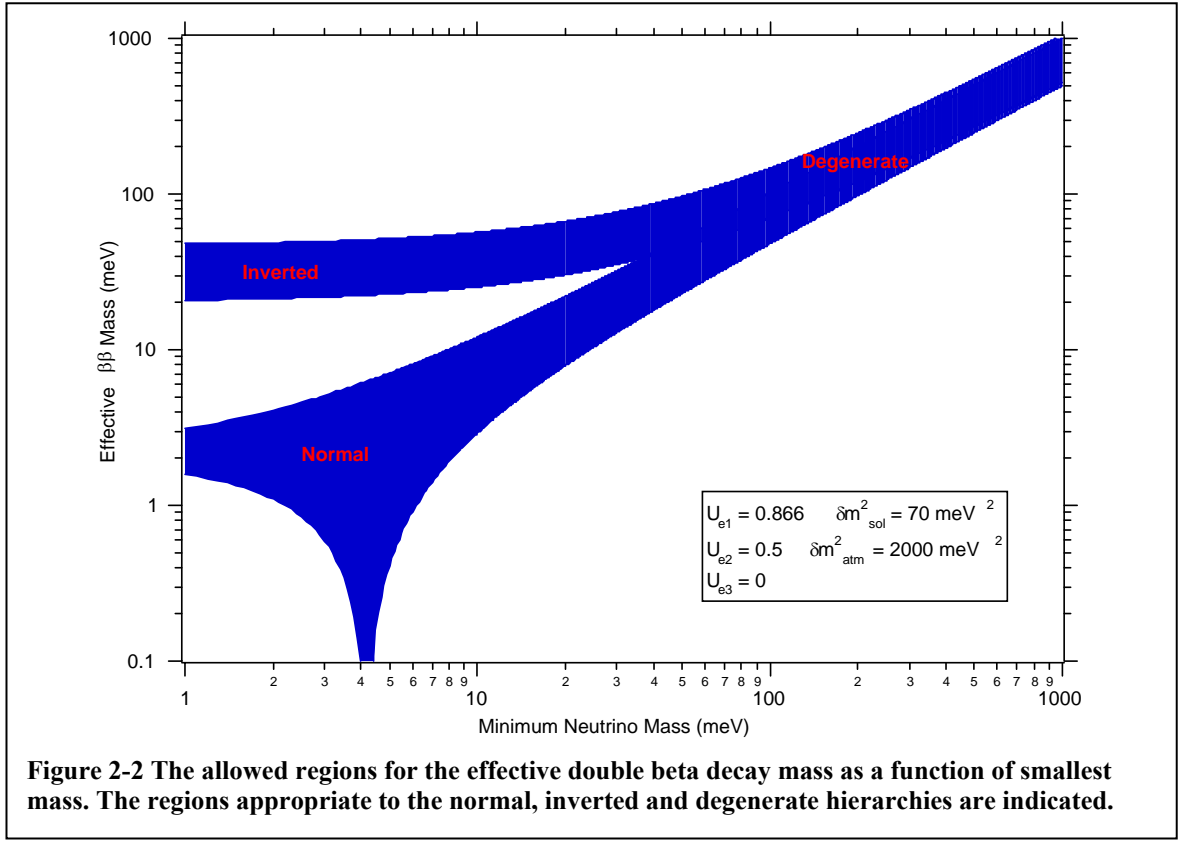

**Figure 2-2 The allowed regions for the effective double beta decay mass as a function of smallest mass. The regions appropriate to the normal, inverted and degenerate hierarchies are indicated.**

Barger *et al.* [Bar02] give equations that relate constraints on the lightest neutrino mass for a given value of $|\langle m_\nu \rangle|$. Using the approximations given above, their equations reduce to the following for normal and inverted hierarchies respectively.

$$|\langle m_\nu \rangle| \leq m_1 \leq \frac{|\langle m_\nu \rangle|}{c_{12}^2 - s_{12}^2} \tag{9}$$

$$\sqrt{|\langle m_\nu \rangle|^2 + \delta m_{AT}^2} \leq m_1 \leq \frac{\sqrt{|\langle m_\nu \rangle|^2 + \delta m_{AT}^2 (c_{12}^2 - s_{12}^2)}}{c_{12}^2 - s_{12}^2} \tag{10}$$

Barger *et al.* also derive similar constraints for the sum of the neutrino masses, $\Sigma \equiv m_1 + m_2 + m_3$, which is important in the consideration of neutrino hot dark matter.

$$2|\langle m_\nu \rangle| + \sqrt{|\langle m_\nu \rangle|^2 \pm \delta m_{AT}^2} \leq \Sigma \leq \frac{2|\langle m_\nu \rangle| + \sqrt{|\langle m_\nu \rangle|^2 \pm \delta m_{AT}^2 \cos(2\theta_{12})}}{|\cos(2\theta_{12})|} \tag{11}$$

Where the plus signs are for normal hierarchy and the minus signs for inverted hierarchy. Equation (11) can be simplified significantly for values of $|\langle m_\nu \rangle|$ achievable in next generation experiments [Avi02]. When $\delta m_{AT}^2 << \Sigma^2$, $\delta m_{AT}^2 \leq 0.005 \; eV^2$ (99.73% CL) and $\cos(2\theta_3)$=0.5, we have:

$$\left|\langle m_\nu\rangle\right| \;\leq\; \frac{\Sigma}{3} \;\leq\; 2\left|\langle m_\nu\rangle\right|. \tag{12}$$

Thus a $\beta\beta$ experiment with a mass sensitivity of $\left|\langle m_\nu\rangle\right|\sim 30$ meV can define the sum $\Sigma$ in the region of $0.1-0.2$ eV. It is evident that next generation neutrinoless double-beta decay experiments are the next important step necessary for a more complete understanding of the physics of neutrinos. In this proposal we describe the Majorana Experiment, show how it could reach the required sensitivity, and give a detailed work breakdown, cost estimate, and list of project milestones.

## 2.2 The Matrix Elements

The observation of $0\nu\beta\beta$ would have profound qualitative physics conclusions associated with it. However, to quantitatively interpret those results in terms of the effective Majorana neutrino mass, one needs a nuclear matrix element. Although the quasiparticle random phase approximation (QRPA) calculations are currently the most popular, it is hoped that the improved shell model techniques that have recently been developed will be brought to bear on this problem as the interest in $0\nu\beta\beta$ intensifies.

Because double-beta decay always results in the emission of two electrons, and because these electrons travel very short distances ($\sim$1 mm) in germanium, $0\nu\beta\beta$ should appear as a sharply defined spectral line at the endpoint energy of the decay, 2039 keV, in a high-resolution germanium spectrometer.

To convert a measured rate of $0\nu\beta\beta$ to a mass value requires nuclear matrix elements, which must be calculated with specific nuclear models. Using such matrix elements, one can compare the values given in Table 1 with the projected sensitivity of the Majorana $^{76}$Ge experiment. Majorana is proposed as a 500 kg Ge array, isotopically enriched to 86% in $^{76}$Ge. The projected sensitivity, discussed in detail later in this proposal, is $T_{1/2}^{0\nu}\geq 4\times 10^{27}\,y$. To do this, we need the nuclear structure factor given in equation (3). It is convenient to define $F_N \equiv G^{0\nu}\mid M_f^{0\nu}-(g_A/g_V)^2\,M_{GT}^{0\nu}\mid^2$ which leads to $\left|\langle m_\nu\rangle\right|=(F_N T_{1/2}^{0\nu})^{-1/2}\,\mathrm{eV}.$

**Table 2-2 Values of the nuclear structure parameter $F_N$ calculated with different nuclear models. The deduced effective Majorana mass of the electron neutrino for each mode is given for a half-life of $4\times10^{27}$ y.**

| $F_N$ (y$^{-1}$) | $\langle m_\nu \rangle$ eV | Reference | $F_N$ (y$^{-1}$) | $\langle m_\nu \rangle$ eV | Reference |
|---|---|---|---|---|---|
| $1.58\times10^{-13}$ | 0.020 | [Hax84] | $8.27\times10^{-14}$ | 0.028 | [Bar99] |
| $2.88\times10^{-13}$ | 0.015 | [Tom86] | $6.19\times10^{-14}$ | 0.032 | [Sim99] |
| $1.12\times10^{-13}$ | 0.024 | [Mut89] | $2.11\times10^{-13}$ | 0.018 | [Sim99] |
| $1.12\times10^{-13}$ | 0.024 | [Sta90] | $1.16\times10^{-13}$ | 0.024 | [Sto00] |
| $1.18\times10^{-13}$ | 0.024 | [Tom91] | $5.22\times10^{-14}$ | 0.035 | [Suh00] |
| | | | $2.70\times10^{-15}$- $3.2\times10^{-15}$ | 0.155- 0.143 | [Bob01] |
| $6.97\times10^{-14}$ | 0.031 | [Suh92] | | | |
| $7.51\times10^{-14}$ | 0.029 | [Suh92] | $1.80\times10^{-14}$- $2.2\times10^{-14}$ | 0.060- 0.054 | [Bob01] |
| $1.90\times10^{-14}$ | 0.059 | [Cau96] | $5.50\times10^{-14}$- $6.3\times10^{-14}$ | 0.034- 0.032 | [Bob01] |
| $1.42\times10^{-14}$ | 0.068 | [Pan96] | $1.21\times10^{-14}$ | 0.073 | [Sto01a] |
| $7.33\times10^{-14}$ | 0.030 | [Pan96] | $1.85\times10^{-14}$ | 0.059 | [Sto01a] |

In 1986, Vogel and Zirnbauer introduced the Quasi-Particle Random Phase Approximation (QRPA) [Vog86]. Since then, there have been many developments and variations, frequently with widely disparate results. Frequently, bounds on $\langle m_\nu \rangle$ are extracted from experimental limits on 0νββ half-lives using nuclear matrix elements from all or many available nuclear models. The results can vary by factors of three or more. This is not satisfactory because it does not account for theoretical progress. An example of the variation in extracted values is clearly seen in Table 2-2.

### 2.2.1 The Shell Model[2]

Germanium-76 is a good isotope for 0νββ studies because the matrix element calculations are more tractable for this relatively low-A isotope. It is anticipated that future shell model calculations for this isotope will be very reliable. The shell model interactions generally are based on G-matrices coming from realistic nucleon-nucleon interactions, with small phenomenological terms that are fit and well constrained by data other than double-beta decay. As single particle energies are also fit, nothing remains to be adjusted for ββ.

A full-shell calculation, in which the sum over huge intermediate spaces of $1^+$ states is done exactly by Lanczos moments techniques, has been done for the neighboring 2νββ nucleus $^{82}$Se by Caurier *et al.* [Cau96] The appropriate model space is $1f_{5/2}$-$2p_{3/2}$-$2p_{1/2}$-$1g_{9/2}$. The corresponding $^{76}$Ge calculation was done as a series, increasing the number (t) of particles allowed in the $g_{9/2}$ shell. Results were obtained for t = 0, 2, 4. Clearly it would be best to allow any $g_{9/2}$ occupation, but that produces a very large model space (at least for the computer capabilities in 1996). But t = 4 is crucial because there are three strongly mixed $0^+$ bands near the $^{76}$Ge ground state, and these correspond to 0, 2, and 4 neutrons being promoted to the $g_{9/2}$ shell. We know these bands strongly mix because of studies of Ge isotopes as a function of neutron number, and show dramatic level-crossing effects in which spectroscopic factors of the ground and excited state "reverse" with the addition of two neutrons. The standard QRPA calculation lacks the essential t = 4 band. The resulting $2\nu\beta\beta$ rates are reasonable, and converge toward the experimental value with increasing t. As expected from the argument above, the change from t = 2 to t = 4 is significant for the $2\nu\beta\beta$ rate.

---

[2] We wish to acknowledge a large contribution to this section by Prof. Wick Haxton, Univ. of Washington.

Excitingly, these calculations can be improved. Progress in standard shell model work has advanced since 1996. New methods [Ohs02] might even be able to handle the full-shell $^{76}$Ge calculation now. These calculations are only as good as the input effective interaction, which generally are determined empirically. In 2002 Honma *et al.* [Hon02] did the analog of Brown-Wildenthal for the $f_{7/2}$-$f_{5/2}$-$p_{3/2}$-$p_{1/2}$ shell, fitting over 600 matrix elements empirically. This is not exactly what is needed for double-beta decay, but it shows that we are getting very close to a Brown-Wildenthal style interaction for $^{76}$Ge.

The use of a shell-model space implies nontrivial wave function normalizations and effective operator contributions due to the neglected high-momentum shells. There is real progress in tackling this problem (see for example Ref. [INT99]) using the theory of effective operators. It also may be possible to "sneak up" on this issue by doing test cases in much lighter nuclei, then applying the deduced effective operators to heavier cases like $^{76}$Ge. Effective operator theory is almost hopeless except in the case of full-shell shell-model calculations. Thus the progress in shell model calculations lays the groundwork for the application of effective operator theory.

Ge is a critical double-beta decay isotope. Generally, calculations for large-A nuclei require severe shell model truncations and employ effective interactions that have not been carefully constrained to data, in the manner of Brown-Wildenthal. Both full-shell calculations and Brown-Wildenthal -style interactions are unlikely for heavier nuclei in the foreseeable future. But such calculations are likely to exist for $^{76}$Ge. If $0\nu\beta\beta$ is observed, the matrix element calculation will be a critical input into the interpretation of the result. It is very likely that confidence in the calculations for this isotope will be high. The future for shell-model calculations is not so bright for the other high-A isotopes frequently considered for $0\nu\beta\beta$ experiments.

### 2.2.2    The Quasiparticle Random Phase Approximation

The practice of using all previously calculated values for $F_N$ to determine a spread, and therefore uncertainty, in the matrix elements ignores recent progress in the field. Doing so results in a factor of 10 uncertainty in $F_N$ and therefore a factor of ~3 in the determination of the neutrino mass. Recent improvements have shown the consistency between the various QRPA calculations. In fact, in addition to the advances in the shell model efforts described above, recent QRPA work is also very encouraging for double-beta decay in general and for $^{76}$Ge specifically.

Previously, an accurate calculation of the $2\nu\beta\beta$ matrix elements has been considered a necessary but not sufficient condition to cross-check the machinery used to determine $0\nu\beta\beta$ matrix elements, because the intermediate nuclear states are very different in the two cases. Recently however, Rodin, *et al*. showed, in the context of QRPA and Renormalized QRPA (RQRPA), that this is not the case [Rod03]. They make a well-documented case that:

> "*When the strength of the particle-particle interaction is adjusted so that the $2\nu\beta\beta$ decay rate is correctly reproduced, the resulting $M^{0\nu}$ values become essentially independent on the size of the basis, and on the form of different realistic nucleon-nucleon potentials. Thus, one of the main reasons for variability of the calculated $M^{0\nu}$ within these methods is eliminated*"

Accordingly, one would conclude that accurate measurements of $2\nu\beta\beta$ half-lives **will** have a very meaningful impact on the predictions of $0\nu\beta\beta$ matrix elements in the same nuclei. Contrary to previous conventional wisdom, accurate $2\nu\beta\beta$ measurements may now be very important in the realm of neutrino physics. A large experiment like Majorana will measure the $2\nu\beta\beta$ half-live very accurately and thus provide this needed benchmark.

Specifically, Rodin *et al*., investigated the dependence of $M^{0\nu}$ on the choice of the single-particle (sp) space by comparing three different, yet each realistic, nucleon-nucleon interactions including: the Bonn-CD [Mac89], the Argonne [Wir95], and the Nijmegen [Sto94] potentials. The result is that $M^{0\nu}$ varies very little over the considered 9 different combinations of sp-space and interactions. The effect of neglecting single-particle states far from the Fermi-level was investigated for $^{76}$Ge, $^{100}$Mo, $^{130}$Te, and $^{136}$Xe. In the case of interest here, $^{76}$Ge, the following three sp-spaces were used: 1) the 9 levels of the oscillator shells N=3 and 4; 2) the addition of the N=2 shell, and finally; 3) the 21 levels from all states in the shells with N=1, 2, 3, 4, and 5. For each change in sp-space, the residual interaction must be adjusted by adding a pairing interaction and a particle-hole interaction renormalized by an overall strength parameter, $g_{ph}$. The value $g_{ph}\sim1$ was found to reproduce the giant Gamow-Teller resonance in all cases. Finally, QRPA equations include the effects of a particle-particle interaction, renormalized by an overall strength parameter $g_{pp}$ that in each case was adjusted to reproduce the known $2\nu\beta\beta$ rate correctly. This final adjustment was found to be key in producing very similar results from all the chosen basis or interaction. Thus the long-standing spread in QRPA calculations can now be considered greatly narrowed.

### 2.2.3 *Effective Majorana Mass Sensitivity*

In the notation of [Rod03], $\langle m_\nu \rangle = 1/|M^{0\nu}| (G^{0\nu} T_{1/2})^{1/2}$. They give $|M^{0\nu}| = 2.40 \pm 0.07$ (RQRPA) and $|M^{0\nu}| = 2.68 \pm 0.06$ (QRPA) with $G^{0\nu} = 0.30\times10^{-25}$ y$^{-1}$eV$^{-2}$. If we choose the round number, $T_{1/2} = 4\times10^{27}$ y for the predicted sensitivity of the Majorana experiment, then the deduced values of the mass parameter corresponding to this half-life are; $\langle m_\nu \rangle = 38 \pm 7$ meV using RQRPA and $\langle m_\nu \rangle = 34 \pm 6$ meV with QRPA. A very similar

value, $\langle m_\nu \rangle = 28 \pm 5$ meV, results from using the matrix elements from the recent paper by Civitarese and Suhonen [Civ03]. It is clear that not only is this spread much smaller than previously assumed, but the matrix element is also rather large. Both of these conclusions are exciting for a double-beta decay experiment using $^{76}$Ge.

Effective Majorana neutrino mass values near 30 meV are well into the range of interest tabulated in Table 2-1. This implies that the Majorana experiment has a sensitivity that reaches well into the interesting range of neutrino mass.

## 2.3 Completed Double-Beta Decay Experiments

The first laboratory search for double-beta decay was made in 1948 by Fireman [Fir48]. The experiment involved a search for coincident pulses in Geiger counters in proximity to a source of $^{124}$Sn and an apparent positive signal was observed. At that time, the Standard Model of Particle Physics did not incorporate parity violation in the weak interaction. Thus this result was assumed to be an observation of $0\nu\beta\beta$ mediated by Majorana neutrinos because, due to phase space, it was expected to have a decay rate 7-9 orders of magnitude greater than the $2\nu\beta\beta$ mode. Subsequent experiments [Law51, Fir52, Kal52] contradicted this positive signal and established lower limits for the $^{124}$Sn half life in the range of $10^{16}$ to $2\times10^{17}$ y.

The existence of double-beta decay was first claimed in a series of geochronological experiments by Inghram and Reynolds [Ing50] in 1950 using $^{130}$Te. These results were confirmed by Takaoka and Ogata [Tak66] in 1966 and again by Kirsten, *et al.* [Kir67a] in 1967. Kirsten, Gentner, and Schaeffer [Kir67b] also reported measurement of double-beta decay for $^{82}$Se in 1967. These experiments relied on mass-spectrometric measurements of the noble gas daughters entrained in very old ores. Excesses of $^{130}$Xe and $^{82}$Kr were used to determine the double-beta decay half-lives from ores that were independently dated by other techniques. While these measurements unequivocally demonstrated that double-beta decay was a real phenomenon, nothing could be inferred about the particular mode of double-beta decay responsible for the buildup of daughter products.

The ingenious utilization of a high-resolution germanium diode gamma-ray spectrometer as both the source and detector for a double-beta decay experiment was introduced by Fiorini and colleagues [Fio67] in 1967. They were able to assign a limit to the $^{76}$Ge neutrinoless double-beta decay mode of $T_{1/2} > 2\times10^{20}$ y.

The first direct laboratory observation of double-beta decay was reported by Elliott, Hahn, and Moe [Ell87] in 1987. They used a Time Projection Chamber to measure the two-neutrino double-beta events from a source consisting of 14 g of 97% isotopically enriched $^{82}$Se contained between thin aluminized Mylar sheets. Their value of $T_{1/2} = 1.1\times10^{20}$ y was in excellent agreement with the geochronological half-life reported earlier for this isotope.

In 1988, Avignone and Brodzinski [Avi88] in a review article reported on the use of an isotopically enriched germanium spectrometer by the ITEP-EREVAN group, and predicted that the combination of large isotopically enriched germanium spectrometers coupled with application of good background-reduction practices would ultimately lead to a sensitivity for the effective electron neutrino mass of a few tens of meV. Interestingly, this prediction corresponds precisely with the now-known requisite mass range based on the atmospheric and solar oscillation results and with the mass range attainable by this proposed Majorana Collaboration experiment.

The first reported measurements of the two-neutrino half-life for $^{76}$Ge were made in 1990 by Vasenko, *et al.* [Vas90] and by Miley, *et al.* [Mil90], later confirmed by Avignone *et al.* [Avi91]. The Russian collaboration measurements were made using the isotopically enriched detector referred to above and were in substantial agreement with the U.S.-based measurement, which was determined from data acquired with two 1-kg natural isotopic detectors. The reported half lives were $T_{1/2} = 0.92 \times 10^{21}$ y and $1.1 \times 10^{21}$ y, respectively. The confirmation experiment utilized one of the small Russian detectors enriched to 86% in $^{76}$Ge. These data were later corrected for the backgrounds from radioactive isotopes created in the Ge by cosmic ray generated neutrons, yielding $T_{1/2} = 1.27 \times 10^{21}$ y [Avi94]. The Heidelberg-Moscow collaboration reported a $2\nu\beta\beta$ $T_{1/2} = 1.77 \times 10^{21}$ y [Gun97]. In 2001 Heidelberg-Moscow collaboration published new value for $T_{1/2} = 1.55 \times 10^{21}$ y [Kla01b]. With uncertainties quoted at about 10%, these two results are in disagreement.

Additional direct measurements of double-beta decay were reported for $^{100}$Mo in 1991 by Elliott, *et al.* [Ell91] and by Ejiri, *et al.* [Eji91] and for $^{150}$Nd in 1993 by Artem'ev *et al.* [Art93] and by Elliott, *et al.* [Ell93]. Since those early days, many other isotopes have had their $2\nu\beta\beta$ ground-state half-lives measured. The virtually identical results for the two-neutrino double-beta decay of $^{100}$Mo to the ground state of $^{100}$Ru, $T_{1/2} = 1.16(1.15) \times 10^{19}$ y, was followed by a direct measurement of the double-beta decay of $^{100}$Mo to the first excited $0^+$ state in $^{100}$Ru by Barabash, *et al.* [Bar95] in 1995. The double-beta decay to the 1130.29-keV state was observed by single-gamma measurements of the cascade de-excitation γ rays at 539.53 and 590.76 keV from a 956-g sample of 98.468% isotopically enriched $^{100}$Mo metal powder. The resulting half-life was determined to be $T_{1/2} = 6.1 \times 10^{20}$ y. The result was confirmed by a γ–γ coincidence experiment by De Braeckeleer *et al.* [Deb01].

Table 2-3 summarizes the best past $0\nu\beta\beta$ half life limits and deduced effective Majorana neutrino mass limits. The most restrictive limits come from the Ge experiments. All the small-scale $^{76}$Ge double-beta decay experiments have now been officially terminated, with the current neutrinoless half-life limit $T_{1/2} > 1.9 \times 10^{25}$ y by the Heidelberg-Moscow collaboration [Kla01b] and $T_{1/2} > 1.6 \times 10^{25}$ y by the IGEX collaboration [Aal02]. The longest half-life bound corresponds to an effective Majorana neutrino mass of the electron neutrino of 0.3 - 1 eV, depending on the theoretical nuclear matrix elements chosen. If one analyzes the data with matrix elements from the most recent QRPA calculations this range is ~0.3-0.5 eV. To become sensitive to a neutrino mass an order of magnitude or more smaller will require a large increase in the scale of a double-beta

decay experiment with further reduction of background; precisely the improvements described in this Majorana Collaboration document.

**Table 2-3 Best reported limits on 0νββ half lives. The mass limits and ranges are those deduced by the authors and their choices of matrix elements within the cited experimental papers. All are quoted at the 90% confidence level except as noted.**

| Isotope | Half-life Limit (y) | $|<m_\nu>|$ limit (eV) | Reference |
|---------|--------------------|-----------------------|-----------|
| Ca-48 | $>9.5 \times 10^{21}$ (76%) | <8.3 | You91 |
| Ge-76 | $>1.9 \times 10^{25}$ | <0.35 | Kla01b |
| | $>1.6 \times 10^{25}$ | $<0.33 - 1.35$ | Aal02 |
| Se-82 | $>2.7 \times 10^{22}$ (68%) | <5 | Ell92 |
| Mo-100 | $>5.5 \times 10^{22}$ | <2.1 | Eji96 |
| Cd-116 | $>7 \times 10^{22}$ | <2.6 | Dan00 |
| Te-128,130 | From ratio of $T_{1/2}$s | $<1.1 - 1.5$ | Ber93 |
| Te-128 | $>7.7 \times 10^{24}$ | $<1.1 - 1.5$ | Ber93 |
| Te-130 | $>1.4 \times 10^{23}$ | $<1.1 - 2.6$ | Ale00 |
| Xe-136 | $>4.4 \times 10^{23}$ | $<1.8 - 5.2$ | Lue98 |
| Nd-150 | $>1.2 \times 10^{21}$ | <3 | Des97 |

## 2.4 The Majorana Background Model

To estimate the sensitivity of the Majorana Experiment, we require a background model. In Section 3.2, we describe the background model in great detail. Here we only summarize the results of that section that are necessary for the sensitivity estimate given below. Table 2-4 summarizes the anticipated contribution to the background due to cosmogenic activities inside the Ge. This contribution to the background has been determined to be the limiting background for the experiment with other backgrounds having a smaller impact. Section 3.2 discusses the background from other such sources. The raw rates used in this table correspond to those determined from previous data [Bro95].

This model incorporates the decay of these isotopes and the anticipated run plan. It also incorporates the anticipated rejection of background due to pulse shape discrimination and detector segmentation.

**Table 2-4 Estimation of sources of activity from early IGEX data within the 3.6-keV region of interest about 2039 keV.**

| Spallation Isotope | $T_{1/2}$ (d) | Rate from [Bro95] | After Construction | Rate During Experiment | Total in ROI | After PSD Rejection | After Seg Rejection |
|---|---|---|---|---|---|---|---|
| $^{68}$Ge | 270.82 | 0.156 | 0.037 | $8.0 \times 10^{-3}$ | 71.3 | 18.9 | 2.6 |
| $^{56}$Co | 77.27 | 0.024 | 0.002 | $1.3 \times 10^{-4}$ | 1.1 | 0.30 | 0.04 |
| $^{60}$Co | 1925.2 | 0.018 | 0.013 | $9.4 \times 10^{-3}$ | 83.7 | 22.2 | 3.1 |
| $^{58}$Co | 70.82 | 0.0024 | 0.0002 | $1.12 \times 10^{-5}$ | 0.10 | 0.03 | 0.00 |
|  |  | cts/keV/kg/y | cts/keV/kg/y | cts/keV/kg/y | Counts | Counts | Counts |
| Total |  | 0.2 | 0.052 | 0.017 | 156 | 41.4 | 5.7 |

## 2.5 Ultimate Sensitivity of the Majorana Experiment

The number of $^{76}$Ge atoms in 500 kg of enriched germanium (86% $^{76}$Ge) is $N = 3.429 \times 10^{27}$. The endpoint energy of the $0\nu\beta\beta$ transition is well known as 2039.006(50) keV [Dou01]. The energy resolution is expected to be ~0.15% FWHM at 2039 keV. A choice of region-of-interest width of approximately 2.8 $\sigma$ is ideal for maximizing signal to background [Aar03]. The resulting energy window of $\delta E = 3.6$ keV is expected to capture 83.8% of the events in a sharp peak at the endpoint. The ultimate sensitivity of the Majorana experiment depends on the signal and background rates. In this section the background rates are evaluated, for simplicity, for the major sources. The details of all the possible background sources are discussed in Section 3.2. In 5 years, we would expect to observe 156 background counts within the energy window.

The next step in estimating the sensitivity of the experiment is to apply two new but easily implemented techniques. The first is an experimentally demonstrated technique to measure the multiplicity of energy depositions by analyzing digitized current pulses using a robust, self-calibrating technique. This method has been shown to accept $\varepsilon_{PSD} = 80.2\%$ of single site pulses (like double-beta decay) and to reject 73.5% of background associated with gamma rays. In fact the reduction factor depends on the type of the decay pattern. The second technique involves the electrical segmentation of the detector crystal to form many small segments within a crystal. A simple Monte Carlo analysis of this configuration was carried out only to count the segments with significant energy deposition and reject events with a multiplicity > 1. This cut accepted $\varepsilon_{SEG} = 90.7\%$ of double-beta decay pulses and rejected ~86% of backgrounds such as internal $^{60}$Co and $^{68}$Ge, which are highly multiple.

Applying the background reduction factors to the simple calculation above, only 5.7 counts of the original 156 counts survive in our 3.568 keV analysis window, a reduction of 96.3% or a factor of 27.3. Because pulse shape discrimination (segmentation) tends to identify radial (azimuthal and axial) differences due to multiple energy deposits, the two cuts are orthogonal and therefore the total rejection is just the product of the individual background rejection factors. Here we have assumed that the reduction rates are the same for the major background sources. In fact, they depend on the decay pattern of the isotope in question and its location. The overall background rate, however, is considered to be close to the present evaluation of 5.7, as discussed in Section 3.2.

For a positive signal at the 90% CL, we would then need to observe 9 counts ($L_c = 9$ actually yields 93.4% CL). This is an additional 3.3 counts over the expected 5.7 background events. Computing the 0νββ half-life must then take into account this number of observable counts, the cut efficiencies, and the fraction of the 0νββ peak found in the analysis window. Thus

$$T_{1/2} = \frac{\ln(2) \cdot N \cdot \Delta t \cdot \varepsilon_{PSD} \cdot \varepsilon_{SEG} \cdot 83.8\%}{3.3} = 4.3 \times 10^{27} \, y \,.$$

Our formulation for the effective Majorana mass of the electron neutrino is

$$\langle m_\nu \rangle = \frac{m_e}{\sqrt{F_N \cdot T_{1/2}}}$$

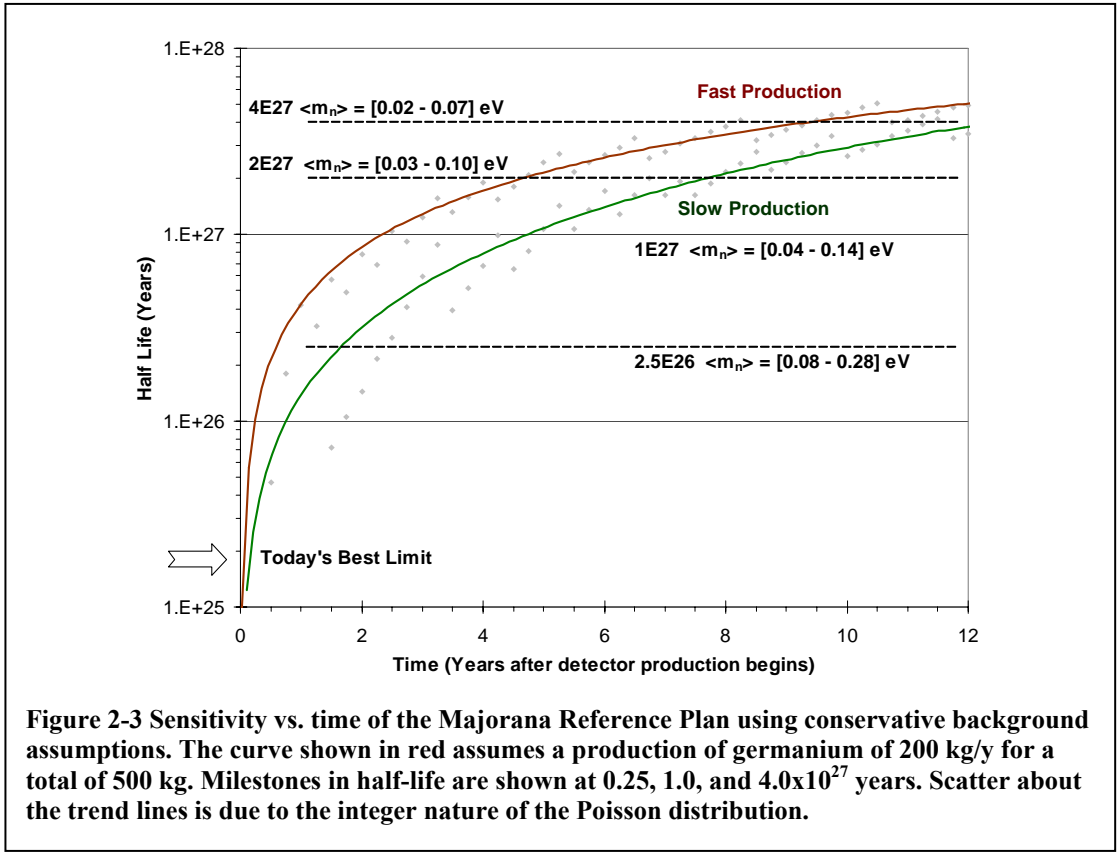

**Figure 2-3 Sensitivity vs. time of the Majorana Reference Plan using conservative background assumptions. The curve shown in red assumes a production of germanium of 200 kg/y for a total of 500 kg. Milestones in half-life are shown at 0.25, 1.0, and 4.0x10$^{27}$ years. Scatter about the trend lines is due to the integer nature of the Poisson distribution.**

where $F_N$ is a set of nuclear factors adopted from various authors[Aal99a]. Using the recent values for $F_N$ as given in Table 2.2, one gets a range of observable effective Majorana neutrino mass from 24 to 31 meV. In fact matrix elements calculated in past scatter in a wider range, and accordingly the observable mass range gets larger. It is noted that the range of the values of $F_N$ does not necessarily indicate the uncertainty of the calculations.

The calculations in this section have covered in some detail the effects of backgrounds on a 2500 kg-y experiment in which the mass is 500 kg and the time is 5 years. It is assumed that the $^{enr}$Ge production rate will be 200 kg/y and that as detectors become operable, we will begin data acquisition. Note that even with the modest mass (~50 kg) of the first year's operation, Majorana would greatly surpass the present limit within a few months of operation.

## 2.6 Other Double Beta Decay Processes

**Search for Double-Beta Decay Transitions to Excited States**

Although the matrix element calculations for $0\nu\beta\beta$ and $2\nu\beta\beta$ are different, they have many commonalities. These commonalities permit studies of the $2\nu\beta\beta$ matrix elements, which can be compared to experiment to constrain the calculation techniques of the $0\nu\beta\beta$ matrix elements that have no direct experimental test. In a similar way, studies of $\beta\beta$ transitions to excited-states in the daughter nuclide allow one to obtain supplementary information about $\beta\beta$. It is very important to note that in the framework of QRPA models, the behavior of nuclear matrix elements with respect to the so-called $g_{pp}$ parameter is completely different for transitions to the ground state (a $0^+$ - $0^+$ transition) and those to excited states ($0^+$ - $0^{+*}$ or $0^+$ - $2^{+*}$ transitions) [Gri92, Suh98, Aun96]. As a result, the decay to excited states may probe different aspects of this calculation method than the decay to the ground states. This additional opportunity for insight into the nature of the matrix elements drives the interest in these decay modes.

Because of smaller transition energies, the probabilities for $\beta\beta$ transitions to excited states are substantially suppressed due to the reduced phase space compared to transitions to the ground state. However, the decay of the excited state emits mono-energetic gamma rays that can be detected in coincidence with the $\beta\beta$ electrons. These gamma rays provide a very clear signature of the decay and can greatly increase the sensitivity of the measurement [Bar90]. In effect, these gammas permit the identification of the daughter in real-time coincidence. In the nuclei $^{100}$Mo, $^{96}$Zr, and $^{150}$Nd for example, the excited-state $\beta\beta$ transition energies are large enough (1903, 2202 and 2627 keV, respectively) that the expected half-lives ($10^{20}$-$10^{21}$ y) are potentially detectable. Currently only $^{100}$Mo has had this transition measured [Bar95,Bar99, Deb01] with half-life of (6-9) $\times10^{20}$ y. Recently additional isotopes, $^{82}$Se, $^{130}$Te, $^{116}$Cd and $^{76}$Ge, have also become of interest to studies of the $2\nu\beta\beta$ decay to the $0^{+*}$ level. (See the recent review by Barabash [Bar00].)

Theoretical estimates of the $2\nu\beta\beta$ to a $2^{+*}$ state have shown that for a few nuclei ($^{82}$Se, $^{96}$Zr, $^{100}$Mo, and $^{130}$Te) the half-lives can be as short as $10^{22}$-$10^{23}$ y [Suh98]. Many of the present experimental limits are approaching these theoretically predicted values. This would mean that the detection of such decays becomes possible using the present and new installations in the near future. Table 2-4 summarizes the theoretical calculations.

Until now, attention was concentrated mostly on the $0\nu\beta\beta$ transition to the ground state of the final nucleus. However, there might be a chance that the transitions to the excited $0^{+*}$ and/or $2^{+*}$ final states are favored experimentally, at least for a particular mechanism for $0\nu\beta\beta$ [Bar00]. Generally speaking, transitions to the excited states are suppressed due to the reduced phase space. However, a lower background due to the multi-particle coincidence might compensate this limitation. This potential advantage depends on the ratio of the corresponding nuclear matrix elements to the excited and ground states and the multi-hit background. If the matrix element values are comparable, the $0\nu\beta\beta$ decay experiment measuring transitions to ground and excited final states could have a similar sensitivity to the neutrino mass. A further motivation for the interest in these excited state transitions was described in [Sim02] where it was shown that it is possible to distinguish among the light and heavy Majorana neutrino mass and R-parity breaking SUSY mechanisms of the $0\nu\beta\beta$ decay by studying the transitions to the first excited $0^{+*}$ states.

**Table 2-5 Theoretical estimates of half-lives for $2\nu\beta\beta$ and $0\nu\beta\beta$ to the $2^{+*}$ and $0^{+*}$ excited states of a daughter nuclei in years. Values without references are taken from [Suh98]. For the $0\nu\beta\beta$ , the half-lives are calculated for $|<m_\nu>| = 1$ eV.**

| Isotope | $2\nu\beta\beta$ $0^+$ - $2^{+*}$ | $2\nu\beta\beta$ $0^+$ - $0^{+*}$ | $0\nu\beta\beta$ $0^+$ - $0^{+*}$ |
|---|---|---|---|
| $^{48}$Ca | $5\times10^{26}$ [Hax84] | | |
| $^{76}$Ge | $5.8\times10^{25} - 5\times10^{26}$ | $1.7\times10^{21} - 1.7\times10^{24}$ | $4.9\times10^{26}$ [Suh00] |
| | | | $(2.4\text{-}4)\times10^{26}$ [Sim01] |
| $^{82}$Se | $1.4\times10^{21} - 3.3\times10^{26}$ | $1.4\times10^{21} - 3.3\times10^{21}$ | $9.4\times10^{26}$ [Suh00] |
| | | | $(4.5\text{--}9)\times10^{25}$ [Sim01] |
| $^{96}$Zr | $3.3\times10^{20} - 7.2\times10^{26}$ | $2.1\times10^{20} - 1.5\times10^{22}$ | $2.3\times10^{24}$ [Suh00a ] |
| $^{100}$Mo | $5.3\times10^{20} - 1.1\times10^{26}$ | $5.4\times10^{19} - 5.5\times10^{21}$ | $7.6\times10^{24}$ – |
| | | | $1.5\times10^{26}$[Sim01] |
| $^{116}$Cd | $1.1\times10^{24} - 7.8\times10^{25}$ | $1.1\times10^{22} - 9.5\times10^{25}$ | $1.3\times10^{27}$ |
| $^{124}$Sn | $6.5\times10^{26}$ | $2.7\times10^{21}$ | |
| $^{130}$Te | $3.2\times10^{22} - 2.8\times10^{24}$ | $5.1\times10^{22} - 1.4\times10^{25}$ | |
| | | [Bar01] | |
| $^{136}$Xe | $4\times10^{23} - 5.4\times10^{24}$ | $2.5\times10^{21} - 3\times10^{21}$ | $4.8\times10^{24}$ – |
| | | | $4.8\times10^{26}$[Sim01] |
| $^{150}$Nd | $7.2\times10^{24} - 1.2\times10^{25}$ | $8.6\times10^{21}$ | |

The SEGA detector (described in Section 4) could be used to search for double beta decay of $^{76}$Ge to the excited states of $^{76}$Se. The anticipated half-life sensitivity is $10^{22}$-$10^{23}$ y for 1 year of measurement. This sensitivity is one to two orders of magnitude higher than present limits on $2\nu\beta\beta$ of $^{76}$Ge to the excited states of $^{76}$Se. With the MEGA detector (described in Section 4), the half-life sensitivity will be $\sim10^{23}$-$10^{24}$ y for a search for double-beta decay of $^{76}$Ge to the excited states of $^{76}$Se. This provides a good chance of detection. In addition the best limit on $0\nu\beta\beta$ to the $0^{+*}$ excited state will be obtained.

At the same time, the MEGA detector can be used to study passive samples sited within the array of high purity Ge detectors. Monte-Carlo simulations have shown that this detector will provide efficiency for $\gamma$–$\gamma$ coincidences on the level 1-2%. This will permit study of the $2\nu\beta\beta$ decay to $0^{+^*}$ excited states in all the isotopes mentioned in Table 2-4 with a half-life sensitivity of $10^{22}$-$10^{23}$ y. This opens the possibility that one might detect the $2\nu\beta\beta$ to the $0^{+^*}$ excited states in $^{100}$Mo, $^{96}$Zr, $^{150}$Nd, $^{82}$Se, and, possibly also in $^{124}$Sn, $^{116}$Cd and $^{130}$Te.

Using the Majorana detector the $0\nu\beta\beta$ of $^{76}$Ge to the $0^{+^*}$ excited state of $^{76}$Se will be investigated with a half-life sensitivity $\sim 10^{28}$ y, which corresponds to a sensitivity to neutrino mass $|<m_\nu>|\sim$50-160 meV (depending on the nuclear matrix element used). In this case it is really possible to have a "zero"- background experiment because of the clear signature of the events. Sensitivity with passive samples (up to 10-20 kg) can reach $T_{1/2}\sim 10^{24}$-$10^{25}$ y.

## $\beta^+\beta^+$, $\beta^+$EC, and EC-EC processes

Contrary to the intensive interest in $2\nu\beta\beta$, the $\beta^+\beta^+$, $\beta^+$EC, and EC-EC modes have attracted almost no attention. The $2\nu\beta^+\beta^+$ processes are much slower than $2\nu\beta\beta$ due to the small phase space, and the Coulomb barrier for positrons. However they are attractive from the experimental point of view due to the possibility of detecting the coincidence signal from four (two) annihilation gammas and two (one) positrons, or the annihilation gammas only. The $2\nu$ECEC-process can have a large decay energy (up to $\sim 2.8$ MeV) but the experimental detection for the transition to the ground state is made difficult by the fact that only X-rays are emitted.

Detection of the two-neutrino mode of these processes would provide additional nuclear matrix element information. Such data are very important in view of the need for cross checks of the theoretical calculations for $0\nu\beta\beta$ . If $0\nu\beta\beta$ is ever detected, the experimental results (or even limits) on $0\nu\beta^+$EC half-lives offer a possibility to determine whether the observed decay is dominated by the neutrino mass mechanism or by right-handed week currents [Hir94]. The next generation of low-background experiments can potentially increase the half-life sensitivity for these decays to $\sim 10^{22}$-$10^{23}$ y. This should be sufficient to detect the $2\nu$ECEC ($0^+$ - $0^{+^*}$) process in $^{96}$Ru, $^{106}$Cd, $^{124}$Xe, $^{136}$Ce, and $^{156}$Dy [Bar94]. A sensitivity of $10^{22}$-$10^{23}$ y can be reached for $\beta^+\beta^+$, $\beta^+$EC, and EC-EC – processes using the MEGA detector. With Majorana, a half-life sensitivity of $10^{24}$-$10^{25}$ y can be reached for the double beta decay for $\beta^+\beta^+$, $\beta^+$EC, and EC-EC -processes. Finally, the very rare single beta decays in $^{96}$Zr and $^{48}$Ca might be measured for the first time with MEGA.

## $^{76}$Ge $2\nu\beta\beta$ Spectrum Shape

The energy carried away by the two electrons $2\nu\beta\beta$ is characterized by a continuous energy spectrum out to the endpoint energy of 2039 keV. The spectral shape is determined, to a first approximation, simply by the phase space available in the decay.

Recoil order corrections to the charged nucleon current correct this spectrum by roughly 10%, and have been calculated recently [Barb99]. Decay modes in which majorons are emitted also distort the spectrum, and a search for this mode can be obtained from a simultaneous fit of the residual spectrum to a combination of conventional $2\nu\beta\beta$ and a possible majoron-emitting mode.

The Particle Data Group reports two-neutrino mode measurements in 10 different nuclei [PDG02], with the most precise measurements quoting uncertainties near 10%. The half-life for $^{76}$Ge has been reported four times by two different groups. These measurements are extremely challenging, in that they require a detailed model of the background to accurately extract the spectrum and the half-life. They also represent our most stringent test of the physics models used to calculate double-beta decay.

Using our fiducial mass of $^{76}$Ge ($3.4\times10^{27}$ atoms) and a mean data-taking time of 5 years, we expect a total number of decays of $7.3\times10^{7}$. This represents roughly an order of magnitude improvement over previous measurements in the statistical precision with which $2\nu\beta\beta$ can be measured. In order to perform higher precision measurements of this process, however, these improvements in the statistical uncertainty of the two-neutrino mode must be accompanied by a corresponding, substantial improvement in background rejection. We note that, at present, essentially all very precise measurements of the $2\nu\beta\beta$ are limited by the systematic uncertainties implicit in subtracting background.

The difficulty in measurements of $2\nu\beta\beta$ to the ground state is that the signal presents itself as a smooth continuum. The spectrum peaks at roughly 700 keV. From previous measurements, we expect the signal to be dominated by $2\nu\beta\beta$ decay above roughly 1 MeV, this may not be the case at lower energies. We also note that the dominant contributors to the backgrounds will probably be different than those relevant to the $0\nu\beta\beta$ mode. In particular, the external backgrounds from the copper cryostat components and lead shielding provide a significant source of potential backgrounds. Hence, in order to improve on the effectiveness of previous measurements, we must control and characterize these external sources of background. As we point out elsewhere, we bring new technology to bear on this problem, as well as an overall strategy for background reduction, which should reduce the backgrounds (relative to the $2\nu\beta\beta$ decays) by well over an order of magnitude above 700 keV. Our background reduction strategy for the $0\nu\beta\beta$ measurement has three primary components:

(1) reducing internal Ge and external material radioactivity through improvements in the detector and material processing (e.g. moving the final stages of detector fabrication and cryostat copper purification underground)
(2) utilizing segmentation to eliminate multi-hit events and provide a degree of self-shielding for a large fraction of the detector volume
(3) using pulse-shape discrimination (PSD) to eliminate multi-hit events.

All of these strategies will provide corresponding improvements to our measurements of the $2\nu\beta\beta$ mode as well.

Of primary concern for the 2νββ measurements are the external γ-ray backgrounds, particularly line radiation arising from the cryostat materials and $^{210}$Bi bremsstrahlung from the lead shielding. We note that, although the efficiency of segmentation and PSD are reduced at lower γ energies, the self-shielding due to the outer portions of our detector array is increasingly effective at lower gamma energies. Two-neutrino double-beta decay in the dead layer will have degraded signals. Since the dead layer comprises about 2% of the volume, the study of spectral effects at the few percent level will need to consider the effect of the dead layer.

We also note that our detector will be ideally suited to measure 2νββ decay to the first $0^+$ excited state, through a coincidence measurement between the beta-induced signal and the 1.12-MeV γ      emitted as the excited state decays to the ground state. Our expected high resolution and segmentation will permit an essentially background-free measurement with roughly $5×10^5$ decays recorded. Such a measurement may well provide a more significant test of the nuclear matrix element calculations, recoil-order charged current calculations and majoron-emitting decay modes than measurements of decay to the ground state.

## 2.7 Other Science Applications of the Majorana Experiment

The Majorana experiment is foremost a neutrino mass experiment. However, we will capitalize on its unique capabilities to realize other interesting physics at little added cost. Several examples are discussed here. Care in the construction of the Majorana apparatus should yield significant sensitivity for both of these purposes.

### Other Science: Majorana as a Weakly Interacting Massive Particle detector

Majorana should be able to contribute significantly to dark matter searches. The Majorana sensitivity should be similar and complementary to that of CDMS-II.

Extensive gravitational evidence indicates that a large fraction of the matter in the universe is non-luminous, or "dark" [Ber01]. However, the nature and quantity of the dark matter remain unknown, providing a central problem for astronomy and cosmology [Kol90, Pee93]. Recent measurements of the cosmic microwave background radiation [Ben03, Pry02, Net02], as well as arguments based on big bang nucleosynthesis and the growth of structure in the universe [Sre00], suggest that dark matter is predominantly made up of non-baryonic particles outside the standard model of particle physics. Supersymmetric particle physics models provide a natural candidate for dark matter: the lightest superpartner (LSP), usually taken to be a neutralino with typical mass about 100 GeV/c$^2$ [Jun96, Ell97, Eds97, Bot00, Bot01, Ell02b, Ell02c]. Analysis of experimental bounds from LEP have been shown to give a lower limit of ~50 GeV/c$^2$ for the LSP [Ell02c, Ell00], although treatment of special cases can be shown to permit a mass a factor 10 below this [Bot03].

More generically, one can consider a class of Weakly Interacting Massive Particles (WIMPs) [Lee77], which were once in thermal equilibrium with the early universe, but were "cold," i.e. moving non-relativistically at the time of structure formation. Their density today is then determined roughly by their annihilation rate, with weak-scale interactions if the dark matter is mainly composed of WIMPs. WIMPs are expected to have collapsed into a roughly isothermal, spherical halo within which the visible portion of our galaxy resides, consistent with measurements of spiral galaxy rotation curves [Kol90]. Direct detection of WIMPS is possible through their elastic scattering from nuclei [Goo85, Pri88]. Calculations of the fundamental WIMP-quark cross-sections require use of a model, usually the Minimal Supersymmetric Standard Model (MSSM) [Jun96]. This interaction, summed over the quarks present in a nucleon, gives an effective WIMP-nucleon cross section. In the low-momentum-transfer limit, the contributions of individual nucleons are summed coherently to yield a WIMP-nucleus cross-section; these are typically smaller than $10^{-6}$ pb. (See for example [Ell01a, Ell01b, Bat01, Bed97, Bal01, Cor00].) The nuclear recoil energy is typically few keV (ionization energy) depending on the WIMP mass, up to tens of keV [Lew96] since WIMP velocities relative to the Earth should be typical of Galactic velocities.

An ultra-low-background segmented Ge detector array designed for double-beta decay has the potential to be used for a WIMP dark matter search. In this section the factors affecting the sensitivity of such a search are summarized, highlighting the additions and complementary studies necessary to achieve this goal without compromising the primary $0\nu\beta\beta$ goal. Conservative sensitivity projections are also made. It is clear that an incremental approach to improving radioactive background levels/rejection and array performance at low energies will be necessary. The results from SEGA and MEGA will be critical in assessing the ultimate dark matter sensitivity. The expected WIMP recoil spectrum in germanium extends from threshold, <1 keV, to ~20 keV (ionization energy), in contrast to the much higher energy $\beta\beta$ signature. It should be noted that in this energy range the ratio of measurable ionization energy for nuclear recoils versus electron recoils of the same underlying recoil energy is ~0.2 at low energy, rising to 0.4. In addition, due to the low energy of the region of interest, additional attention must be paid to screening detector materials for their contribution to background in this window. Several attractive features that the Majorana experiment displays as a WIMP detector are listed as follows:

1) *Close-packing (self-shielding) and segmentation of the crystals* will contribute to reducing the gamma-ray background in the low-energy region where the WIMP signal is expected. Single isolated nuclear recoils are expected due to WIMP interactions, whereas $\gamma$ rays generally interact more than once in the detector ensemble, allowing them to be rejected in a large, spatially divided device like Majorana. SEGA will allow these background rejection capabilities to be better characterized.

2) *Segmentation* also lowers detector capacitance, reducing the energy threshold and increasing the acceptance of the WIMP signal. Thresholds as low as 0.75 keV are achieved in segmented HPGe, a considerable reduction from a customary 5-10 keV in unsegmented large diodes.

3) *The spatial information revealed by pulse-shape analysis* (PSA) may help eliminate surface events such as low-to-medium energy betas or other surface contamination, already a limiting background in some WIMP detectors [Kud01]. The feasibility of this approach and its relevance to this detector application must be studied at depth during SEGA: no attempt to exploit PSA in the low-energy region has been made by this collaboration yet.

4) *Majorana's ability to reject low-energy neutron events* is less evident but potentially important. In a typical deep underground location the dominant neutron flux arises from ($\alpha$, n) and natural fission in rock, and to a lesser extent from hard neutrons originating in $\mu$ spallation in rock and shielding. The main concern here is from neutrons with energies above ~200 keV and a typical flux ~ $10^{-6}$ n/cm$^2$/s [Bel99, Ste01]. The referenced energy spectrum dies off rapidly above ~5 MeV. The maximum recoil energy imparted by a neutron to a Ge nucleus is ~1/18 of the incident energy, with only a few percent going into ionization, the rest being lost to phonons. This causes the neutron recoil signal to concentrate below ~60 keV ionization energy. Neutron recoils are identical to those expected from WIMPs. They constitute the limiting background in any WIMP detector, unless a rejection method or substantial neutron shielding can be applied.

An estimate shows that the present low energy signal in IGEX detectors (0.05 counts/keV/kg/day) is indeed compatible with an origin in neutron-induced recoils. This same observation that neutron recoils may already be limiting WIMP searches has been emphasized by the EDELWEISS collaboration [Ste01]. The viability of using additional external shielding in Majorana (neutron moderator and active muon veto) without affecting $\beta\beta$ performance, or physical access to the detectors, will be studied with a full GEANT geometry that is under development. Experimental data from SEGA will be used to validate the simulation, which can then guide the final shielding structure.

Energetic (50–600 MeV) "punch-through" neutrons generated by cosmic-ray interactions in surrounding rock can easily penetrate traditional moderator shielding. This source of neutrons can be reduced by locating the experiment at a deep site [Gai01]. However, a significant veto against neutron-induced events can be achieved in Majorana by monitoring event multiplicity. Considering that the mean free path between recoils in Ge for the neutron energies of concern is ~5 cm, the finely grained segmentation and close packing of Majorana detectors should allow the identification of a large fraction of neutron events by their characteristic multiple-site interactions. This promising feature of Majorana merits a dedicated Monte Carlo analysis. CDMS-I relies on this same consideration to tag neutron events [Abu00] at a depth of only 24 m.w.e. where the high energy neutron flux is much higher. Majorana should exhibit a better neutron rejection ability from its larger target mass.

Another worthy advantage of Majorana as a WIMP detector is the large exposure to be collected. For an apparatus like this, with a planned 500 kg target mass and 5-year data collection, the best WIMP sensitivity originates not from the standard signal-to-noise analysis method (i.e., comparing the expected WIMP signal in a spectral region with the background by means of a suitable statistical estimator), but from an absence of temporal modulations in the background that could otherwise be assigned to a time-dependent WIMP signal. A known example is the yearly modulation in scattering rate and deposited energy expected from the combined movement of Earth and Sun through an isotropic WIMP galactic halo [Dru86].

The improved sensitivity in the modulation analysis is brought about by the progressive reduction in statistical background fluctuations that comes with an increasing exposure. Several authors have discussed this approach to data analysis [Ceb01]. A stable detector gain over long periods of time (years) is a necessary condition for its applicability. In the case of unsegmented HPGe this has been already demonstrated for periods of ~2 years [Dru92]. It is nevertheless our goal to corroborate this crucial point during SEGA and MEGA using segmented devices and Majorana's DAQ system. This system will ultimately be designed to monitor detector acceptance stability directly, at the ~0.1% level for the low-energy bins.

For the time being, a first Monte Carlo calculation of the minimum detectable modulated background fraction after a 2500 kg-y exposure has been performed, using the statistical estimator proposed by Freese [Fre92]. In order to obtain sensitivity projections from this Monte Carlo it is necessary to make a working hypothesis about Majorana's achievable background in the energy region between a few keV and a few tens of keV. In the interim until the SEGA background measurements and dedicated Monte Carlo simulations are completed, a flat 0.005 counts/keV/kg/day from detector threshold $(0.5 - 1$ keV) to 20-keV ionization energy is assumed. It must be emphasized that this represents just one order of magnitude improvement with respect to the most recent IGEX data. This is believed to be a conservative premise in view of the anticipated background rejection capabilities discussed above.

Some preliminary estimates have been made of the contributions from radioactive cosmogenic activation products to this energy region, based on the period of crystal exposure (~60-90 days) at sea level shown in Table 3-3, leaving ample room for improvements. [Bau01] The cosmogenic background rates for natural Ge [Avi92, Col92] should be taken as a conservative upper limit for Majorana. Activation rates for $^{76}$Ge are roughly one order of magnitude smaller due to the higher neutron spallation-reaction energy thresholds [Col00], with a possible exception for tritium production (see Table 3-2). This represents a clear advantage vis-à-vis other large-mass WIMP detectors planning to use natural Ge. The majority of the cosmogenics contribute activity well below 0.005 counts/keV/kg/day, however, we will summarize those that will need to be monitored.

Cosmogenic $^{68}$Ge will be expected to accumulate at a rate of 0.5-1 atom/kg/day following its complete removal producing 86% $^{76}$Ge enriched detectors. $^{68}$Ge (270 day half-life) undergoes decay generating peaks at 10.4 (1.2) keV following the Ga K(L) shell electron capture with BR of 88%(10%) respectively (see Fig. 3-5). Taking the mid values for exposure, and production rate, 75 days of sea level exposure creates 51 atoms $^{68}$Ge/kg. If these crystals are underground for 1 year, the $^{68}$Ge will decay (60% reduction), resulting in a background contribution of 0.05(0.005) cts/kg/day at the K(L)-shell peak energies. The 1.2-keV peak region is near background projection, while the 10.4 keV peak will be a factor 10 above the projection. However, the K-peak region can be bracketed and rejected without significant effect on the dark matter sensitivity. Furthermore, preliminary investigations of vetoing $^{68}$Ge decays by correlating them with the subsequent positron decay (89% BR) of $^{68}$Ga (68 minute half-life) to $^{68}$Zn from the same segment of the detector indicate that a significant further reduction (>5) of the lines can be made. This will be studied further in the background Monte Carlos.

Tritium will also be cosmogenically regenerated in the detectors following its elimination at the time of crystal growth. There is some uncertainty in the sea level cosmogenic production rates with the values shown in Table 3-2 (~110-140 atoms $^{3}$H/kg/day) taken as conservative upper limits. The tritium beta end–point (12.3 year half-life) occurs at 18.6 keV with a peak in the differential spectrum at 3 keV of 0.005 cnts/keV/d/300 atoms $^{3}$H. In order to achieve the target background this will require <2 days above ground exposure during/after crystal growth. It is clear that tritium creation in the detectors and possible contamination during production will have to be closely controlled. We will perform studies to obtain accurate $^{3}$H cosmogenic production rates, methods for detector production underground, and final detector transportation under a few meters-water-equivalent (mwe) of shielding in order to minimize the direct limitation of the dark matter sensitivity due to this contaminant.

Although not a cosmogenic source, we also raise the issue of $2\nu\beta\beta$ background (~$10^{21}$ year half-life in $^{76}$Ge) for dark matter. The differential spectrum (in enriched 86% $^{76}$Ge) for the electron recoils falls below $10^{-4}$ events/keV/kg/day for energies <60 keV, and so it is not a concern at the projected dark matter sensitivity. However, in p-type Ge detectors it is estimated that less than ~2% of the Ge will form a dead layer in proximity to the outer contact. The $2\nu\beta\beta$ background occurs in the enriched crystals at a rate of 10 decays/kg/day in the range 0-2 MeV. Preliminary studies of how higher energy events (<~0.2 /kg/day) originating in the dead layer, but reaching the active volume, may produce partial energy signals that pile up at low energies indicate that this will be well below the target background 0-20 keV of 0.1 events/kg/day. In addition to this dead layer contribution, $2\nu\beta\beta$ events near the crystal edge may only deposit a few keV before exiting the detector. These effects will be simulated in further detail when the choice of detector and size of the dead layer are better known, however, they do not appear to be a limitation.

**Majorana dark matter sensitivity similar to and complementary with CDMS-II**

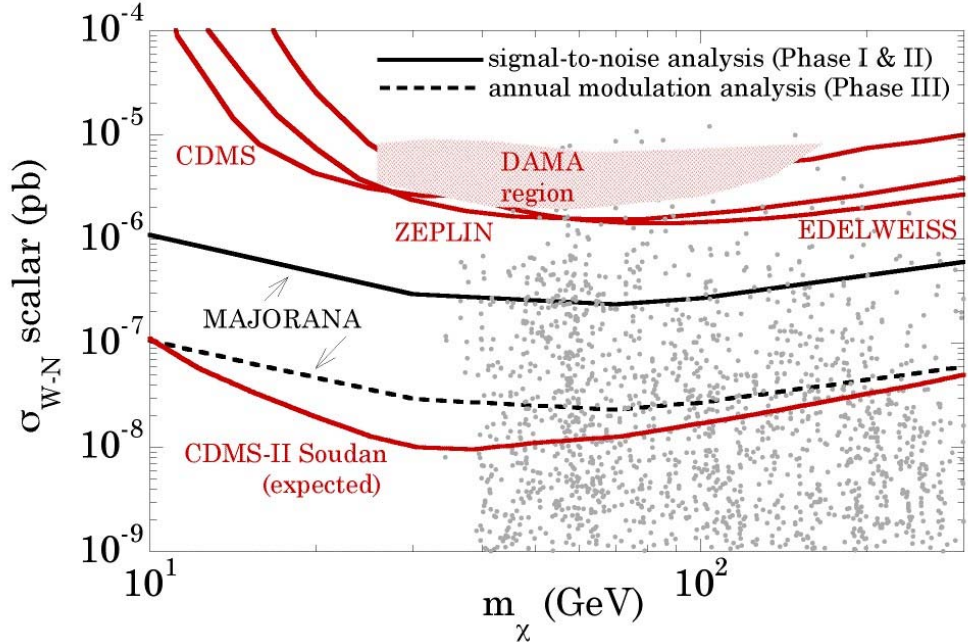

**Figure 2-4 Projected 95% C.L. Majorana WIMP limits for an assumed low-energy background of 0.005 counts/keV/kg/day, just one order of magnitude lower than in present unsegmented single HPGe detectors. Calculated for an ionization energy threshold of 1 keV, achievable via segmentation. "Signal-to-noise" limits are within reach after modest exposures < 1 kg-y (i.e., during SEGA and MEGA). "Annual modulation" limits are calculated for the total exposure of 2500 kg-y. Present DAMA [Bel02], CDMS, EDELWEISS, and ZEPLIN limits (For references, and a complete list of Dark Matter search results see [Gai03]) and expected CDMS II limits are offered as a reference. The shaded region is presently favored by DAMA to explain an unconfirmed WIMP annual modulation in its signal. Dots represent the location in this phase space (spin-independent scattering cross section vs. WIMP mass) of plausible supersymmetric neutralino WIMP candidates, using the same parameters as in [Col00]. Even under these very conservative background assumptions, the expected WIMP Majorana sensitivity is comparable to the most promising cryogenic projects.**

With this conservative approach, the expected sensitivity via annual modulation analysis approaches CDMS-II projections (Fig. 2-4) after collection of the planned 2500 kg-y exposure, if a threshold ~1 keV is achieved. In addition, if the neutralino scattering cross-section resides close to the limit of sensitivity for both experiments, ~$10^{-8}$ pb (Fig. 2-4), Majorana may detect the annual modulation signature, something that the much smaller CDMS-II future total exposure (~10 kg-y) is unable to achieve. CDMS expects to disentangle a WIMP signal from neutron backgrounds using different targets (Si and Ge), for which WIMP and neutron responses differ. The observation of both responses would be highly complementary in making the argument for neutralino dark matter a convincing one.

Finally, if the present DAMA [Bel02] annual modulation claim were to survive the test of time, Majorana would not only confirm it as a >50-sigma effect, but also reveal a second WIMP signature: the tiny, ~0.1% daily rate modulation arising from the coupling of the rotational speed of the Earth (~0.45 km/s near the equator) to orbital and solar speeds through the halo [Col99]. Unfortunately, for cross sections any lower than in the DAMA favored region of Fig. 2-4, an exposure even larger than 2500 kg-y would be required to detect this.

While its main goal is to measure the effective Majorana mass of the neutrino, the singular characteristics of the Majorana detector make it a promising tool in the quest for dark matter. The projected WIMP sensitivity is competitive even under the conservative background assumptions made. It must be noted that the "signal-to-noise" limits depicted here do not rely on a long exposure, and it is expected to make immediate improvements over the existing Ge detector dark matter limits during SEGA operation. Thereafter, the new low-energy background information and associated Monte Carlo studies will be used to project (and then execute) further incremental improvements in the sensitivity of the experiment.

**Other Science: Exploitation of Majorana Data for Solar Axion Searches**

The Majorana Experiment will have 500 times the mass, twice the energy range, will run 10 times as long, and should be able to reduce the background over that of SOLAX, a previous germanium-based axion search, by at least a factor of 50. This should translate into a bound on the axion-to-two-photon coupling constant of $\sim 10^{-10}$/GeV [Ira00]. This would be about as sensitive as the bound set by Raffelt using the population distribution of red giant stars [Raf96] , and would represent the most sensitive laboratory search for axions of mass > 0.01 eV.

The theoretical motivation and history of experimental searches for axions has been recently reviewed by Rosenberg and van Bibber [Ros01]. Quantum chromodynamics (QCD) is very successful in describing many features of the strong interactions. However, the complete QCD Lagrangian contains some symmetries that do not survive quantum effects. Classically, complex terms that break these symmetries can be rotated away if the fermion fields have chiral invariant interactions. At the quantum level, however, such transformations involve a phase angle ($\theta$) that is not arbitrary. Although it must be near zero so as not to introduce a T-violating term, the transformation that brings the quark-matrix to a real, diagonal chirally invariant form does not have a small phase angle ($\theta$). Since QCD respects CPT symmetry, this phase leads to CP-violation, which predicts an electric dipole moment a factor of $10^{11}$ larger than the experimental upper bound [Pec89].

Peccei and Quinn solved this problem by recognizing that the quark mass-matrix is a function of vacuum expectation values (VEVs) of weakly coupled scalar fields ($\varphi$). The VEVs are determined by minimization of the associated potential $V(\varphi)$. They assumed that the Lagrangian has a global U(1) chiral symmetry under which the determinant of the mass-matrix changes by a phase fixed only by instanton effects that spontaneously break the global U(1) symmetry. This results in an additional phase that cancels the offending one that leads to the large CP-violation [Pec77].

Spontaneous symmetry-breaking processes naturally produce Goldstone-bosons. The Goldstone-boson arising from the breaking of the Peccei-Quinn symmetry is called the axion. In two independent papers Weinberg [Wei78], and Wilczek [Wil78] pointed out that these axions could have physically observable and important properties.

The conventional wisdom says they could possibly couple to electrons, to photons, or directly to hadrons. Accordingly, they might have been produced in the Big Bang, and therefore are candidates for cold dark matter (CDM). They might also be produced in stellar burning and in stellar collapse, etc.

The Peccei-Quinn axion is the most plausible solution to the strong CP problem found to date. This fact continues to motivate experimental searches. The technique presented below is one initiated by members of the Majorana collaboration and is an interesting side application of the Majorana array of detectors.

The first technique aiming at the detection of solar axions was suggested by Sikivie in 1983 [Sik83]. It involves Primakoff axion-to-photon conversion in an intense transverse magnetic field, in what is called a magnetic helioscope. This technique is highly efficient for very light mass axions and an experiment operating at CERN uses a 10-m long magnet with a transverse magnetic field of 10 Tesla. This experiment (CERN Axion Solar Telescope, CAST) will reach the maximum sensitivity that the helioscope technique can offer using existing or conceivable magnet technology [Avi01]. The projected sensitivity is better than astrophysical constraints based on the lifetime of red giants [Raf96]. This technique is nevertheless limited to axion masses up to about 0.1 eV. This limitation is due to the requirement that axion and photon wave functions stay in phase throughout the magnet (coherence loss) [Zio99]. In order to search for solar axions with masses > 0.1 eV it is necessary to fill the magnet bores with a gas that will act like plasma, effectively slowing the speed of the photon, allowing it to remain coherent with the slower massive axion. However, this addition to the technique has its own limitations [Zio99]. For axion masses larger than ~1 eV the needed gas density would require a pressure of 15 atmospheres and hence absorb the axion-induced photons (the signal) before they can reach the detectors. For masses beyond this range one needs a different experimental technique.

To address this problem, several members of the Majorana collaboration, at the time leading the SOLAX collaboration, designed a technique using an ultra-low background germanium detector to detect photons coherently converted by Primakoff scattering off the crystalline-Ge planes at times when the line of sight from the detector to the Sun makes an angle with one of the planes that fulfills a Bragg coherence condition. Creswick et al. [Cre98] developed the theory describing the expected conversion rate. A complete description of how such data are analyzed was published in the proceedings of AXION-98 [Avi99]. An experiment was performed in the Hiparsa iron mine in Sierra Grande, Argentina, during which 1.94 kg-years of data were collected. Each event in the energy region of interest was marked with the exact Julian time. For each day of every year, a pattern of the expected times for Bragg coherence was calculated for use in the analysis of the data. The resulting lower bound on the axion-to-two-photon coupling constant was 2.7×10$^{-9}$/GeV. A complete description of the experiment and of the data analysis was published in Physical Review Letters [Avi98].

The SOLAX experiment effectively served as a demonstration of the principle of detecting axions with single crystals. In SOLAX only the (100) crystal axis direction was known and the data had to be analyzed for every degree of rotation about this symmetry axis of the detector, which was along the radius of the Earth. The Majorana Experiment will have 500 times the mass of the SOLAX experiment, with crystal planes fixed as desired.

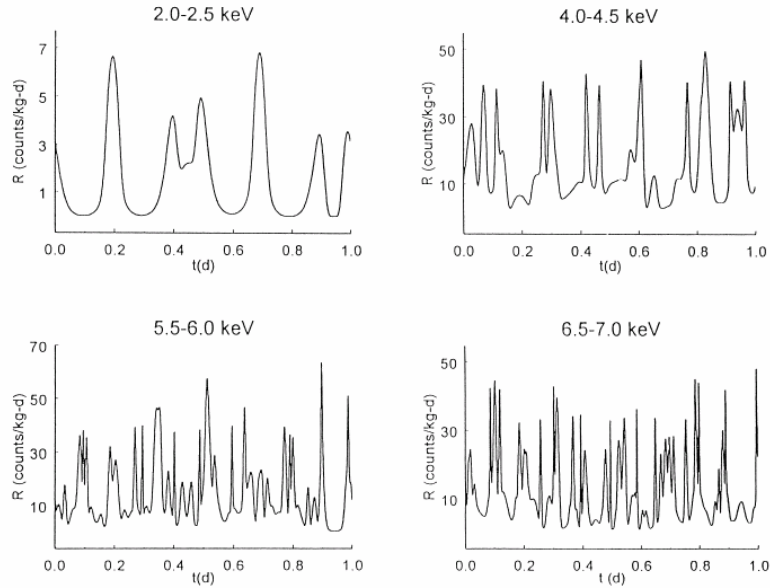

**Figure 2-5 Theoretical prediction [Avi99] of the count rate of photons converted (g$_{\alpha\gamma\gamma}$ = 10$^{-8}$ GeV$^{-1}$) from axions incident at a Bragg angle, for a detector at Sierra Grande, Argentina.**

In a granular experiment like Majorana, the axes can be oriented in a variety of ways so that background can be subtracted.

There are two significant improvements that can be made in the quality of the solar axion data obtained with the Majorana Experiment, relative to that obtained by SOLAX. First, in SOLAX the low-energy background was high due to microphonic noise and cosmic-ray neutrons associated with an overburden of less than 1,000 mwe. Secondly, the pulse-shape discrimination technique used in the SOLAX experiment was a crude, first generation technique. Recent developments have resulted in very sophisticated digital techniques for pulse-shape discrimination. The digital electronics planned for the Majorana Experiment, described elsewhere in this proposal, should allow an energy threshold below 1 keV. Compared with the 4-keV threshold of the SOLAX experiment, this implies Majorana will be sensitive to significantly more of the critical low-energy fraction of the signal.

As summarized above, the Majorana Experiment will have much more mass, cover a crucial part of the axion response energy range, gather much more exposure, and have a much lower background than SOLAX. This translates into a bound on the axion-to-two-photon coupling constant of about 10$^{-10}$/GeV [Ira00].

To improve on this expected Majorana bound will be extremely difficult and expensive. Nevertheless, there are other experiments being planned that involve hundreds of kilograms of germanium detectors, bolometers, and scintillators. By careful application of the data analysis technique in [Avi98] it is possible to add the parameters resulting from the maximum likelihood analyses of the different experiments. It can be shown that this procedure is legitimate, and that combining experimental results this way does not depend on the location of the experiment, the crystal structure, or the orientation.

This technique will allow exploration of a significant portion of the axion model space. Its main advantage is that it is sensitive to axion rest masses well beyond 1 eV, surpassing any other existing laboratory methods in this mass range.

## Other Science: Supernova Neutrinos

Recently Horowitz [Hor02] has noted that the next-generation dark matter and double-beta decay experiments will have target masses large enough to observe neutral-current elastic scattering of the target nuclei by supernova neutrinos. The Majorana detector might expect to see a few tens of events.

## Other Science: Electron Lifetime

The Ge detectors in the Majorana Experiment will contain approximately $5 \times 10^{28}$ electrons of which about $3 \times 10^{27}$ are in the k-shell. Furthermore there are a great number of electrons in the shield surrounding the Ge. This great number of electrons offers the opportunity to look for electron decay. Two signatures are possible: Observation of the Doppler-broadened, 255.5-keV $\gamma$ ray originating from the bound electron decay to $\nu$ plus $\gamma$ ray, and the search for the x rays resulting from the relaxation of the atomic shell following a k-shell electron disappearance. Due to the excellent energy resolution of Ge detectors and the low levels of background expected for the detector array, Majorana should have good sensitivity to search for this process. Previous mean life limits on these two processes are $\tau = 4.6 \times 10^{26}$ y [Bac02] and $6.4 \times 10^{24}$ y [Bel99].

The potential sensitivity of Majorana will depend on the efficiency of detecting the $\gamma$ rays and x rays in the detector array and what levels of background are present at the two energy regions of interest. The efficiency will require a detailed simulation of the response of the array. We have not focused on what levels of background one might expect at these particular energies. However, one would expect to place significantly more sensitive limits on these processes.

## Other Science: Further Implications of $0\nu\beta\beta$

If $0\nu\beta\beta$ is observed it implies that neutrinos are massive Majorana particles [Sch82]. Even so, other mechanisms may mediate the process giving rise to a finite decay rate even in cases where the neutrino mass may be small. As a result, limits on the $0\nu\beta\beta$ decay rate provide stringent limits on many proposed extensions to the standard model of particle physics. The recent reference [Pre03] provides a nice overview of many of these non-standard model processes and their context with respect to $0\nu\beta\beta$ and is a useful guide to the literature. For example, a heavy right-handed neutrino arising in the left-right symmetric model might contribute to the process [Ver02, Moh75, Sen75]. Alternatively, lepton-number violating interactions arising in R-parity violating supersymmetric interactions involving the exchange of charged-lepton superpartners might mediate the decay instead of a neutrino [Moh86, Ver87, Hir96, Hir00]. Furthermore, the process may also be mediated by an exchange of leptoquarks [Hir96a]. The indicated references describe the limits that can be placed on these and other extensions to the standard model from the experimental limits on $0\nu\beta\beta$.

The possibility that leptogenesis may provide an explanation for the baryon asymmetry of the Universe is very intriguing [Fuk86]. Neutrinos are massive particles and the seesaw mechanism [Li82, Kay82, Kay82a] can motivate why neutrinos are so much lighter than their charged partners. This mechanism would also result in heavy right-handed Majorana neutrinos in addition to the light left-handed Majorana neutrinos one usually considers in the context of double-beta decay. These heavy neutrinos would be present in the early universe and, as it cools, they would decay into leptons and scalars via a Yukawa interaction coupling to the left-handed fermions and Higgs. The decay of these Majorana particles violate lepton number, so if they decay out of equilibrium, they can result in a net lepton number for the Universe so long as CP is also violated. Later on this net lepton number is converted to a net baryon number by non-perturbative sphaleron processes. (See Refs. [Pil99, Buc00] for a review of the topic.) Therefore the observation of $0\nu\beta\beta$ would indicate that neutrinos have many of the necessary characteristics for leptogenesis.

### 2.8 Educational Outcomes

The Majorana Project contains elements of several disciplines, and can be expected to produce advanced academic degrees on several fronts. The project opportunities for undergraduate and graduate students in physics, and mechanical, electrical, and computer engineering cover many diverse challenges. The Majorana Collaboration institutions have produced many successful Ph.D. and Master's degree students in science and technology areas closely related to the Majorana Project, and are cultivating graduate and undergraduate students now in anticipation of a number of exciting degrees.

We anticipate that students from our several organizations will work at some combination of their home institutions, the collaborating National Laboratories, and the experiment location during the course of their degree work. Several physics Ph.D. and/or Master's topics can be predicted with certainty:

*Master's level topics*
- Digital Filter Models for Optimal Low-Energy Threshold Operation of the Majorana Experiment
- Optimization of HPGe Detector Segmentation for Background Rejection and Process Yield
- Monte-Carlo Analysis of Detector Segment Self-Shielding for the Majorana Experiment
- Suppressing Cosmic Muon Induced Neutrons in an Underground Laboratory Scenario
- Identifying Low-Energy Backgrounds in an Ultra-Low Level Germanium Spectrometer

*Doctoral level topics*
- Confirmation/Denial of DAMA Dark Matter Mass Result Based on the MEGA Experiment
- Annual Modulation Dark Matter Sensitivity of the Majorana Experiment
- New Limits on Existence of Solar Axions from MEGA Data
- Precision Re-measurement of 2ν Double-Beta Decay of $^{76}$Ge Using Multiplicity Cuts
- Measurement of the 2ν Double-Beta Decay to the Excited State of $^{82}$Se, $^{96}$Zr, $^{100}$Mo, $^{130}$Te, or $^{150}$Nd
- Measurements or limits on the rates of 2νEC-EC, EC-β$^+$, or β$^+$β$^+$ in various isotopes
- New Bound on 0ν Double-Beta Decay with the Emission of a Majoron
- New Limits on Existence of Solar Axions from Majorana Data
- New Limits/Measurement on Majorana Mass of Electron Neutrino
- New limits on the lifetime of the electron
- New limits on the existence of the Goldstone boson, the Majoron

The Majorana Experiment will also provide many opportunities for the educational development of students in a non-traditional or cross-disciplinary way. A significant number of Master's theses and Doctoral dissertations are expected to accompany the collaboration's progress toward and through its final stage. A brief list of some possible degree titles follows:

*Master's level topics*
- Mechanical and Thermal Design and Analysis of an Ultra-Low Background Cryostat for the Majorana Experiment (mechanical engineering)
- Signal Routing for the Majorana Project: Ultra-Low Background Transmission Lines with Low Thermal Conductivity (electrical engineering, physics)
- Monte-Carlo simulation of the Majorana Integrated Active and Passive Shield (physics)
- A Control System and Data Server for the Majorana Installation (physics, computer science)
- Time-Correlation Analysis of Data from the Majorana Double-Beta Decay Experiment (physics, mathematics)
- Failure Prediction for the Majorana Apparatus (physics, mathematics)
- Optimizing Dark-Matter Sensitivity for the Majorana Experiment (physics)
- Shield Mechanical Design and Optimization for the Majorana Experiment (mechanical engineering)
- Failure Prediction of Solid State Systems Based on Regular Time Series Data (statistics)
- Alternate Cooling Methods for HPGe Detectors (physics, mechanical engineering)

*Doctoral level topics*
- Process Control and Material Quality Monitoring for the Electroforming of Ultra-Low Background Copper (chemistry, chemical engineering, physics)
- Pulse-Shape Discrimination for Background Rejection in the Majorana Segmented Detector Array (physics, statistics)
- A High Bandwidth Charge-Integrating Preamplifier Suitable for Ultra-low-background, Cryogenic Sensor Signals (electrical engineering)
- Interaction Localization with HPGe Detector Segmentation and Pulse-Shape Discrimination (physics, electrical engineering)
- Surface preparation methods for alternative detector segmentation

## 2.9 Outreach Program

Aspects of the Majorana Project can easily be presented to inspire the interest of the general population in science. However since the experiment will be sited deep underground, it is doubtful, although not infeasible, that tours of the laboratory itself will be available. Instead we envision kiosks or posters at visitor centers near the laboratory site that are mostly passive, but occasionally would be manned by members of the collaboration. The NUSEL proposal, for example, includes an extensive outreach program that includes a visitor center. The SNOLab location is near the Science North educational facility that has included many presentations on the Sudbury Neutrino Observatory. (See Section 3.11 for a brief discussion of NUSEL and SNOLab.) Some specific examples of educational topics for use in outreach include: Relative levels of radioactivity in various environments, half-lives, and applications of low-level background counting or products.

A discussion of the relative level of the activity in the human body (~12000 Bq $^{40}$K) compared to the initial $^{68}$Ge activity in the Ge crystals of our experiment (~500 decays/day for 500 kg) could form the cornerstone of a lesson on activity in the environment. This would make the point that radioactivity is everywhere and that the levels in our experiment are remarkably low. It could lead into the usual discussion of the typical exposures a person receives each year and how that compares to dangerous levels. This will contribute to the National discourse on the requirement for radiological remediation of DOE/NNSA legacy sites.

Two-neutrino double-beta decay remains the longest measured half-life of any process. Thus the science of the Majorana Project naturally leads to a presentation on half-lives. The comparison of half-lives to the age of the universe ($10^{10}$ y) for $^{76}$Ge (~$10^{21}$ y), $^{238}$U (~$10^{10}$ y) and shorter-lived activities such as our primary $^{60}$Co background (278 d) can make the point succinctly.

Low level counting and low radioactivity products are becoming important in our society beyond just pure science. Low level counting has applications for national security and whole body counting, for example. The semi-conductor industry requires low-activity lead to make solder because $\alpha$ decays can cause single upset failures in sensitive electronic components. These topics will also elucidate the importance of this field of research to the public.

## 3.0   The Majorana Experiment Configuration

The desired outcome of the Majorana Experiment is the discovery of the effective Majorana mass of the electron neutrino, and the approach is measuring the rate of zero-neutrino double-beta decay ($0\nu\beta\beta$). Although our Reference Plan for Majorana (described in this Section) is founded on established technologies, there is potential for some engineering optimizations. Therefore the Collaboration is conducting two initial experiments (SEGA and MEGA) to determine the optimum configuration to fully exploit the new background suppression techniques. Each of these experiments comes with distinct physics goals, and will serve to prepare the Collaboration in terms of analysis and acquisition software, specialized copper electroforming, detector manufacturing, and efficient contracting for the enriched material for creating the full set of germanium detectors.

### 3.1 Summary of the Reference Plan

In this sub-section, we delineate the Reference Plan components. The purpose is to provide the reader with a short summary of the plan in one place. In the subsequent sub-sections, we motivate and discuss each aspect of the Reference Plan in greater detail and consider the possible variations that are under consideration.

In the Reference Plan, we propose to:

- purchase 525 kg of intrinsic Ge metal, enriched to 86% in isotope 76, from the ECP in Russia
- surface ship this Ge to a detector manufacturing company in North America to have Ge crystals, suitable for detector fabrication, produced
- quickly, deliver these crystals to a collaboration-supplied underground detector fabrication facility
- at the underground facility, produce approximately 500 1-kg, n-type, segmented Ge detectors with each segmentation geometry consisting of 2-4 segments
- install these detectors into Cu cryostats that have been electroformed underground
- install these assembled cryostats into a ~10-cm thick "old" Pb shield that is contained within a ~40-cm thick common Pb shield
- incorporate an active, neutron and cosmic ray anti-coincidence detector (a veto system) into the Pb shield
- electronically read out the Ge detector signals with one high-bandwidth electronic channel per crystal and one low-bandwidth electronic channel per segment
- use commercial digitizers based on CAMAC technology for the data acquisition electronics.

## 3.2 Overview of the Majorana Design

*Schematic Setup*

The origin of the technology for the measurement of $^{76}$Ge double-beta decay goes back to the clever introduction of the internal source technique by Fiorini [Fio67]. This method allows the experimenter to use a high-resolution germanium γ-ray spectrometer to measure the radiation that is emitted from the germanium itself. Initially, natural germanium spectrometers were simply shielded from environmental backgrounds to achieve the first limits. The limitation of this technique for application to $^{76}$Ge double-beta decay is the quantity of background signals observed at the desired detection energy of 2039 keV. While the 0νββ energy is above most ubiquitous environmental radiation, it is does not exclude all backgrounds. However, over time, sources of radioactivity have been identified and removed resulting in greatly improved sensitivity to longer half-lives.

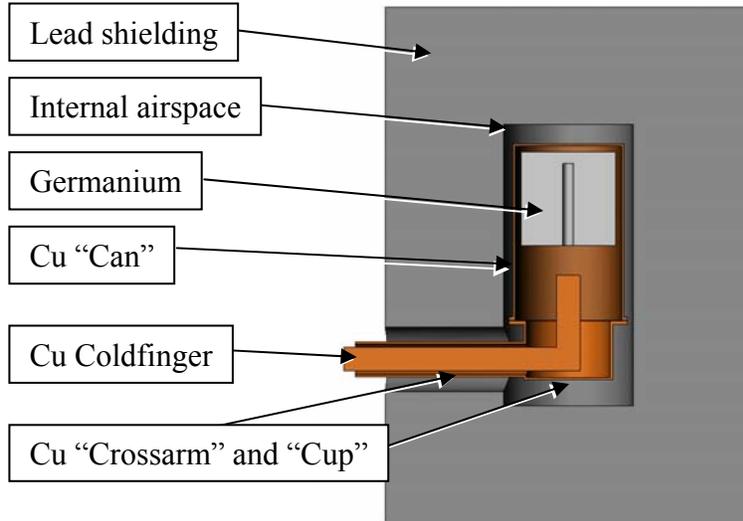

**Figure 3-1 Typical germanium detector arrangement. The volume of Ge is about 400 cc (2.1 kg) and the volume of Cu is Can: 97 cc (0.8 kg); Crossarm + Cup: 63 cc (0.6 kg) + Coldfinger: 82 cc (0.7 kg); Total: 242 cc (2.1 kg).**

We begin our discussion with a schematic setup, shown in Fig. 3-1, in which the generic arrangement of germanium, structural materials (largely copper) and lead shielding are shown. The detector may be configured at the time of production either as p-type or n-type, depending on the majority impurities in the material. Ge detectors of p-type have an outer 0.2-0.5 mm thick "dead" layer of highly doped germanium that is insensitive to radiation. An n-type detector also has this dead layer, but on the surface of the inner hole which provides the electrical contact and therefore provides little shielding against external radiation. Dead layers have the effect of screening the detector from low-energy radiations and alpha particles from external or surface-layer activities. Thus, low-energy backgrounds in the detector are either within the inner materials, or on the innermost surfaces, or the result of scattered radiations emitted from within the dead layer (e.g. 2νββ). Higher energy backgrounds, however, are usually penetrating γ rays originating from the germanium, copper, the inner region of lead, or cosmic ray muons and muon induced neutrons and γ rays penetrating the bulk of the lead. Since in this arrangement copper[3] surrounds the germanium, any radiations emanating from the lead must pass through a small amount of copper to reach the germanium.

Since the earliest use of germanium in the internal source configuration, several improvements to this schematic design have been added. In approximate order these include:

➢ Cosmic veto shielding
➢ Underground locations
➢ Detector construction using screened low-background materials (copper)
➢ Detector construction using electroformed copper
➢ Ultra-pure shielding materials
➢ Low-cosmic-ray exposure germanium and copper
➢ Germanium enriched from 7.8% to 86% in $^{76}$Ge
➢ Signal processing
➢ Coming soon: Detector segmentation

The point of this recapitulation (see Fig. 3-3) is that a ~20-year R&D program by Majorana collaboration members has preceded this proposal as we struggled to understand each successive background. For instance, the same germanium crystal was used for the upper curves in Fig. 3-3. On the occasions when this detector was rebuilt and returned underground, the effects of cosmogenic activation were observed and quantified. Similarly, the effort to produce ultra-pure support structures has resulted in materials rivaling the germanium itself in radiopurity [Bro95].

*Traditional Cryostat Design vs. Alternate Cooling*

The design of the Reference Plan is dependent on our collaboration's understanding of the origin of the signals at 2039 keV, as explained in the background model section below. Knowledge of the basic background sources, when combined with the equation for the $T_{1/2}$ of the decay mode, motivate our Plan.

Considering only proportionalities (i.e. neglecting units and constants), the generic functional forms of $T_{1/2}$ and $|<m_\nu>|$ are given by:

$$T_{1/2} = \frac{MT}{C}$$

$$\left| \langle m_\nu \rangle \right| = \frac{1}{\sqrt{T_{1/2}}}$$

where M is the mass of $^{76}$Ge, T is the effective counting time, and C is the sum of counts attributable to the decay of interest. If backgrounds dominate such that no signals are seen at 2039 keV, The $T_{1/2}$ limit would approximately be:

---

[3] The Collaboration uses copper, rather than the industrial standard of aluminum, for the detector enclosure because of the ability to highly purify copper through electroforming. This technique has been in use by the Collaboration for many years.

$$T_{1/2} > \frac{MT}{\sqrt{B}}$$

where B is the number of background counts in the region of interest. If the dominant backgrounds are in the germanium, as we have shown for previous experiments, the number of background counts is proportional to the product of M and T and the functional form of the $T_{1/2}$ limit simplifies to:

$$T_{1/2} > \sqrt{MT}$$

So doubling the mass of the experiment increases the $T_{1/2}$ by only 40%, and decreases the effective neutrino mass limit by 20%. Since the cost of the enriched material is a significant fraction of the experiment's cost, doubling the mass of the experiment from 500 kg to 1000 kg could greatly increase cost for a small return in this background scenario.

On the other hand, if one assumes there is no background in the germanium itself, but there is a limiting and constant background rate (b) in the support structures around the active detector, then the background would be $B=bT$, and the $T_{1/2}$ limit would simplify to:

$$T_{1/2} = M\sqrt{\frac{T}{b}}$$

And it would behoove the planner to construct an experiment with high mass and exceedingly small amounts of structural materials. This is the focus of most alternate cooling techniques that have been proposed.

We have demonstrated that small amounts of cosmogenic materials will be in the detector crystals, even if manufactured underground. Furthermore we have shown that support structures for the detectors can be readily made without unacceptable background contributions. Therefore we conclude that the experiment should be designed with a moderate enriched Ge mass and a long (inexpensive) run time. In addition, special emphasis should be placed on electronic suppression methods for backgrounds and in construction techniques that promote very long-lived detectors.

In conclusion, since it has been shown that cosmogenic isotopes have played the most important role in previous germanium experiments and support structures have not, the Reference Plan of the Majorana Experiment is to plan for and minimize cosmogenic background sources, then concentrate on the next most serious background sources. In the event that an alternative cooling/shielding arrangement is found which does not compromise the gains in cosmogenic rejection or neutron suppression, it will be entirely possible to adopt these methods without causing harm to the already-manufactured crystals.

### *Majorana Description*

The Majorana Experiment is a complete next-generation double-beta decay experiment, using a large quantity of enriched materials, state-of-the-art detector fabrication, deep underground facilities, cutting-edge instrumentation and data analyses, and acquisition of complete systems-status data (herein referred to as "state-of-health" data). Construction and operation will be done in extremely clean facilities, and all materials will undergo extensive radiological and mass-spectrometry screening. We have great confidence in the technologies that we outline for our Reference Plan. Although we will always look to improve upon that design, this proposal presents a default configuration based on proven technologies. We do discuss the various places where the design is being considered for improvement. In this subsection we outline the Reference Plan and in the remainder of this section, we provide greater detail about this configuration.

The apparatus will consist of modules constructed from electroformed copper, each containing many germanium crystals. The Reference Plan is to house about 55 kg of crystals per cryostat, arranging cryostats in pairs such that 500 crystals of about 1.05 kg each would comprise the 500 kg of germanium in the total experiment. The organization of crystals and cryostats can be altered if other design criteria require it. Surrounding the cryostats is a thick shield of lead, a neutron-absorbing blanket, and an active cosmic-ray veto shield.

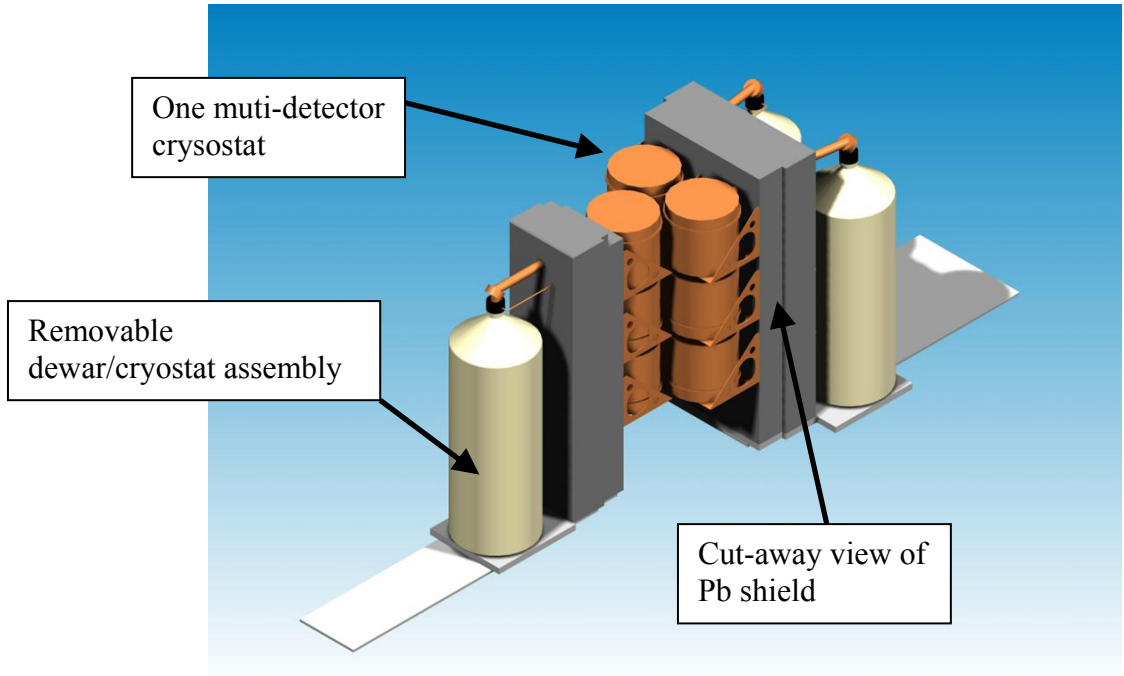

**Figure 3-2. Majorana apparatus.**

This modularity allows the gradual commissioning of the apparatus such that early results could greatly influence the ongoing manufacturing process. Once the crystals are produced and underground, it is possible to avoid the introduction of internal contaminants, so repeated repackaging to take advantage of emerging cryogenic techniques is possible. This gives the opportunity for any background problems not inherent to the Ge at low energy to be mitigated to improve dark matter searches. Modularity also facilitates maintenance as individual modules can be taken off-line to allow repair or upgrades with minimal downtime for the system.

The gradual production of detectors also allows stock materials to be located underground for years in advance, meaning that the short-lived cosmogenics seen in germanium and copper will eventually become undetectable even at detector commissioning.

### 3.3 The Background Model

*Philosophical underpinning*

The background model of an experiment like $^{76}$Ge double-beta decay is critical to the design and execution of the experiment. The model predicts the sources of background and therefore guides the pre-commissioning efforts in detector and acquisition system design and development. In particular, the background model leads to a sensitivity calculation. Since mistakes in this model can cause delays and extra costs, we must consider unlikely or unexpected backgrounds and should over-design methods of coping with backgrounds, to assure the success of the experiment.

The background model can be based on empirical findings or on first principles. It is usually more credible to have an empirically established background, although first-principle background estimates and their derived sensitivities are not without merit. Here, both pathways from established experiment and from first principles are discussed.

*Background in Ge or elsewhere*

The basic experimental geometry is the germanium mass as the fiducial volume, surrounded by a modest mass of copper and small plastic parts, all contained within a shield. Thus, we first consider background within the germanium crystal, because backgrounds from this source have very high probability of registering in the data acquisition system. Furthermore, because summing is highly favored inside the Ge crystal, backgrounds arising from the germanium itself tend to be smoothed and distorted. This complicates the spectral identification and analysis. After the consideration of the germanium, backgrounds from other materials and cosmic-ray backgrounds are considered. In this section we will build the Majorana background model.

### 3.3.1 Germanium Double-Beta Decay Backgrounds

*History of Majorana collaborators on background reduction*

Since 1980, various combinations of Majorana collaborators have worked to identify and remove sources of radioactivity in germanium spectroscopic measurements [See Section 2.]. This work has identified materials, techniques, and measurement strategies to reduce the signal level per mass of detector and unit time. During the early years the effects of cosmic ray activity and gross primordial activity were addressed by canonical muon vetos and vacuum jacket parts remanufactured from known low-activity materials. Later, electroformed copper was introduced to reduce the primordial and cosmogenic activities in ordinary copper. At this time the cosmogenic activity in the germanium was first noticed. Eventually, the contribution of $^{68}$Ge, a cosmogenic isotope in enriched germanium was identified as the largest remaining background. At about this time the two-neutrino double beta decay was identified by the comparison of ordinary germanium detectors vs. enriched detectors: the only feature in common for the two detectors was that above a certain energy, after the subtraction of a well known $^{68}$Ge component, the activity of the two detectors per atom of $^{76}$Ge was the same.

*Cosmogenic Backgrounds in Ge*
A germanium detector that has been exposed to cosmic ray neutrons at the earth's surface will contain radioactive isotopes that can produce backgrounds. Since the spectrum of neutrons at the earth's surface extends to very high energy, we might expect that every isotope of equal or lower mass number could be created. However, because thresholds increase and cross sections decrease with $\Delta A$, we expect large $\Delta A$ reactions to be increasingly rare. Note that certain highly stable light ejecta (e.g. tritium) can be preferentially produced and may also be a problem.

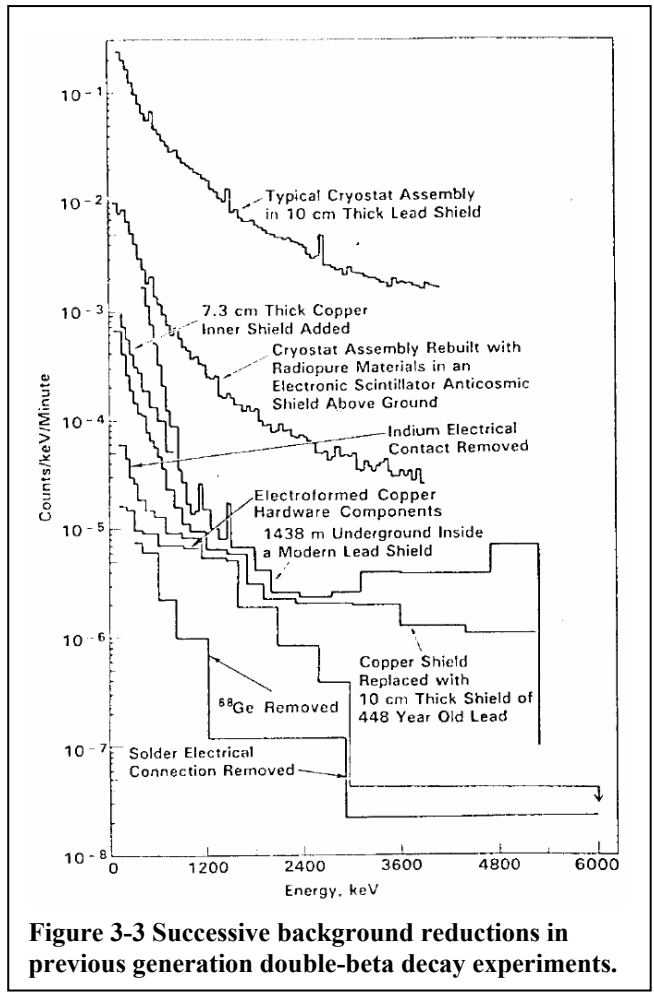

**Figure 3-3 Successive background reductions in previous generation double-beta decay experiments.**

While we might directly calculate the rates of production at the Earth's surface of a variety of isotopes, our philosophy of relying on our own empirical data leads us to investigate previous results for evidence of cosmogenic activity sufficiently serious to warrant the effort. We easily find that it must be taken seriously.

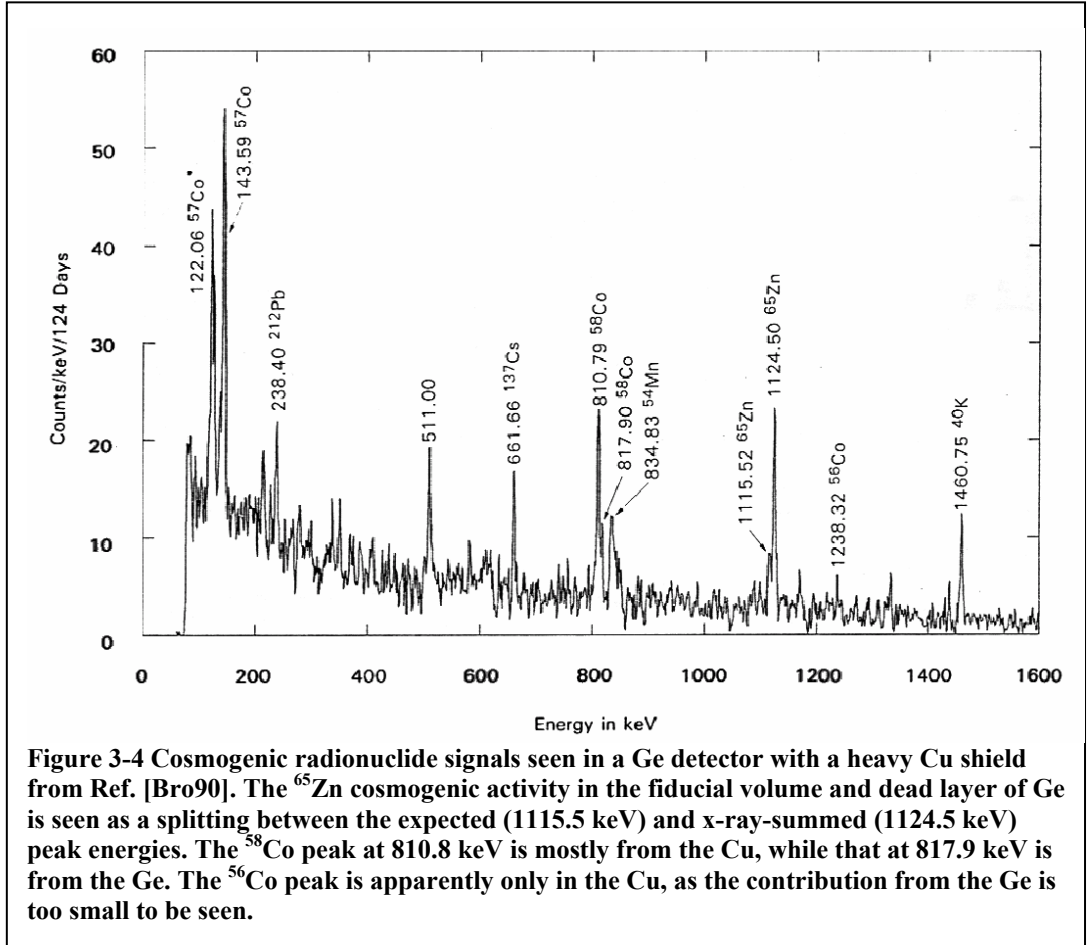

**Figure 3-4 Cosmogenic radionuclide signals seen in a Ge detector with a heavy Cu shield from Ref. [Bro90]. The $^{65}$Zn cosmogenic activity in the fiducial volume and dead layer of Ge is seen as a splitting between the expected (1115.5 keV) and x-ray-summed (1124.5 keV) peak energies. The $^{58}$Co peak at 810.8 keV is mostly from the Cu, while that at 817.9 keV is from the Ge. The $^{56}$Co peak is apparently only in the Cu, as the contribution from the Ge is too small to be seen.**

Figure 3-4 shows several radioactive isotopes in Ge and Cu. While these isotopes are clear signatures of cosmic neutron reactions on germanium, easily visible via their summing with electron capture (EC) x rays, they do not themselves pose a serious threat given their modest lifetimes or total decay energies. However, they do illustrate the existence of the cosmogenic problem, where $\beta^-$ or $\beta^+$ decays cannot because the summing of $\gamma$-ray energy with the continuous energy deposition blurs otherwise easily identifiable peaks.

**Table 3-1 Isotope data on cosmogenic isotopes in Figure 3-4.**

| Isotopes | $T_{1/2}$ | $Q_{EC}$ | $Q_\beta^-$ |
|---|---|---|---|
| $^{56}$Co | 77.26 d | 4566 keV | |
| $^{57}$Co | 271.79 d | 836.1 keV | |
| $^{58}$Co | 70.82 d | 2307.4 keV | 381.5 keV |
| $^{65}$Zn | 244.26 d | 1351.4 keV | |

Unlike the above examples, $^{68}$Ge and its daughter $^{68}$Ga do represent a problem. The half-life of $^{68}$Ge is 270.82 days and its electron capture endpoint is Q$_{ec}$ = 106 keV. The daughter $^{68}$Ga has a very short half-life of 67.6 minutes but a problematic positron endpoint of 2921.1 keV. This pair of isotopes is a serious potential background owing to the long half-life of the parent and the total energy of the daughter, as shown in Fig. 3-5. The pure EC decay of $^{68}$Ge produces only gallium x rays after the k-shell electron is absorbed in the decay. These x rays sum to the binding energy of the k-shell electron at

10.367 keV. This was easily observable in a previous double-beta decay experiment. [Bro90]. Figure 3-6 shows this peak and Fig. 3-7 shows the low energy portion of the full spectrum. The intensity of the 10.367-keV peak normalizes the Monte Carlo simulation of the $^{68}$Ga decay. The measured spectrum in Fig. 3-7, with the experimentally normalized $^{68}$Ge removed, corresponds to the lowest curve in historical background reduction shown in Fig. 3-3. Evidently, cosmogenic $^{68}$Ge can contribute to the region of interest about 2039 keV. In fact, about 2.3 counts per keV arise at 2039 keV from 10,000 $^{68}$Ge decays.

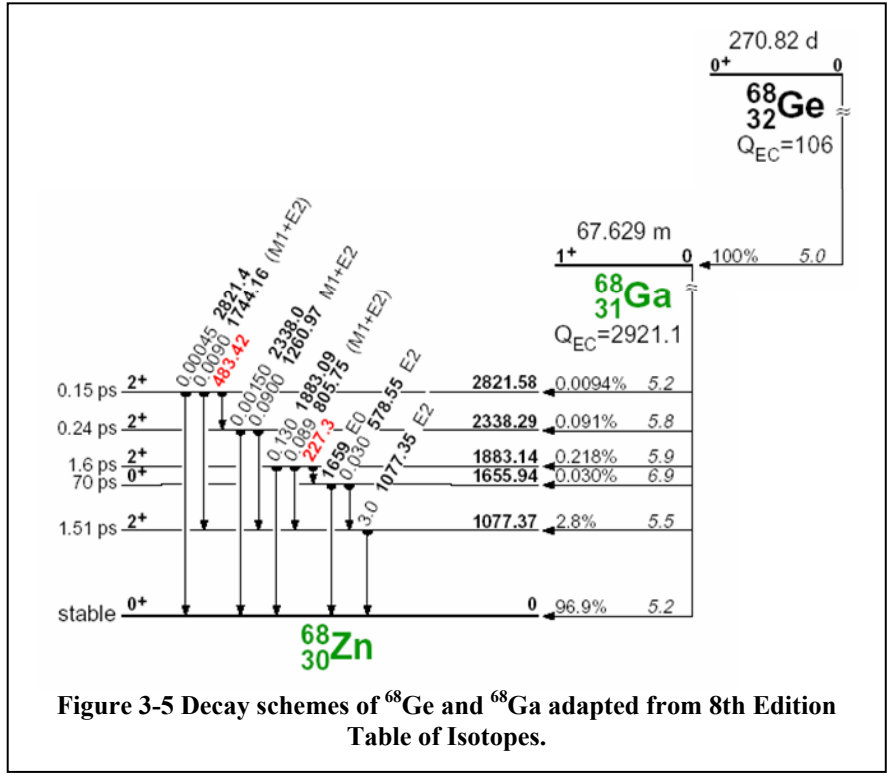

**Figure 3-5 Decay schemes of $^{68}$Ge and $^{68}$Ga adapted from 8th Edition Table of Isotopes.**

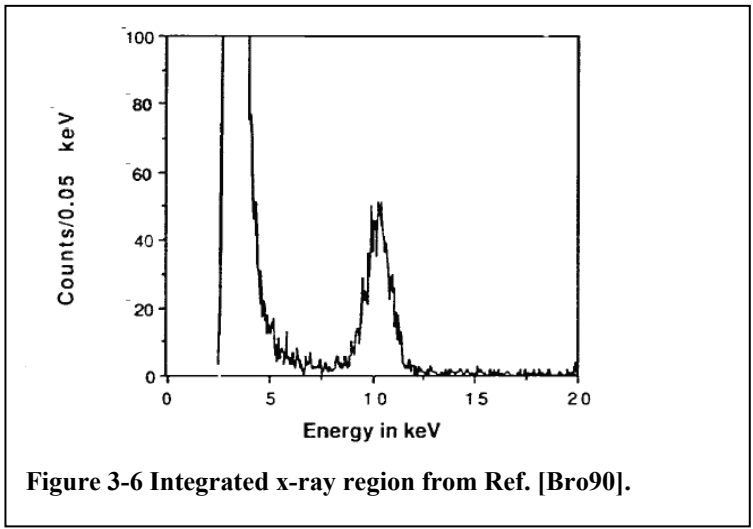

**Figure 3-6 Integrated x-ray region from Ref. [Bro90].**

Gamma rays from the 1077.37 keV state may in principle be observed when they arise from the dead layer (3-4% of a 1-2 kg mass detector). However, in the data shown in Fig. 3.4 this could have amounted to only about 3 counts in a 2-keV wide channel; below the detection limit.

In Fig. 3-7 it is apparent that the majority of counts above 1500 keV are due to the $^{68}$Ge→$^{68}$Ga decay series. The 270.82-day half-life would have allowed 61% of the $^{68}$Ge to decay in one year. Thus a double-beta decay experiment whose limiting background was $^{68}$Ge would improve in half-life sensitivity by about 79% per year.

In most previous experiments, new Ge detectors spent a considerable amount of time (days to weeks) underground before commissioning as a double-beta decay system. However in some cases, new detectors were rushed into service immediately after delivery underground. In these cases, the effects of short-lived isotopes could be seen and absolutely cement the case for cosmogenic spallation. In one particular case, the Ga x rays could be seen to decay with the 3.26-day half-life of $^{67}$Ga. In this case, the spallation product $^{67}$Ge (18.9 min) decayed to $^{67}$Ga (3.26 d), which in turn produced Zn x rays at 9.659 keV. (This line is unresolved from Ga x rays at 10.367 keV.) After the short-lived $^{67}$Ge is gone, the x-ray line resumes the 270.8-day $^{68}$Ge decay characteristic.

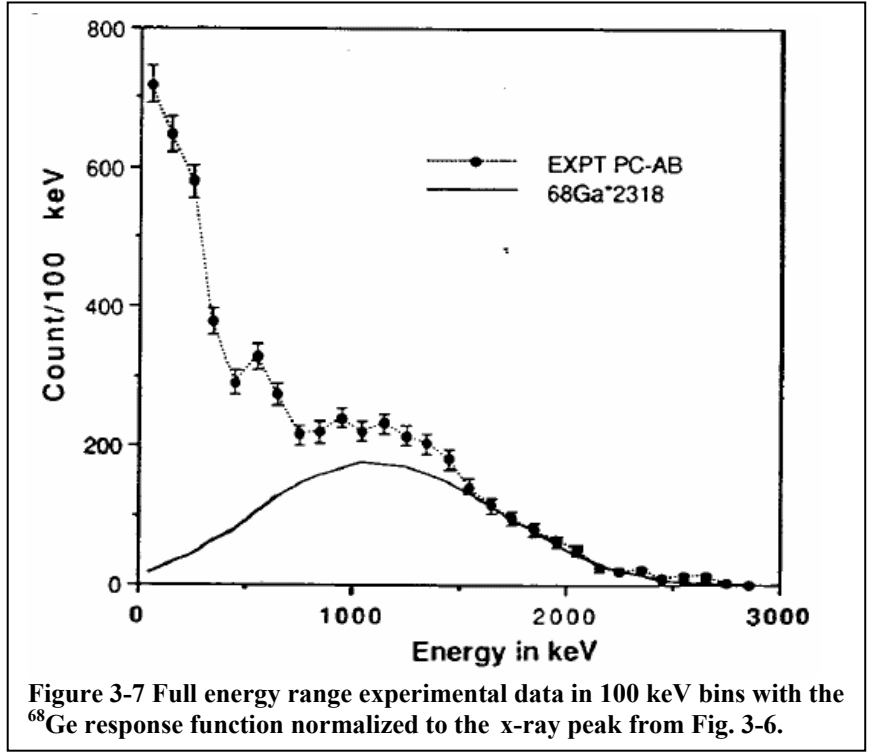

**Figure 3-7 Full energy range experimental data in 100 keV bins with the $^{68}$Ge response function normalized to the x-ray peak from Fig. 3-6.**

The effect of cosmogenic $^{60}$Co is as important as $^{68}$Ge. In fact, owing to the 5.2 y half-life (See Fig. 3-8), $^{60}$Co may be more important. To γ-ray spectroscopists the 1332.5-keV and 1173.2-keV peaks of $^{60}$Co are very familiar. Even the 2505-keV sum line is fairly commonly observed. However, when summed with the 318.1-keV β emitted in the decay to the 2505.8-keV state, the decay signature becomes a continuum inside the germanium crystal and therefore difficult to identify. The decay of $^{60}$Co produces about 1.8 counts per keV near 2039 keV per 10,000 $^{60}$Co decays when the decay occurs within the crystal.

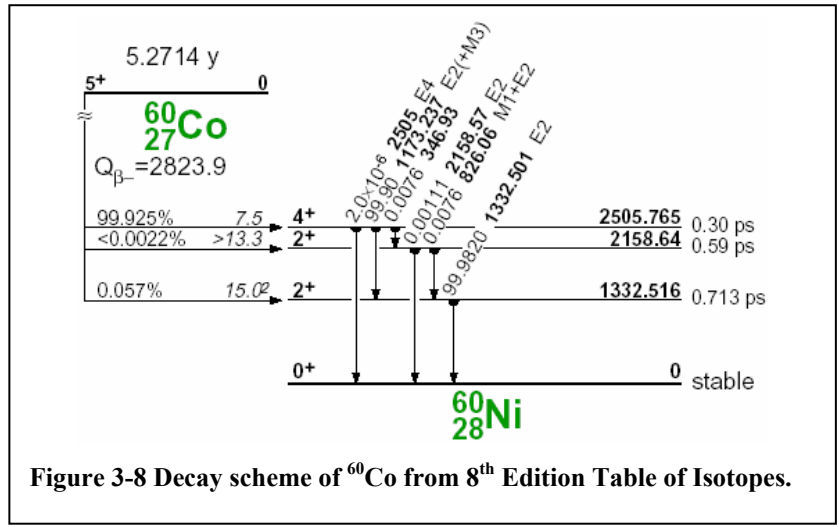

**Figure 3-8 Decay scheme of $^{60}$Co from 8$^{th}$ Edition Table of Isotopes.**

This isotope is produced in very unusual multiple-nucleon reactions on the various Ge isotopes ($^{70}$Ge(n,6n5p)$^{60}$Co to $^{76}$Ge(n,12n5p)$^{60}$Co). (See Appendix 3 for a pictorial representation of the chart of the isotopes in this region.) Surprisingly, these isotopes have similar thresholds and cross sections for the production of $^{60}$Co. Furthermore, the production process of cosmogenic isotopes must be understood to have a complete background model because the complex spallation ejecta (e.g. 6n5p and 12n5p) may be in any configuration. As a result, stable or long-lived products, such as $^4$He and $^3$H would be found.

*Computation of Rates in Ge*

To calculate the activity of cosmogenic isotopes in Ge, we assume it is entirely the result of exposure to high-energy neutrons arising from cosmic ray interactions with the atmosphere and earth. For this calculation, we need to know the neutron flux and production cross-sections as a function of energy. The neutron spectra observable on the earth's surface have been measured and published [Lal67, Hes59]. These neutron spectra have significant numbers of neutrons at very high energies. The variation between these spectra and those computed for a 2000-meter water equivalent location [Gai01] is striking. (See Fig. 3-9.)

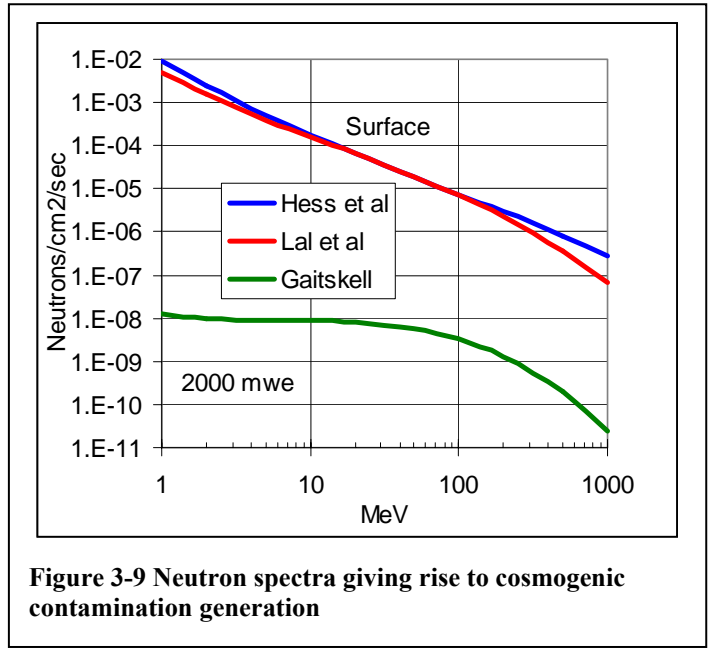

**Figure 3-9 Neutron spectra giving rise to cosmogenic contamination generation**

The required cross section data for these reactions has been calculated using ISABEL [ISABEL]. ISABEL uses direct and evaporative simulation to estimate the production rate of a variety of products from a projectile-target-energy starting point. Cross sections were calculated for neutron energies between 4 and 1000 MeV on stable Ge targets from $^{70}$Ge through $^{76}$Ge. These cross sections were then convoluted with the neutron spectra to predict the number of atoms of a specific isotope. Results were calculated even for very unusual reactions such as $^{76}$Ge(n,12n5p)$^{60}$Co. Some validation of this code has been published [ISABEL].

Careful observation of low background detectors irradiated by surface neutrons provides a valuable empirical estimate and cross check of the production rate. For such data to be cleanly interpreted, the detector must have been underground long enough for the isotopes of interest to decay below the minimum detectable activity (MDA), brought to the surface for a known time, then reintroduced underground and commence operation promptly. The observations can then be used to estimate the activation per unit time above ground. The results can then be compared to the ISABEL calculations described above.

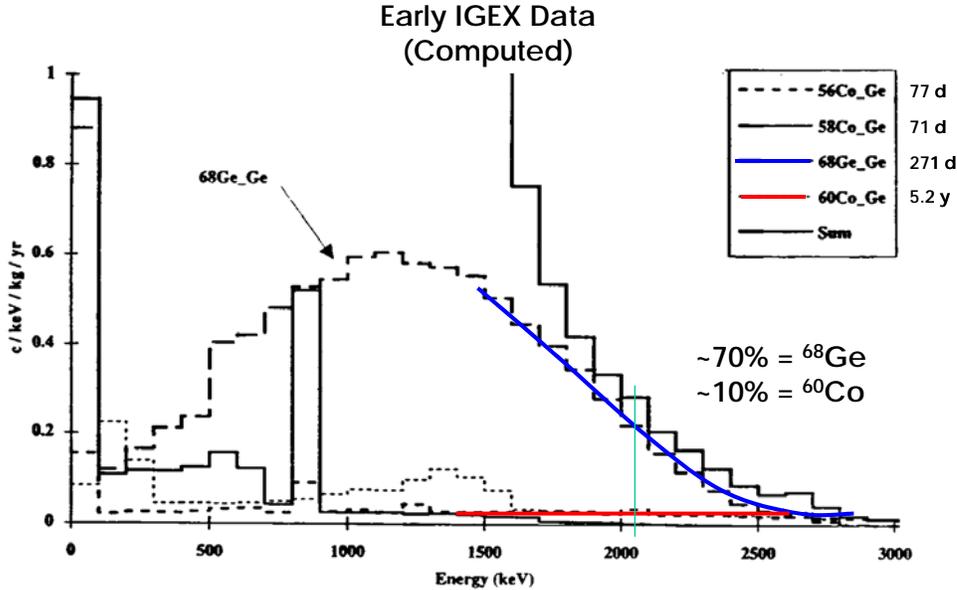

**Figure 3-10 The computed spectral contributions for an enriched detector.**

**Table 3-2 Calculated and experimental production rates in natural germanium, Calculated rates in enriched germanium assuming 86% $^{76}$Ge and 14% $^{74}$Ge. Units are atoms per day per kilogram (from [Avi92]).**

| Isotope | Natural Germanium | | | Enriched Germanium | |
| --- | --- | --- | --- | --- | --- |
| | Lal et al | Hess et al | Experiment | Lal et al | Hess et al |
| $^{3}$H | ~178 | ~210 | | ~113 | ~140 |
| $^{54}$Mn | 0.93 | 2.7 | $3.3 \pm 0.8$ | 0.37 | 1.4 |
| $^{57}$Co | 1.70 | 4.4 | $2.9 \pm 0.4$ | 0.28 | 1.0 |
| $^{58}$Co | 2.30 | 5.3 | $3.5 \pm 0.9$ | 0.59 | 1.8 |
| $^{65}$Zn | 24.6 | 34.4 | $38 \pm 6$ | 3.12 | 6.4 |
| $^{68}$Ge | 22.9 | 39.0 | $30 \pm 7$ | 0.54 | 0.94 |

A low background detector was operated underground for 32 months then sent to the surface for 5 months for a modification (removal of a solder bead, see Fig. 3.3.). Upon return underground, it was operated for 234 days. The observable electron capture isotopes measurement is reported in Table 3-2 along with a computed estimate of tritium.

The $^{68}$Ge entry in Table 3-2 is from the analysis of a detector assumed to be at equilibrium operated underground long term.

Table 3-2 shows generally good agreement between the experimental and computed values, so the predicted activities in enriched material are considered reliable. Also, the assumption that the neutron exposure accounts for all the cosmogenic activity is supported, within the ~30% uncertainty of the measurement.

Using these experimental values for the expected quantity of cosmogenic activities in the (non-enriched) detector, we can compute the expected background for a specific activation scenario and compare to experiment. One such application of this approach is the prediction of the initial backgrounds in an enriched detector after introduction underground. For this prediction, it was assumed that $^{68}$Ge concentration in the detector was at surface equilibrium since the enriched material had been on the surface for many $^{68}$Ge half-lives. The contributions by isotope are shown in Fig. 3-10. It can be seen that initially $^{68}$Ge is a dominant contributor, but that $^{60}$Co is also present. The calculation predicts that the $^{68}$Ge and $^{60}$Co contributions would be equal in 2-3 years and therefore $^{60}$Co is a serious contamination.

While the ISABEL calculations have an estimated uncertainty of about a factor of two, using them to estimate the activation rates in the enriched material based on the measured activation in natural material is reliable. When we compare the calculated and measured spectra in Fig. 3-10, we find that essentially the entire count rate in the detector above 750 keV is due to cosmogenic activity (Fig. 3-11), at least during the initial operation.

Recently, we have recalculated these spallation rates using a more modern ISABEL

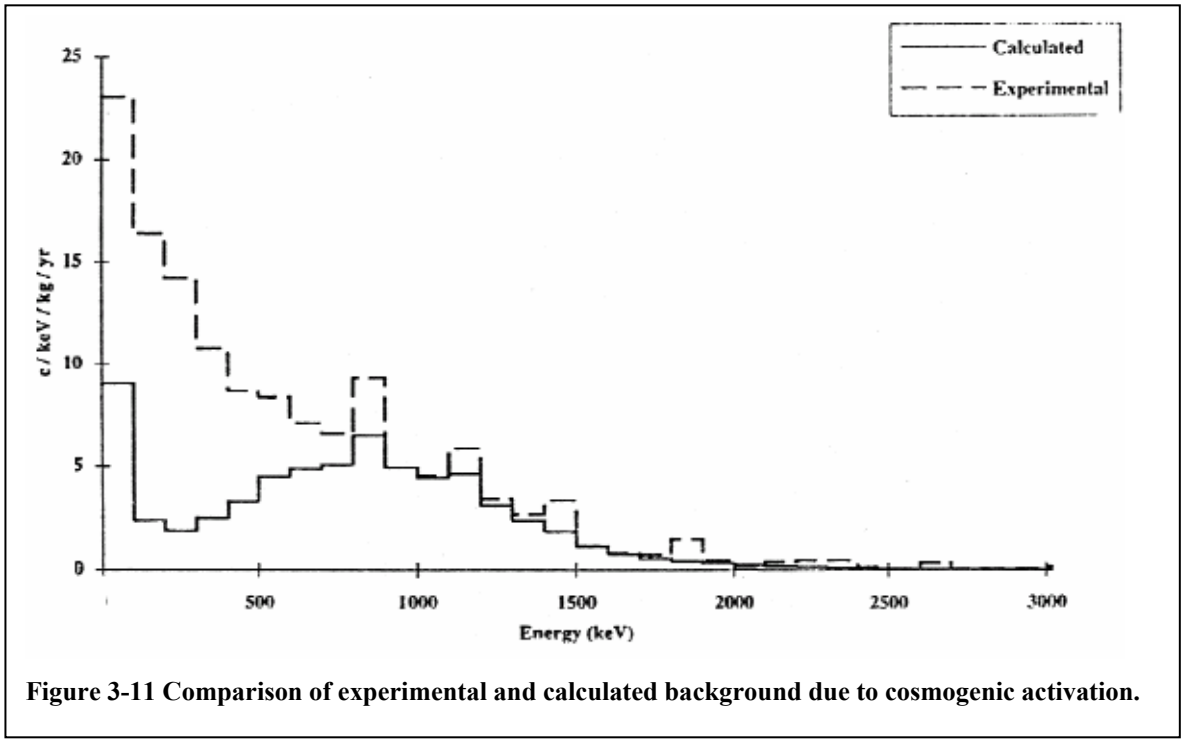

**Figure 3-11 Comparison of experimental and calculated background due to cosmogenic activation.**

implementation accounting for the angular momentum of the nucleus during the evaporative steps. We conclude that some of the rates calculated in [Avi90] may have been low. However, they could not have been two times greater and still be compatible with the experimental data. Thus we will consider the enriched activation rates from Table 3-2 to be accurate within a factor of two. We therefore take as a simple rule that the dangerous cosmogenic species are produced on the Earth's surface at roughly 1 atom per day per kilogram. Below, we discuss methods of minimizing or eliminating the major cosmogenic components of the background.

**Table 3-3 The steps in a simplified detector production process showing optimistic nominal durations. The importance of the indicated exposures is a subjective description based on the duration of the exposure and its timing relative to the purification steps.**

|   | Process | Time (Days) | Importance of exposure |
|---|---------|-------------|------------------------|
| 1 | Mine and refine germanium ore | | Not App. |
| 2 | Chemical processing (into gas form for separation) | | Not App. |
| 3 | Separation process (sequential separation in centrifuges) | | Not App. |
| 4 | Enriched gas storage (Collection of batch quantity) | 15-30 | Moderate |
| 5 | Chemical processing (into pure metal for zone refining) | 1-7 | Low |
| 6 | Transport | 30 | Moderate |
| 7 | Zone refining / Crystal pulling | 7-15 | Moderate |
| 8 | Detector mounting | 1-7 | High |
| 9 | Transport | 3 | Moderate |
| 10 | UG testing and operation | Years | Very Low |

For the purposes of our Majorana background model, we must understand the history of the Ge used to produce a Majorana detector. The basic steps in the production of a detector are shown outlined in Table 3-3.

The hypothetical duration of each step is suggested in a scenario in which the process has been tailored for Majorana significantly (but still above ground). Cleaning processes such as enrichment and zone refinement negates production of certain isotopes early in the process. Thus, the importance of each exposure has been rated subjectively based on the duration and the timing of the process.

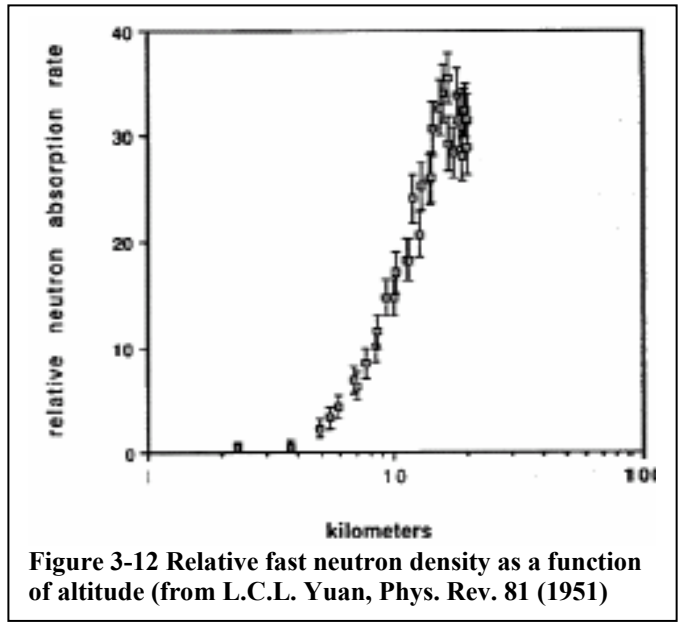

**Figure 3-12 Relative fast neutron density as a function of altitude (from L.C.L. Yuan, Phys. Rev. 81 (1951)**

### Neutrons at Altitude

The worst-case exposure in the scenario described in Table 3-3 is about 90 days. The effects of this exposure can be increased or decreased depending on strategy. One strategy, for instance, would be to fly the material from the enrichment location, taken to be in Russia and thereby cut the 30-day shipment to 2 days. However, the neutron flux, while roughly energy invariant with altitude, increases dramatically. In fact at the typical altitude of a commercial flight, 35,000 feet (10.7 km), the rate of spallation may increase to 150 times that on the surface [Hes59], negating any advantage or possibly increasing the exposure.

Fortunately, other mitigations may be more successful and many operations may be performed underground at low or no additional cost and most operations can be run underground at a modest investment.

### $^{68}$Ge vs. Enrichment

The production of cosmogenic isotopes is a function of enrichment because of the strong dependence on ΔA of the threshold for the cross section. Given this dependence and the possible variation in the composition of the enriched material, a quality standard can be established related to the $^{68}$Ge present in the enriched material at the end of the preparation process. Of course, the total quantity of $^{76}$Ge is an important value as well. But assuming that the enrichment in $^{76}$Ge is close to the preferred value (about 86%), we must consider the remaining constituents. It can be seen from Fig. 3-13 that a modest (~10-20%) admixture of $^{74}$Ge will contribute about half of the $^{68}$Ge in the detectors because cross section is about 10 times larger. However, the $^{70}$Ge must be at least a thousand times less than the $^{76}$Ge, due to the differences in the threshold and cross section. An example of the contributions of stable isotopes to the rate of $^{68}$Ge production is shown in Table 3-4.

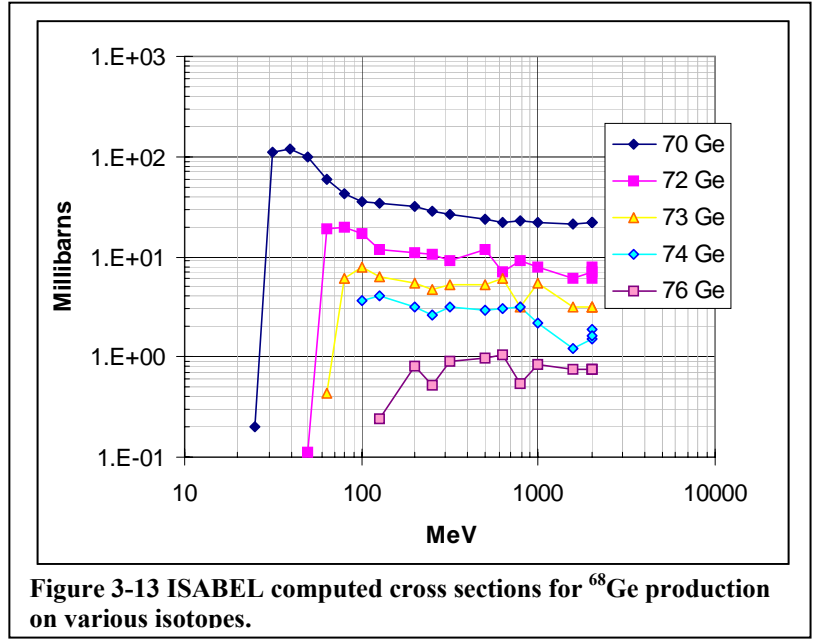

**Figure 3-13 ISABEL computed cross sections for $^{68}$Ge production on various isotopes.**

Inspection of Table 3-4 shows that the effect of greater enrichment is less $^{68}$Ge production and from this table, we determine our nominal enrichment preference of 86% $^{76}$Ge. As noted above, uncertainties in the ISABEL cross sections indicate that the absolute production rates could be somewhat different. Regardless, the activation rate might be reduced more cost effectively by shallow (20 mwe) underground storage. This is not a major concern for $^{60}$Co as this isotope is starkly reduced by the zone refining and crystal pulling processes late in the production process.

### 3.2.2 Primordial (Potassium, Uranium and Thorium)

Having considered the effects of cosmogenic activation in Ge, we turn to primordial radioactivity as there are several aspects worthy of discussion. First, the absence of high-energy (5-8 MeV) signals due to $\alpha$ decay in previous experiments constrains the level of these activities. Appendix 2 displays a table of the $\alpha$-decays from the $^{238}$U and $^{232}$Th decay chains and in most cases, the $\alpha$ is emitted without associated $\gamma$-ray or x-ray emission. Figure 3-14 shows the high-energy part of a spectrum accumulated during about one-half year of counting with a natural detector. From this spectrum we can estimate that the $\alpha$-decay rate for $^{238}$U, is less than 1 count per year. This limit corresponds to no more than about $5 \times 10^9$ $^{238}$U atoms in the kg crystal or a contamination level limit of about $6 \times 10^{-16}$ atoms U/atoms Ge for this decidedly conservative estimate. In fact, given that no observed events are present at known $\alpha$ energies from either chain (except $^{210}$Po), the contamination level must be at least 100 times less than this limit. Furthermore, if the chain were in equilibrium with a shorter half-life member like $^{226}$Ra, the limit would be about $10^6$ lower. Therefore, the U and Th content in the Ge is less than required to meet our sensitivity goals.

**Table 3-4 The $^{68}$Ge spallation rate for several individual Ge isotopes, the total spallation rates for the composition of several enrichment scenarios, including natural. There is little difference between any scenario with less than 0.1% $^{70}$Ge.**

| Ge Isotopes | $^{68}$Ge Spall Prod. per Isotope | Enrichment Scenarios | | | | | |
|---|---|---|---|---|---|---|---|
| | | Natural Enrichment | Enriched with Some Low A Contamination | Sub Nominal Enrichment | Nominal | Super Nominal | Highly Enriched |
| $^{70}$Ge | 94.20 | 20.50% | 3.10% | 0.07% | | | |
| $^{72}$Ge | 15.78 | 27.40% | | | | | |
| $^{73}$Ge | 6.31 | 7.80% | | | | | |
| $^{74}$Ge | 3.04 | 36.50% | 12.40% | 13.22% | 14.00% | 10.00% | 5.00% |
| $^{76}$Ge | 0.59 atom/kg/d | 7.76% | 84.50% | 86.71% | 86.00% | 90.00% | 95.00% |
| $^{68}$Ge Spallation Rate per Composition | atom/kg/d | 25.28 | 3.80 | 0.98 | 0.94 | 0.84 | 0.72 |

### Semiconductor Argument

The purity required for germanium to operate as a diode limits the total amount of primordial background a germanium detector can possess. For example, germanium is usually processed into $GeO_2$ but with only a purity typical of chemical processing, which can vary but is controllable. It is then reduced and zone refined. The zone refining process passes a melt region created by RF inductance back and forth through a roughly cylindrical mass of germanium metal. The temperature gradient of the melt zone sweeps elements of different melting temperatures to the end regions of the ingot. This first zone refinement results in an electrical impurity level (number of electron donors) of about $10^{13}$ electron donors /cc. This is the material that the detector manufacturers traditionally receive from the suppliers. The Ge is then zone refined again to a purity level of about $10^{11}$ donors/cc. This is the material that is introduced into the crystal puller. The crystal is then pulled into a single crystal with a concentration of donors of a few times $10^{10}$ donors/cc.

If we assume that these contaminants are typical of normal crustal abundance of the earth, the contribution of U and Th would be around a few parts per million of the electron donors, which brings us to $<10^4$ /cc or a few times $10^{-18}$ U/Ge. Assuming equilibrium with the long-lived parent, this could yield at most $2.8 \times 10^{-4}$ decays per kg-y of $^{238}$U, which corresponds to $<0.7$ decays per 2500 kg-y. Even fewer of these events could be in the region of interest. We therefore can regard this as completely negligible.

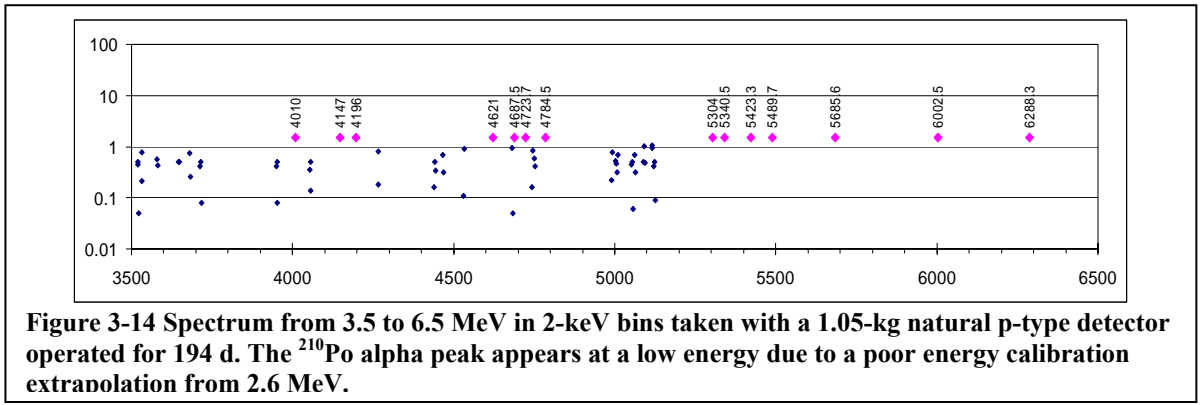

**Figure 3-14 Spectrum from 3.5 to 6.5 MeV in 2-keV bins taken with a 1.05-kg natural p-type detector operated for 194 d. The $^{210}$Po alpha peak appears at a low energy due to a poor energy calibration extrapolation from 2.6 MeV.**

*Potential Surface Contamination*

We can also constrain the contribution of primordial surface contamination using the 5.3 MeV alphas of $^{210}$Po apparent in Fig. 3-14. The peak counts are doubtless due to $\alpha$ particles striking the inner, active surface of the p-type detector with essentially no energy loss. This could be from $^{210}$Po in the inner contact or $^{210}$Po deposited on the active surface as a consequence of the deposition of airborne $^{210}$Po atoms during detector assembly. The partial energy deposition events (from ~3 MeV to 5.3 MeV) may have arisen due to $^{210}$Po decay just below the surface of the inner contact assembly or surface deposition atoms located on the passivated detector surface, which has only a very thin inactive layer. If these counts (25 total) were due to surface effects and if the efficiency was 100%, this would represent a contamination of only 39 atoms in the 0.56 y counting period shown. This could be largely eliminated by providing a moderate flow of radon-free air to the small dust-free hood used for detector assembly, but in any case, a surface contamination will decay away with a 138-day half-life. If a contaminant is in the inner contact of this detector (stainless wire) it contributes 0.03 counts/keV/kg/y or 62 counts/keV in the ROI in 2500 kg-y. As a note, this type of SS contact is no longer used.

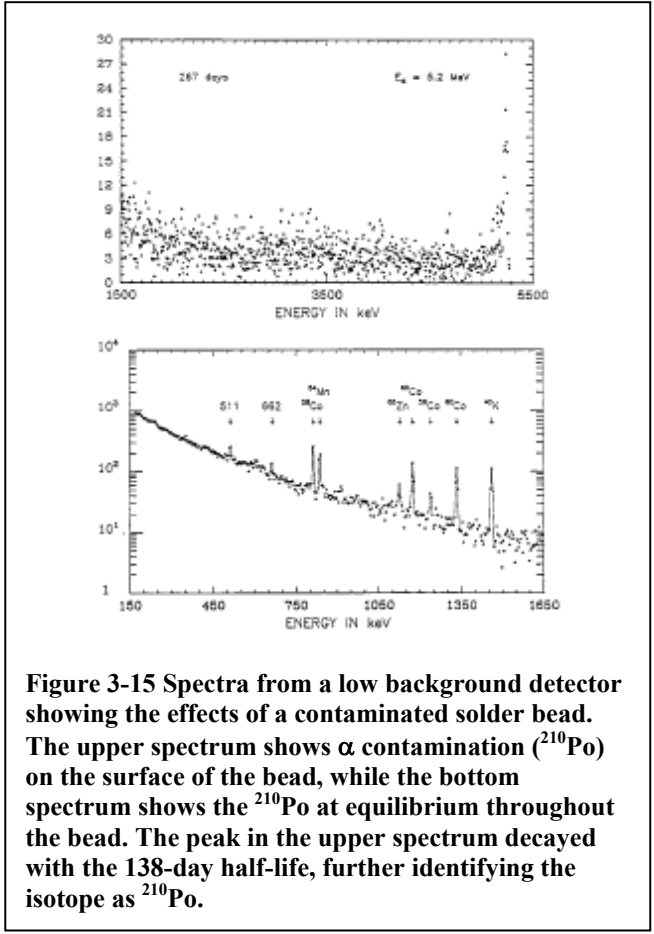

**Figure 3-15 Spectra from a low background detector showing the effects of a contaminated solder bead. The upper spectrum shows α contamination ($^{210}$Po) on the surface of the bead, while the bottom spectrum shows the $^{210}$Po at equilibrium throughout the bead. The peak in the upper spectrum decayed with the 138-day half-life, further identifying the isotope as $^{210}$Po.**

A clearer example of α signals on the active surface of a germanium detector can be observed in the earlier case of $^{210}$Po on the surface of a solder bead in the central hole of a p-type detector (Fig. 3-15). This type of contact assembly is also no longer used.

Alpha emanations are present in the natural decay chains both above and below the gas phase isotopes ($^{220}$Rn and $^{222}$Rn). Although the dead layer on a detector crystal will reduce the sensitivity to surface α emissions, the existence of a surface layer of material containing U or Th on the active surfaces (the central hole in a p-type crystal or the outside surface of a n-type crystal) would be heralded by high energy lines and continua. These effects are seen in Fig. 3-15 in the two spectra taken before the removal of the contaminated solder bead. Where α-emitting atoms are directly on the surface, a clear peak can be seen. Due to energy loss, if the alpha contamination is distributed within a part in view of an active surface of the detector, a rather flat continuum can be observed up to the expected energy of the alpha emission.

A sharp peak representing 100% detection efficiency should be observed at the energies denoted by diamonds in Fig. 3-14. The complete absence of alpha peaks above this energy essentially eliminates the possibility of U/Th contamination on the active surface other than $^{210}$Po.

### 3.2.3   Backgrounds in Cu

Copper has been a major component of ultra-low background Ge systems. It is a superior material because Cu electroplating is somewhat analogous to the zone refining process and allows strong reduction of primordial and cosmogenic contaminants. The copper is dissolved from a sacrificial electrode across a copper sulfate bath and onto a negatively charged form. After the copper is transferred to the anode, any impurities from the feed stock copper are left in the bath. A recirculation system removes large particles from the bath and a barium scavenging system removes the radium accidentally introduced via the ultra-pure starting reagents or the cover gas in the bath. Great improvements have been made in the past by recrystalization of the reagents. Additional improvements can be made by underground electroplating, sequential reagent recrystalization, and possibly by sequential electroplating, in which raw commercial electrolytic copper stock would be electroplated into high-purity feed stock underground, then finally electroplated into finished parts. This approach would also allow complex machining on high purity stock, a process that heretofore has resulted in unwanted surface cosmic neutron exposure.

### Cosmogenic Activities in the Copper

The most important contribution from cosmogenic isotopes in copper is $^{60}$Co. It has a lower activation threshold than, say, $^{70}$Ge because of the lower $\Delta A$ needed to produce it from $^{63}$Cu and $^{65}$Cu. (See Appendix 3.) For instance, $^{63}$Cu(n,2p3n)$^{60}$Co has a $\Delta A$ of 3 vs. 16 for $^{76}$Ge(n,5p12n)$^{60}$Co. Production rates of $^{60}$Co in Cu should therefore be higher than Ge for the same exposure to surface neutrons.

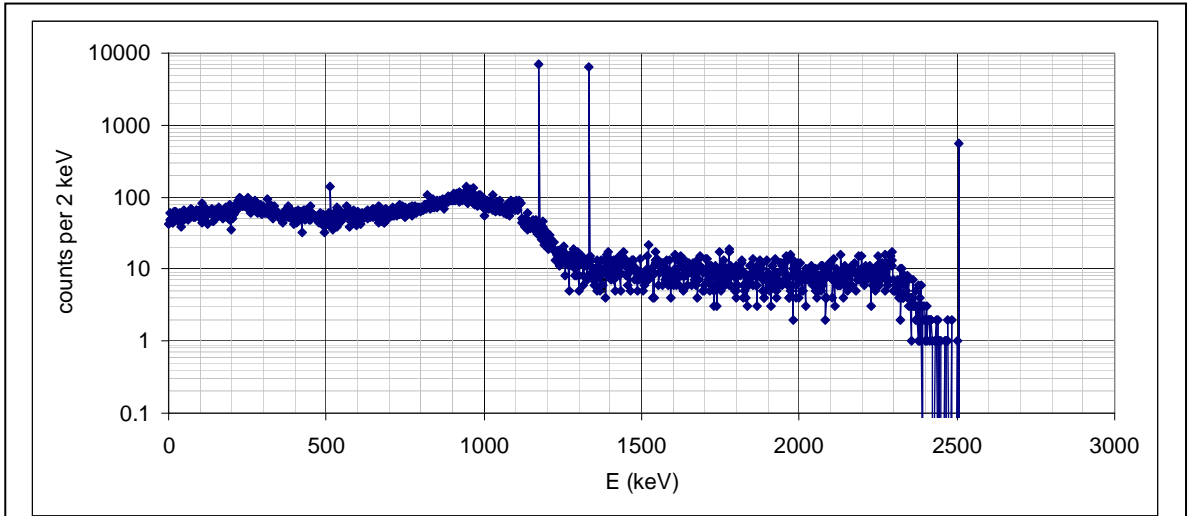

**Figure 3-16 Monte Carlo simulation using EGS4 of 170000 $^{60}$Co decays in a copper jacket surrounding a single 2 kg Ge crystal. About 35 events fall within an 8 keV window around 2039 keV. Parts further from the crystal than this jacket will have a lower interference probability.**

Unlike the $^{60}$Co in the Ge crystal, $^{60}$Co located in the copper makes a familiar spectral shape as shown in Fig. 3-16. Only scattering from the sum energy peak at 2505 keV has a reasonable chance of interfering with the 2039-keV region of interest for double-beta decay. (See Fig. 3-8.) Therefore two separate gammas of $^{60}$Co must interact in the detector to produce a background event. Given that these two gammas are emitted with essentially random relative angles, this summing from Cu contamination is somewhat suppressed. In fact, the copper jacket directly around the detector crystal can contribute only about 0.25 counts/keV per 10,000 $^{60}$Co decays near the region of interest. Also, the effects of crystal-to-crystal or segment-to-segment anti-coincidence suppression are not considered here. This additional suppression that will be present in the Majorana multi-crystal, multi-segment design, will be considered below in the computation of sensitivity.

Historically, $^{60}$Co activation in Cu has been reduced by electroplating and minimizing the subsequent exposure above ground. These approaches could be maximized in the Majorana Experiment, where the bulk creation of detectors affords an economy of scale to locate production processes underground. Since the production of copper from electroplating and finishing using clean machine tools is not expensive or particularly hazardous, this activity can be easily done underground. This should immediately eliminate the [60]Co threat from Cu materials.

*Natural Activities in the Copper ([208]Tl and [214]Bi)*

As stated above, the primordial contamination in electroformed copper has been mitigated by several chemical treatments on the plating bath. Recrystalization of the bath into $CuSO_4$ crystals reduces the contamination because the crystals tend to exclude contamination on formation. Thus the use of reconstituted electroplating bath material then results in a significantly cleaner copper. This procedure could be repeated, but has not been to date. Second, a barium scavenge in the bath filtration system is a very specific procedure for the reduction of [226]Ra. The exchange of radium and barium sulfate diminished the free radium in the bath below detectability in long, deep underground measurements. These procedures have been documented [Aal99c].

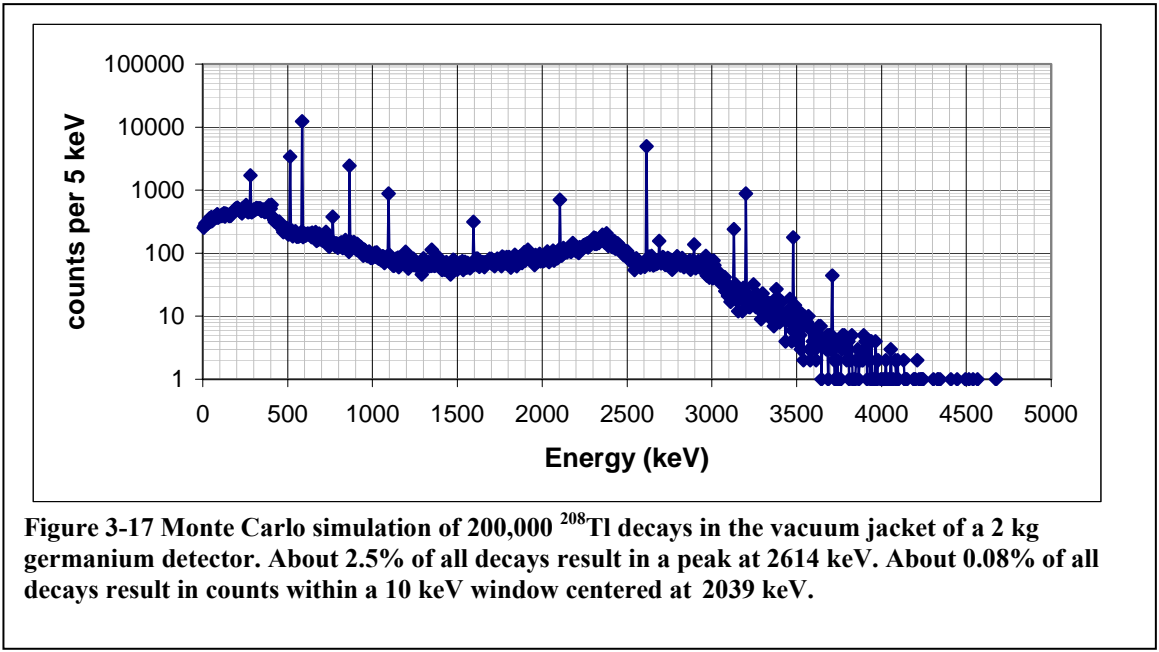

**Figure 3-17 Monte Carlo simulation of 200,000 [208]Tl decays in the vacuum jacket of a 2 kg germanium detector. About 2.5% of all decays result in a peak at 2614 keV. About 0.08% of all decays result in counts within a 10 keV window centered at 2039 keV.**

A prime example of a primordial background in copper is [208]Tl. This can originate from [232]Th or many of the daughters in that decay chain. In particular, [208]Tl can arise from solid daughters of [220]Rn, a noble gas in the [232]Th decay chain. The gaseous origin is important; if ordinary room air enters the detector chamber, decays of [220]Rn (or [222]Rn) will produce ionized atoms which have a high (~50%) probability of becoming attached to the chamber walls. These atoms will decay eventually producing [208]Tl (and [214]Bi).

Clearly, if we assume that the two main contaminant primordial chains are in equilibrium in copper and of the same magnitude, the 2614-keV γ ray from [208]Tl would be the most easily observable.

The efficiency at 2204 keV is about 20% larger than at 2614 keV and typically [238]U is about 5 times more concentrated in many environmental samples than [232]Th. Therefore, the 2614-keV line would be about twice as intense as the 2204 keV line. The [nat]Ge detectors (2 "twin" detectors of 1.05 kg) operated for about 1 kg-y had 23 background-subtracted counts within 2 full widths of 2614 keV, and 6.6 background-subtracted counts at 2204 keV. After corrections for detector efficiency (assuming the material was within 1 cm of the crystal), the ratio between the [208]Tl and [214]Bi was 4.3, in reasonable agreement with the estimate above. Therefore it appears that primordial U and Th were present very close to these detectors, possibly the copper. Primordial background reduction in copper has been refined since these detectors were built. Also, it is quite possible that the chains are not in equilibrium in the copper due to the chemical action in the electroplating bath.

A Monte Carlo simulation of [208]Tl in the copper surrounding a 2 kg detector shows that about 80 events/keV per 10000 [208]Tl decays will contribute in the region near 2039 keV. By comparison, about 2.5% of all decays contribute to the peak at 2614 keV. This complex decay scheme also should lend itself easily to segment-to-segment and crystal-to-crystal suppression techniques.

**Table 3-5 Potential Problems from Primordial Contaminants in Copper. Some very small branching fraction gamma rays above 2039 keV in [214]Bi have been neglected.**

| Isotopes | Chain | $Q_\beta$ | Gamma | Branch |
|----------|-------|-----------|-------|--------|
| [208]Tl | [232]Th | 5001 keV | 2614 keV | 99% |
| [214]Bi | [238]U | 3272 keV | 2204 keV | 4.86% |
| | | | 2118.5 keV | 1.14% |
| | | | 2447.9 keV | 1.5% |
| [234]Pa | [238]U | 2197 keV | 2072 keV | 0.004% |
| [228]Ac | [232]Th | 2127 keV | 2029 keV | 0.0019% |

Although [208]Tl is a complex decay scheme with a significant, high-energy γ ray, [214]Bi is even more complex. A number of decay paths exist from the [214]Bi ground state to the ground state of [214]Po, and a number of them have high-energy γ rays. Those decay paths listed above in Table 3-5 are about a factor of ten more intense than the next most intense in the energy range of interest.

Just using the top three most-probable decay paths, we have simulated the contribution of [214]Bi per decay to the double-beta decay energy window. About 0.1 cnts/keV per 10000 decays of Bi-214 contribute near the region of interest, about a factor of 25 less than [208]Tl. In both cases, the magnitude of the contribution would be much lower for more distant copper parts such as the vacuum or structural components. Also, the suppression factor from coincidences between segments or crystals would decrease the contributions greatly.

Importantly, these isotopes cannot contribute to the background at 2039 keV without a significant peak at 2614 keV or 2204 keV. Conversely, a low limit or measured value on the intensities of these lines guarantees no significant contribution to the double-beta decay background. For instance, in the case of the pair of natural 1.05 kg detectors, the integral of the 2204-keV line was 6 counts and the $^{214}$Bi contribution to the 2039-keV region could not have been larger than 6/30 = 0.2 counts, if the source was the copper. Similarly, the 23 counts observed at 2614 keV indicate at most 5 counts in the 2039-keV region could be due to $^{208}$Tl in copper.

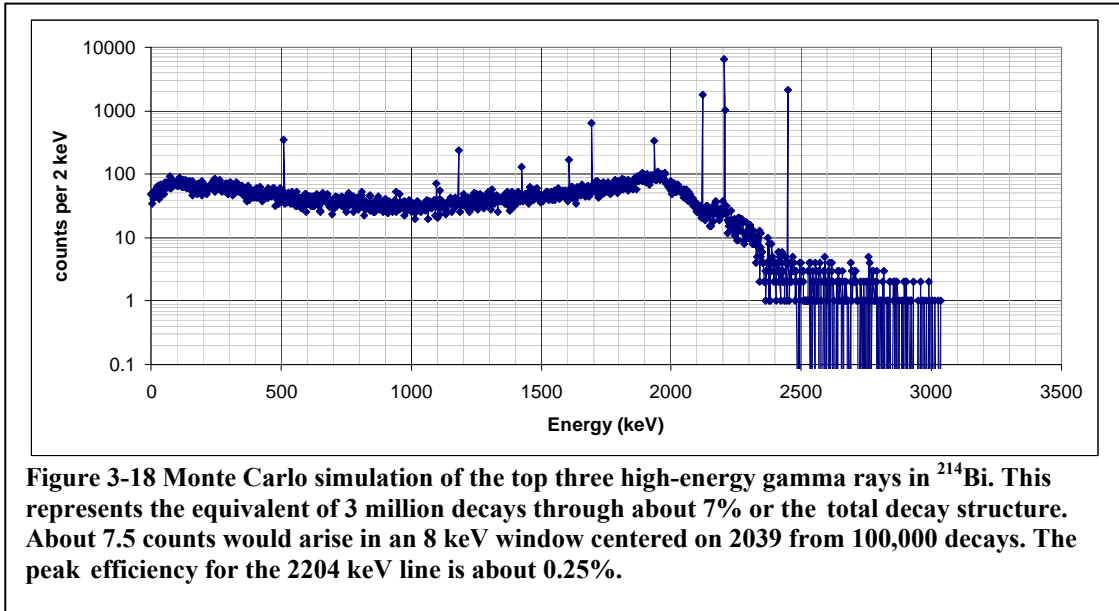

**Figure 3-18 Monte Carlo simulation of the top three high-energy gamma rays in $^{214}$Bi. This represents the equivalent of 3 million decays through about 7% or the total decay structure. About 7.5 counts would arise in an 8 keV window centered on 2039 from 100,000 decays. The peak efficiency for the 2204 keV line is about 0.25%.**

In reality, some of this activity could be due to radon daughter decays in the air space around the detector within the lead shield, and these figures therefore do not represent the best limit on the primordial contamination in the copper. However, we have established several facts: (1) $^{208}$Tl is of more concern than $^{214}$Bi, and (2) the contribution to the region of interest from these isotopes can be constrained by count rates in the 2614 keV and 2204 keV peaks, if the activity is in the copper or very near the crystals.

A much more stringent limit on the amount of primordial contamination in the copper can be created by electroplating and radioassaying a thick copper tube (essentially a Marinelli-beaker geometry), thereby greatly increasing the contribution from the copper and decreasing the contribution from other materials via the shielding provided by the copper. Such an experiment was carried out in the development of the barium scavenge technique. Prior to the implementation of this technique, the measurement of a multi-kg copper Marinelli beaker observed the products of $^{226}$Ra [Bro95] in a long count underground using one of the twin 1.05 kg natural-abundance detectors. A second measurement on an 8-kg copper Marinelli beaker provides our current worst-case estimate of copper contamination: 9 μBq/kg $^{228}$Th and <25 μBq/kg $^{226}$Ra after the introduction of the barium scavenge. The positive detection of $^{228}$Th results from attributing all possible counts to the copper. It is possible that the copper contamination levels are much lower. This experience and others as described in this section lead us to believe that we can produce copper as clean as needed in an underground setting.

**Table 3-6 Measured activities in electroformed copper.**

| Isotope | Chain | Activity | Conc (g/g) | Chain Conc (g/g) | Relevant Daughter |
|---|---|---|---|---|---|
| $^{226}$Ra | $^{238}$U | <25µBq/kg | <7.1×10$^{-19}$ | <2.1×10$^{-12}$ | $^{214}$Bi |
| $^{228}$Th | $^{232}$Th | 9 µBq/kg | 3.0×10$^{-22}$ | 2.2×10$^{-12}$ | $^{208}$Tl |

To produce this copper, we need to establish an underground copper manufacturing location free of dust, with the ability to exclude radon in production and storage, and an ultra-low background counting system to perform checks on materials and parts. Within this modest facility, we should be able to perform repeated purification steps to the purity level required.

### 3.3.4    Backgrounds in Pb

Lead is frequently used as shielding. Historically, in double beta decay experiments the inner 10 cm of shielding has been from ultra-low background Pb, while the remainder of the Pb shield was merely screened for alpha activity.

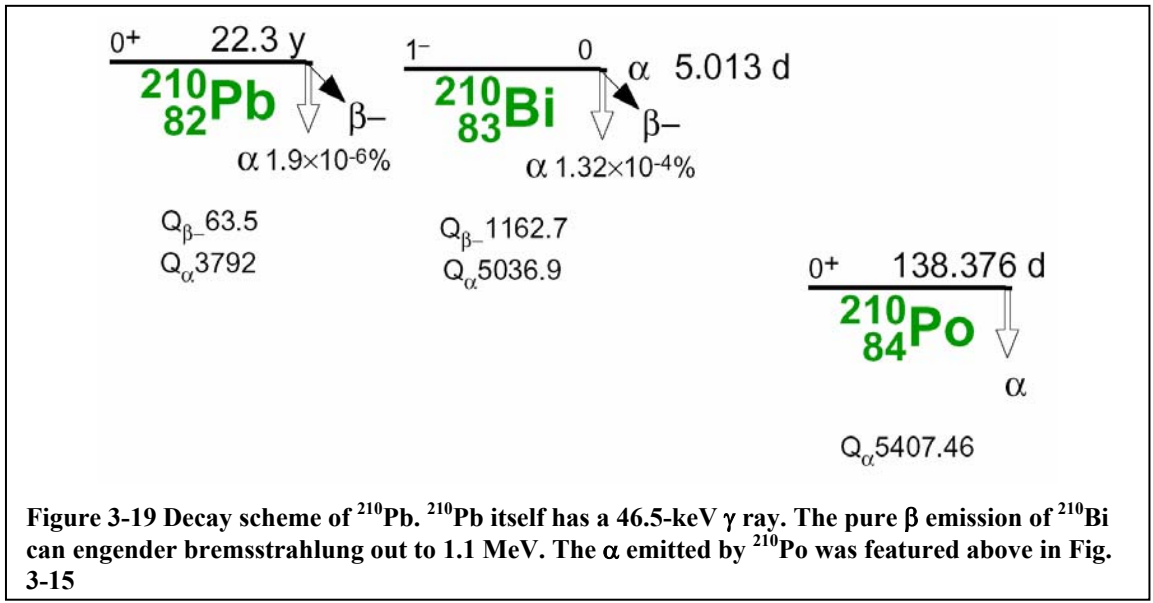

**Figure 3-19 Decay scheme of $^{210}$Pb. $^{210}$Pb itself has a 46.5-keV γ ray. The pure β emission of $^{210}$Bi can engender bremsstrahlung out to 1.1 MeV. The α emitted by $^{210}$Po was featured above in Fig. 3-15**

Lead ore is frequently found close to uranium deposits, so the source may need to be carefully chosen. In addition, Pb is frequently contaminated by the use of reagents and processing materials containing uranium and thorium. As a result, after chemical processing to remove other elements, the lead may be contaminated with $^{210}$Pb. The decay scheme of $^{210}$Pb (see Fig. 3-19) yields no problems for zero-neutrino double-beta decay, assuming the lead is not in direct contact with an active surface of the detector.

From the IGEX data indicating a background rate of 0.2 counts/keV/y/kg near the 2-MeV region of interest, one can derive an upper limit on the U/Th contamination of the Pb that neighbors the detectors. The Canfranc detector [Aal99c] had 6 kg of Ge inside 2.5 tons of old lead. A $^{208}$Tl decay in the Pb has a $1\times10^{-5}$ probability of producing a count in the region of interest. This translates to an activity of about 1 $\mu$Bq/kg. At this level, the Pb would produce a negligible contribution to the background. Furthermore, if the Pb turned out to be significantly higher in radioactivity, the inner layer could be replaced with electroformed copper.

Lead is not very vulnerable to activation from cosmic neutrons. Table 3-7 shows a partial listing of isotopes with half-lives longer than 30 days which might be produced by spallation from Pb. As we saw previously in Fig. 3-13 above showing the variation in the threshold and cross section vs. $\Delta$A, long before we get to a $\Delta$A of 40, the realistic possibility of spallation is unrealistic. In any case, the potentially produced isotopes have little impact above 500 keV.

The lead outside the inner shield area need not be as clean as the inner shield. Ten cm of Pb permits only about 0.5% transmission of gamma rays in the 2-MeV range. If the proper quantity of ultra-pure Pb is not available, it is possible to replace the inner Pb with electroformed Cu. Copper is not prohibitively expensive to produce. The transmission fraction at 2 MeV is about 2% for Cu, or about 4 times larger than Pb.

The contamination level of Pb can only be determined by long underground counting, much as the Cu case. However, it is possible, for about $1000 per sample, to test the $\alpha$ contamination using an above ground $\alpha$ counting procedure using semiconductor-grade reagents, produced by multiple sub-boiling-distillation (SBD) passes. This procedure has been used to qualify the outer shielding in the past.

**Table 3-7 Potential Spallation Isotopes of Lead**

| Isotope | T1/2 | Max E$_\gamma$ | Delta A |
|---------|------|------|---------|
| Pb 205 | $1.40\times10^7$ y | | 1 |
| Tl 204 | 3.78 y | 82 | 2 |
| Hg 203 | 46.6 d | 279 | 3 |
| Pb 202 | $3.00\times10^5$ y | 85 | 4 |
| Au 195 | 183 d | 211 | 11 |
| Ir 194m | 171 d | 687 | 12 |
| Os 194 | 6 y | 82 | 12 |
| Ir 192 | 74 d | 612 | 14 |
| W 188 | 69 d | 290 | 18 |
| Os 185 | 94 d | 880 | 21 |
| Re 184 | 38 d | 903 | 22 |
| Re 183 | 70 d | 291 | 23 |
| W 181 | 121 d | 152 | 25 |
| Hf 181 | 43 d | 482 | 25 |
| Ta 179 | 664 d | 65 | 27 |
| Lu 177m | 161 d | 418 | 29 |
| Hf 175 | 70 d | 432 | 31 |
| Lu 174 | 3.3 y | 1318 | 32 |
| Lu 174m | 142 d | 1318 | 32 |
| Lu 173 | 500 d | 635 | 33 |
| Hf 172 | 683 d | 125 | 34 |
| Tm 171 | 701 d | 66 | 35 |
| Tm 170 | 129 d | 84 | 36 |
| Yb 169 | 31 d | 307 | 37 |
| Tm 168 | 87 d | 821 | 38 |
| Ho 166m | $1.20\times10^3$ y | 830 | 40 |
| Ho 166 | 26 h | 1749 | 40 |
| Ho 163 | 33 y | 53 | 43 |

### 3.3.5   Other Materials

A short suite of materials has been qualified for use in germanium double-beta decay experiments in the past. Firstly, we require no special low-background materials outside the bulk Pb shield, only a low dust situation to prevent contamination of the inner shield during maintenance. Materials inside the shield not produced from electroplated Cu include thermal insulators and structural parts (plastic), wires, electrical contacts (metalized plastic), and vacuum seals (indium).

The collaboration does not produce these materials. Rather, they are carefully selected from commercial sources then screened in large quantities. As an example, a small coaxial cable was selected for use by the above-ground counting of hundreds of meters of cable. The required purity depends on the mass of the material to be used. In the IGEX detector system, only 30-40 cm of cable was required, or only about 2 g of material. Depending on the configuration, 20 to 50 times as much will be inside the multi-crystal cryostats of the Majorana Experiment. Above ground screening will not suffice for the increased purity check required. It should be noted that several options exist should no commercial cable prove sufficient, including collaboration manufacturing of cable underground.

In some circumstances, large quantities of a material may not be available for screening. This is particularly true for the field effect transistors employed as the front end of the preamplifier. These devices are already produced specially with no mounting (i.e. not in metal cups with potting compound and large leads). These devices have 'flying leads' and are intrinsic to a small quantity of semiconductor-grade silicon. Because the mass is so low, the only precaution taken historically with these devices has been gross screening in an above-ground counter. They are placed behind a disk of Pb within the vacuum jacket for γ-ray and thermal considerations. The availability of an ultra-low level apparatus underground for screening these low mass parts will allow more careful selection of parts, especially from the viewpoint of surface contamination introduced

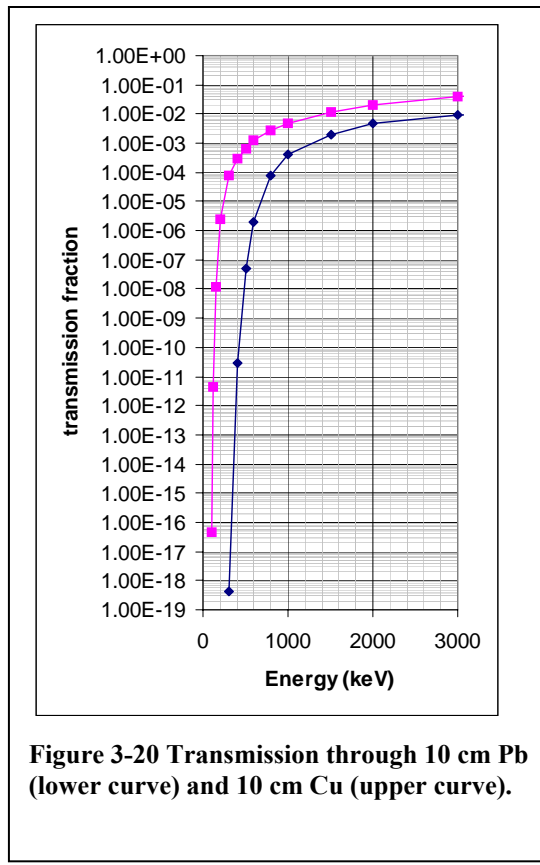

**Figure 3-20 Transmission through 10 cm Pb (lower curve) and 10 cm Cu (upper curve).**

from handling.

Bulk materials inside the shield (i.e. inside the cryostat) have approximately the same purity requirement as the Cu. The expense of these materials is not sufficient to prohibit bulk counting. Care must be exercised to prevent surface contamination of items that naturally develop static surface charges, however, as these tend to attract the charged radon daughters attached to normal aerosols.

### 3.3.6    Sensitivity Calculations

Given that we have established the significant contributors to the background at 2039 keV in our germanium array, we now estimate the sensitivity the Majorana Experiment might ultimately achieve. However, the estimates of the current and achievable backgrounds of radioactivity and radiation do not include the capability of the system to reject background through data analysis means. First, the 500 kg of detector mass will be in many discrete crystals. In addition, these crystals may be segmented such that the independent segments have ~500 g of Ge or less. In addition, the individual segments will be instrumented such that the time evolution of each individual pulse can be used to extract multiplicity and location information, thus further suppressing backgrounds. (Most background radiation interactions with the crystals take place in multiple locations within a 1-μs interval, while double-beta decay deposits energy at only one resolvable location.) Finally, temporal coincidence may be used to eliminate backgrounds arising within or in close proximity to the crystals. The usefulness of these background suppression methods can be estimated against a particular design. To this end, we have simulated several reasonable designs.

We present two approaches to sensitivity calculation: first, a calculation based solely upon previously attainable contamination levels. This approach suffers from near worst-case levels because it assumes that the Majorana construction plan will not successfully implement the various planned improvements. One positive advantage of this approach is the empirical basis of the understanding of the sources of background and the excellent potential for mitigation with suppression techniques.

Our second approach is to estimate the backgrounds from first principles and determine the levels of contamination we propose to achieve in the Majorana construction plan. From these estimates we can also project sensitivity. While this estimate will show the full potential of the background removal techniques, it is not based on previous successes but on anticipated performance. However, these two approaches should reasonably bound the sensitivity of the Majorana Experiment, from no risk to limited risk.

#### Efficacy of Background Cutting Techniques

Both these sensitivity-estimate approaches have been implemented in a grand simulation scheme working on one module of the apparatus, assumed to contain 57 one-kg crystals segmented into 500 g units. This simulation includes the entire decay scheme of several isotopes, such as $^{208}$Tl, $^{60}$Cu, and $^{68}$Ge. The background suppression can be separated into several separable causes: radial pulse shape discrimination, segmentation suppression, and crystal-to-crystal suppression.

Time series analysis, as distinct from pulse shape analysis or prompt (~1 μs) coincidence, uses the fact that many isotopes decay in succession. As long as the overall count rate is low compared to the decays rate of contaminants in or on the crystals, one may look for those other signals outside the energy window of interest to flag and remove signals in our analysis region about 2039 keV.

### Impact of Uncertainties in Worst Case

In Table 2-3 above, the background estimate assumed no primordial isotopes contributed. The entire rate is assigned to backgrounds of cosmogenic origin and its magnitude is scaled from the IGEX measurements described above in this subsection. While this assumption is the case when adequate care is exercised in crystal preparation and copper production, this conclusion also arises from the fact that the cosmogenic activity computed for the enriched detector accounted for the entire signal in the region of interest. (See Figs. 3-7, 3-10 and Table 3-2.) However, there are experimental uncertainties and of course, uncertainty in the germanium reaction calculations, as noted above. In the following we describe variations on the background model that permit fractions of the background rate to be assigned to other sources. Essentially here we are taking a worst-case cosmogenic contamination and considering if this activity arose from other forms.

To provide a simple estimate of this impact, we take the entire uncertainty, 25% of the raw experimental rate in Table 2-3, to be from long-lived primordials and not cosmogenics. This would result in 892 total counts in the Region of Interest. Applying the same suppression factors for PSD and segmentation, 32 counts would remain from this source. This compares to about 7 counts from all sources, namely cosmogenic, in the Table 2-3 estimate. The increase in background from 7 to 40 counts would raise the effective mass sensitivity by the fourth root of the ratio of 40/7, or a factor of 1.54. This is not really acceptable, but would only change the ultimate sensitivity from $|<m_\nu>| \sim 50$ meV to $|<m_\nu>| \sim 70$ meV. Therefore, even this unanticipated worst-case scenario is not a show stopper.

If the source of these counts were electroplated copper, this would be expected to decrease by at least one order of magnitude in a cleaner, multi-crystallized plating bath, which would change the net background counts from 7 to 10, with no real impact on the mass sensitivity. In addition, this estimate would likely have to be revised down by a factor of 5 to represent the lower copper to germanium mass ratio in the Majorana Experiment over IGEX detectors.

Some thought has been given to the chemical treatment of the germanium crystals in the manufacturing process. Several surface treatments need to be examined for the primordial isotope contamination potential. Because of the concentrated nature of the materials used to treat the surface, we anticipate that screening in an ultra-low level underground facility should eliminate this possibility quickly or allow identification and substitution of cleaner materials.

### Sensitivity to Contaminations in Other Materials

Reference [Bro95] estimated that maximum activity present in the copper. It is possible the activity detected did not reside in the copper. The activity could not have originated from the lead shielding (since it did not decrease when 2.54 cm of copper was placed between the detector and the lead), but it is possible that the activity resided in the small parts inside the detector (electronic amplifier front-end consisting of a silicon FET, cable, bonding agents and interconnects) or in the surface coating of the germanium detector. If we take the activity observed (9 μBq/kg for 8 kg of Cu) and assume that it resided in either the small parts or the surface layer, we see interesting results. Table 3-8 compares the potential contributions from these sources to the contribution from cosmogenic contributions in enriched germanium similar to the early IGEX experiment extrapolated to 2500 kg-y of Majorana operation with pulse shape and segmentation background suppression.

Table 3-8 essentially recaps the descriptions of potential backgrounds above. Interestingly, this simple calculation and more sophisticated calculations agree that the activity level in the copper ❶ needs to be about 1 μBq/kg to contribute negligibly. We regard this as a low risk development. If this activity is in the germanium dead-layer ❷, it must also be about 1 μBq/kg. Fortunately, existing underground screening capabilities of the Collaboration are sufficient for this screening study. If the materials used to establish

**Table 3-8 Comparison of cosmogenic and other backgrounds. Based on 2500 kg-y exposure, the cosmogenic figures are deduced from early IGEX data, the others that are non-zero (❶,❷, and ❸) are based on the measured values attributed to Cu in [Bro95]. These cannot be simultaneously true, but point to the heightened susceptibility to the small, close-in parts. The legacy rate is the anticipated decay rate at the time that data taking begins.**

| Background | Legacy Rate | Raw ROI counts | After PSD | After Segmentation |
|---|---|---|---|---|
| Cosmogenic in Ge | | | | |
| $^{68}$Ge | 3.93x10$^{-03}$ | 35.7 | 9.295 | 1.90 |
| $^{56}$Co | 6.43 x10$^{-05}$ | 0.57 | 0.15 | 0.03 |
| $^{60}$Co | 7.15 x10$^{-03}$ | 63.25 | 16.9 | 3.08 |
| $^{58}$Co | 5.60 x10$^{-06}$ | 0.05 | 0.015 | 0.00 |
| | cts/keV/kg/y | counts | counts | counts |
| | | | | |
| Subtotal | 1.12 x10$^{-02}$ | 99.475 | 52.72 | 5.01 |
| | cts/keV/kg/y | counts | counts | counts |
| ❶ Primordials in Cu | 3.60 x10$^{-04}$ | 113.53 | 34.06 | 6.95 |
| | Bq/40 kg | counts | counts | counts |
| ❷ Primordials in Ge surface layer | 1.52E$^{-07}$ | 229.45 | 68.83 | 14.05 |
| | Bq/g | counts | counts | counts |
| ❸ Close In Parts | 7.20 x10$^{-07}$ | 2724.71 | 817.41 | 166.83 |
| | Bq/g | counts | counts | counts |
| Cosmogenic in Cu | negligible | 0 | 0 | 0 |
| Primordials in Pb | negligible | 0 | 0 | 0 |
| Primordials in bulk Ge | < 2.90 x10$^{-4}$ | <1 | 0 | 0 |
| | decay/kg/y | decay per | | |

p-n junctions in detectors are indeed impure, they can be purified using ordinary chemical separation techniques. A more difficult case arises for the small, close-in parts ❸.

We estimate these to total about 20 g of material. Half of this mass (cable) is already screened to the required contamination level and can be screened much lower. Thus, the cable can be screened to eliminate this background risk. The ~8 g of contacts and wires, can be screened in ~500-g batch, resulting in an acceptable contamination limit. The silicon die (representing the final 2 g of close-in parts) can be screened in a 100-g batch reaching a sensitivity level about a factor of 5 above the requirement using existing detectors operated underground by the Collaboration. Therefore, the silicon parts will require counting in the Multi Element Gamma Assay (MEGA) device under construction by the Collaboration in order to screen the level of contamination required. It should be noted, however, that the silicon FET is expected to be as pure as the bulk germanium and at least no worse than the doped germanium surface. Thus, the only potential difficulty is with the interconnect materials: wire and epoxy. If the epoxy is found to be a problem, we have used a silver based material in the past that was found to be sufficiently clean and capable of being produced by the Collaboration from ultra-pure materials.

**Table 3-9. Background rates predicted based on first principles. A 2500 kg-y exposure is assumed. Screening of materials can greatly reduce the potential background using an underground facility (level 1 remediation) and screening with an advanced system, the Multi-Element Gamma Assay (MEGA) system (level 2 remediation). The results in this table are conservative because the activity levels used here are higher than the desired levels discussed in the text.**

| Background | Computed Rate | Total ROI Counts | After PSD | After Segmentation | After Remediation Level 1 | After Remediation Level 2 |
|---|---|---|---|---|---|---|
| | | | | | Undergnd. Screening | MEGA Screening |
| Cosmogenic Ge | | | | | | |
| $^{68}$Ge | $4.62 \times 10^{-03}$ | 41.98 | 10.93 | 2.23 | 2.23 | 2.23 |
| $^{56}$Co | $1.29 \times 10^{-04}$ | 1.15 | 0.30 | 0.06 | 0.00 | 0.00 |
| $^{60}$Co | $9.39 \times 10^{-04}$ | 8.31 | 2.22 | 0.40 | 0.00 | 0.00 |
| $^{58}$Co | $1.12 \times 10^{-05}$ | 0.10 | 0.03 | 0.01 | 0.00 | 0.00 |
| | cts/keV/kg/y | | | | | |
| Subtotal | $1.12 \times 10^{-02}$ | cts | cts | cts | cts | cts |
| | cts/keV/kg/y | 51.54 | 13.48 | 2.70 | 2.23 | 2.23 |
| Primordials in Cu | $3.60 \times 10^{-04}$ | 113.53 | 34.06 | 6.95 | 0.70 | 0.07 |
| | Bq/40 kg | counts | counts | counts | counts | counts |
| Primordials in Ge surface | $1.52 \times 10^{-07}$ | 229.45 | 68.83 | 14.05 | 0.94 | 0.09 |
| | Bq/g | counts | counts | counts | counts | counts |
| Close In Parts | $7.20 \times 10^{-07}$ | 2724.71 | 817.41 | 166.83 | 1.67 | 0.17 |
| | Bq/g | counts | counts | counts | counts | counts |
| Cosmogenic in Cu | NA | 0 | 0 | 0 | 0 | 0.00 |
| Primordials in Pb | NA | 0 | 0 | 0 | 0 | 0.00 |
| Primordials in bulk Ge | $< 2.90 \times 10^{-4}$ | <1 | 0 | 0 | 0 | 0.00 |
| | decay/kg/y | decay per 2500 kg-y | | | | |
| Sum | | 3120 | 934 | 190 | 5.5 | 2.6 |

To set an overall worst case scenario, we must combine the cosmogenic ROI counts based on early IGEX data with some combination of ❶, ❷, and ❸. Since these materials are relatively easily screened and mitigated, we will take ❶ to serve as an estimate of a partially mitigated primordial contribution. The ROI background sum in this estimate is then about 12 counts for a 2500 kg-y exposure. This will be compared to potential signal levels and an estimate of background calculated from first principles below.

### *Based on Predicted Future Capability*
If we assume the surface exposure described in Table 3-3, we need only consider the cosmogenic impact of 84 days of exposure for $^{68}$Ge and 15 days for $^{60}$Co. In addition, if we use the contributions from other materials listed in Table 3-8 except with the mitigations discussed in the previous paragraph, we have the results of Table 3-9.

Regarding the case of level 2 remediation, we assume a screening capability at levels attained in double-beta decay facilities such as MEGA that are about a factor of 100 more sensitive than good measurements on the Earth's surface. It is also assumed that no cobalt isotopes remain in germanium crystals produced underground. It is interesting to note that after 2500 kg-y, the $^{68}$Ge is essentially all decayed away. A subsequent 2500 kg-y operation would have essentially no cosmogenic background, if the detectors were fabricated underground. If primordial backgrounds are identified and eliminated, perhaps a second experimental campaign could be conducted with greatly suppressed backgrounds.

**Table 3-10. ROI counts in 2500 kg-y of operation as a function of double-beta decay half-life. Germanium enriched to 86% in $^{76}$Ge is assumed.**

| $^{76}$Ge 0ν Half-Life | ROI counts |
|---|---|
| 1.00x10$^{+25}$ | 1180.59 |
| 3.16x10$^{+25}$ | 373.34 |
| 1.00x10$^{+26}$ | 118.06 |
| 3.16x10$^{+26}$ | 37.33 |
| 1.00x10$^{+27}$ | 11.81 |
| 3.16x10$^{+27}$ | 3.73 |
| 1.00x10$^{+28}$ | 1.18 |

The projected background sum in the worst-case scenario presented in Table 3-8 is 12 counts based on the cosmogenics and the primordials residing in the copper. Estimates made on first principles in Table 3-9 indicate 2.6 counts of background. In Table 3-10, we can see that a signal to noise ratio of 1 would be obtained at a 0νββ half-life of 1×10$^7$10$^{27}$ y or 5×10$^7$10$^{27}$ y respectively. These correspond to $|<m_v>| \sim 75$ meV and $|<m_v>| \sim 35$ meV, respectively. The approximations are to remind the reader of the uncertainty associated with the matrix elements.

The conclusion is that under fairly widely disparate assumptions, the Majorana Experiment will reach the desired effective Majorana neutrino mass sensitivity. In reality, these results probably bound the mass sensitivity attainable in 2500 kg-y of operation.

## 3.4 Monte Carlo Simulations

Monte Carlo studies of the Majorana apparatus will lead the final stages of design and are essential in interpreting the results of the measurement data. The very limited known list of historically observed radio-isotopes serve to simplify the simulation problem. The great majority of the materials used in the experiment include Pb, Ge, and Cu. On the other hand, the apparent complexity of the potential detector configuration (as many as 4000 segments in ~500 crystals grouped in crystal modules) is surprisingly simple. Simulations run by Collaboration members have shown that this complexity is tractable and we have studies of the background impact from several configurations.

The goals of the simulation effort can be stated as follows:

What is the response of individual crystals to several sources of radiation?
- Natural isotopes in Cu and Pb and other materials
- Cosmogenics in Ge and other materials
- Cosmic muons
- Cosmic muon secondary neutrons
- Fast neutrons and nuclear recoils
- Rn in the volume surrounding the crystals
- $2\nu$ double-beta decay
- $0\nu$ double-beta decay
- Calibration sources

What is the collective response to these sources (i.e. self shielding effects)?
What are the effects of proposed background cutting measures?
- Pulse-shape discrimination
- Simple segmentation
- Segmentation with inferred position/multiplicity info

These questions have implications on $0\nu\beta\beta$ (2039 keV) and, at low energies, for dark matter. Validation of the results of this approach can be obtained by comparison of $2\nu\beta\beta$ and calibration simulations to experimental data obtained in SEGA and MEGA, and ultimately with Majorana itself.

The basics of the simulation, i.e. single-crystal energy deposition responses, have been developed previously using a code based on EGS4 [Mil94]. More recently, a complete pulse simulation combined EGS4, a pulse formation code based on finite element analysis, and SPICE circuit simulation.

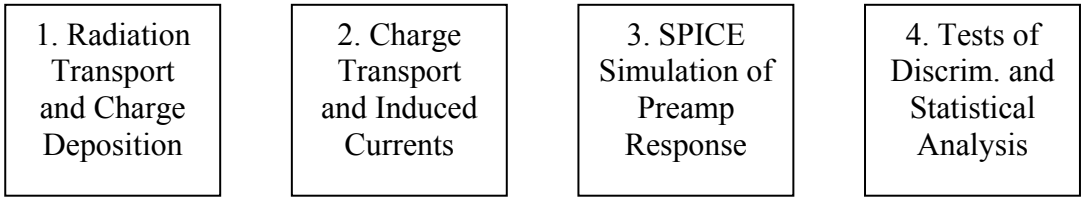

**Figure 3-21. Complete pulse simulation approach.**

Early development of pulse-shape discrimination was guided by this approach [Aal99b]. Comparisons between simulated and experimental data helped to elucidate those aspects of crystal properties and variations in electronic instrumentation causing large variations in observed pulse shapes. The observation of the large variations led to the development of a purely observational, self-calibrating pulse-shape discrimination approach.

The effects of segmentation have been addressed by evaluating a simple segmentation cut using energy deposition location details from the EGS4 simulation output. These elementary results are quite promising. It is anticipated that finer localization than the actual segmentation could be inferred from simple signal processing of adjacent-segment signals [Vet00]. However, to go beyond the simple segmentation-rejection results mentioned above, signals from real segmented detectors must be analyzed, in order to produce realistic refinement of the sensitivity calculation. The results of SEGA will address this need.

Several of the results of interest have already been produced, for example the response of the pulse-shape discrimination method to internal and external sources of radiation, including $0\nu\beta\beta$ decay. However, because multi-crystal depositions will also strongly identify signals as non-single-site, these methods have been expanded to include the complete Majorana apparatus response. The multi-crystal geometries and the ability to transport hadrons and score their energy depositions are needed. These features have been implemented by adopting GEANT4 for simulation. Besides allowing more detailed physics models, GEANT4 has the ability to accept geometry files created by the 3-d solid modeling mechanical engineering software (SOLIDWORKS) used to design the Majorana detection hardware. Thus, extremely complete geometry files may be created, shared, and maintained with modest effort.

Natural backgrounds generally include U and Th daughters, and $^{40}$K. (See Appendix 2: Natural radioactivity decay chain data.) Of particular note are gamma rays from $^{208}$Tl and $^{214}$Bi owing to their energy sufficient to reach the 2039 keV energy region-of-interest. Special cases exist like $^{210}$Pb and $^{210}$Bi in the lead shielding, which pose a difficulty for low-energy measurements.

Cosmogenic radioactive isotopes in Ge are mainly $^{68}$Ge and $^{60}$Co. Shorter-lived cosmogenics, such as $^{58}$Co and $^{56}$Co need to be simulated also because during early operation their presence can be used to validate the Monte Carlo and test electronic background-suppression methods. Cosmogenic activity in the copper should be non-existent due to the underground electroforming location and the underground storage of copper anode material.

Neutrons present a two-pronged problem: low-energy spontaneous fission or ($\alpha$,n) neutrons ($<E> \sim 1$ MeV) and high energy muon-secondary neutrons ($<E> \sim 100$ MeV). For the high-energy neutrons, simulations will help optimize the shielding configuration of the experimental setup.

The results of the simulations done to date are discussed below in Section 3-5 in the context of the detector configuration and manufacture.

## 3.5 Isotopic Enrichment of $^{76}$Ge

The major requirement of the Majorana Experiment is the procurement of isotopically enriched germanium. Previous double-beta decay experiments have simply borrowed the enriched material from Russian collaborators. Due to losses in the processing and manufacturing, certain complications can arise in the eventual return of the material. The scale of the next generation of double-beta decay experiments, both in mass and duration preclude the possibility of borrowing such large quantities of material.

The Majorana Collaboration has begun investigating the aspects of enrichment both as relating to science and project management. ITEP has played a crucial role in the investigations, largely because of their previous involvement in enrichment activities. Several discussions with the Electrochemical Plant (ECP) of Zelenogorsk have elucidated the main challenges in the enrichment process, which are simply financial and not technical. In other words, the Collaboration has been assured that the ECP can produce germanium of the same purity and enrichment as has been provided for several double-beta decay experiments in the past, including IGEX and Heidelberg-Moscow. Therefore, we have determined that at least one facility can produce the required material.

### Production Facilities

So far the Majorana Collaboration has investigated only one producer of isotopically enriched materials. Now that we have established that one supplier exists, we will explore other options. It may be possible for example, to produce germanium at a net higher rate by contracting with more than one enrichment facility. We will, with the aid of a National Laboratory procurement office, perform a detailed market study of the question. We also are investigating other potential technologies for the enrichment of Ge.

### Enrichment Costs and Schedule

A delegation of the Majorana collaboration traveled to Moscow in October of 2001 and met with the representative of the Electro-Chemical Plant and members of his staff at the Institute for Theoretical and Experimental Physics (ITEP). At that meeting a Memorandum of Understanding was signed, cost estimates and scenarios for production rates and delivery were provided.

The ECP, a centrifuge facility, presently has capacity to produce 30 kg/y of material to the standards required for double-beta decay. Provided an investment is made in increased production capacity, matched by ECP, this capacity can be increased to 50 kg/y. They have recently quoted the Majorana collaboration that prices for germanium produced under the 30 or 50 kg/y rates are greater than for larger production rates. (See Appendix 1: Enrichment)

ECP has offered that for a modest investment, the production capacity could be increased to 100 kg/y. Given that the first year or two would be at 'low' production rates (20-50 kg/y), a total production period of seven years would result. This is quite a long commissioning period and not our preferred option

The ECP has offered that for a somewhat larger investment, capacity could be increased to, an annual production of 200 kg. This production rate means that only two years of main production would be required in addition to the startup period. This is our Reference Plan, on which we base the estimates of the time evolution of the sensitivity. A total of four to five years of production would substantially speed the commissioning of the Majorana apparatus. ECP estimates that production under these conditions would be the most cost effective.

### Shipping

While the Reference Plan includes shipping via sea from Russia to a point of entry into the US, it may not be in the scientific best interest for the material to be shipped in this way. The high-energy neutrons that cause the main cosmogenic backgrounds ($^{68}$Ge and $^{60}$Co) are produced in cosmic-ray reactions in the atmosphere. This high-energy neutron flux is about 15 times more intense at the cruising altitude of aircraft than at ground level. Thus, 12 hours at 12,000 meters would be the equivalent of 8 days exposure at sea level. While land/sea transit would be much slower and involve some complicated border crossings, the neutron exposure would be half or less, and could be much less if clever shielding and placement were pursued.

Because a shipment of a 3-month production at 200 kg/y would be 50 kg of metal (actually more of oxide), the shipments will be escorted to help prevent loss of such valuable material. Assuming an oxide density of about 2.0 g/cc, the volume of a quarterly shipment would be a cube of about 30 cm, or one cubic foot. Some shielding of this amount of material should be possible. If a factor of two to three in neutron exposure reduction could be accomplished, the air transport would be competitive with land/sea with fewer complications of customs and border crossings.

### Taxes, Duties, Customs

While the ECP quote includes Russian taxes, duties and customs, we wish to test the procedures on an initial, relatively small shipment. We also plan to use National Laboratory expediting procedures and contacts to estimate the potential for difficulty in this area. Further, the Collaboration will explore the tax exemption of isotopes in general in the US.

### Isotopic Enrichment and Material Processing Flow Diagram

Although the natural isotopic abundance of $^{76}$Ge is 7.83%, the Heidelberg-Moscow [Kla01] and IGEX [Aal02] collaborations imported a combined 36 kg of Ge, isotopically enriched to 86% in $^{76}$Ge, from the Soviet Union. In total about 12 detectors were fabricated from this material. The IGEX collaboration operated 6 detectors with a total fiducial mass of about 8 kg. Most of the material used in the IGEX experiment came from the clean laboratory at the ECP, where the germanium is isotopically enriched. These experiences demonstrate that the facilities in Russia are capable of providing clean $^{enr}$Ge from which detectors for the study of 0νββ can be fabricated. The Russian enrichment technology is available and ready to produce hundreds of kilograms of the ~86% $^{76}$Ge material required for the Majorana Project.

The Germanii Plant resides in Krasnoyarsk only a few tens of kilometers from the ECP in Zelenogorsk. This plant has the facilities to convert the enriched germanium, from the GeO$_2$ provided by the ECP to metal and then zone refine it to industry "intrinsic" standard (~$10^{13}$ impurities/cm$^3$). This is the usual purity level accepted by the detector manufacturers who then further zone refine it to approximately $10^{11}$ impurities/cm$^3$. From this material, they pull the single Ge crystals with impurity levels between 2 and 3 times $10^{10}$ p- or n- type impurities/cm$^3$ depending on the type of detector to be produced.

The proposed scenario for isotopic enrichment, recovery in oxide form, reduction to metal, and first zone refinement is depicted in Fig. 3-22. The waste products from this phase can be reprocessed without leaving the Germanii Plant. The final product can be shipped to Saint Petersburg, Russia and then onto the United States by the fastest ship available. There a courier will drive it directly to the detector manufacturer where it would be processed as described in the next section.

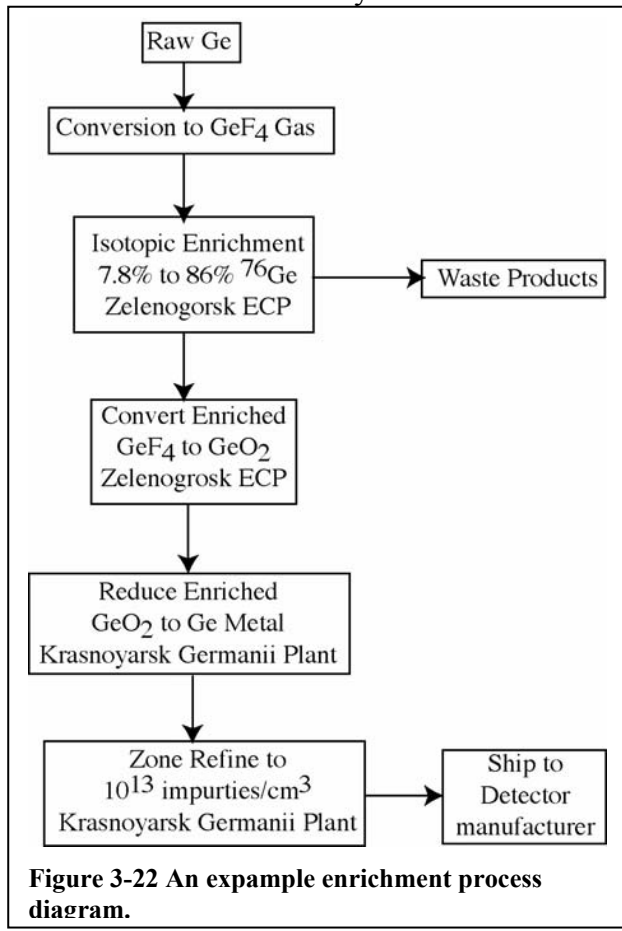

**Figure 3-22 An exampte enrichment process diagram.**

### 3.6 Detector Manufacture

*Detector Parameters*

The detector-array geometry examined in this section is our Reference Plan configuration and is shown in Fig. 3-23. It is a stack of 3 planes of one- or two-kg detectors each within a close packed geometry. We used this geometry to determine the relative value of different segmentation schemes. We investigated similar arrangements of 3 planes of 19, 1-kg detectors and 3 planes of 7, 2-kg detectors. The three most important potential sources of radioactivity were investigated using a GEANT-4 based Monte-Carlo simulation. The following contaminations were investigated:

1. $^{68}$Ge internal to the detector crystal generated by cosmic ray neutrons during the entire time that the germanium ore and germanium have been above ground.
2. $^{60}$Co internal to the detector crystal generated by cosmic ray neutrons during the time period after the crystal was pulled.
3. $^{208}$Tl from the decay chain of $^{232}$Th and $^{214}$Bi from the $^{238}$U chain, external to the crystal; possibly in the shielding and cryostat parts.

For each potential design, the code was run for 100 different detector segmentation schemes with ten different azimuthal segmentation geometries (pie shaped sections) with ten axial segmentations (disk shaped sections) each. The choice of detector parameters resulted from a study of available technologies, production feasibilities, background suppression figures of merit (FOM), and cost. Here we summarize the results of extensive Monte-Carlo simulations of the FOM for suppressing various backgrounds in various locations, including cosmogenic isotopes in the germanium.

Figures of merit were computed for four different detector locations within these test arrangements: external detectors in the top or bottom plane; external detectors in the central plane; internal detectors in the top or bottom plane; and internal detectors in the central plane. Figure 3-23 shows the detector layouts. The FOM is the factor by which the half-life sensitivity is increased, where FOM $\alpha$ (signal)/(background)$^{-1/2}$. For example, a FOM of 2 implies that the background has been reduced by a factor of 4 and the half-life sensitivity has been doubled.

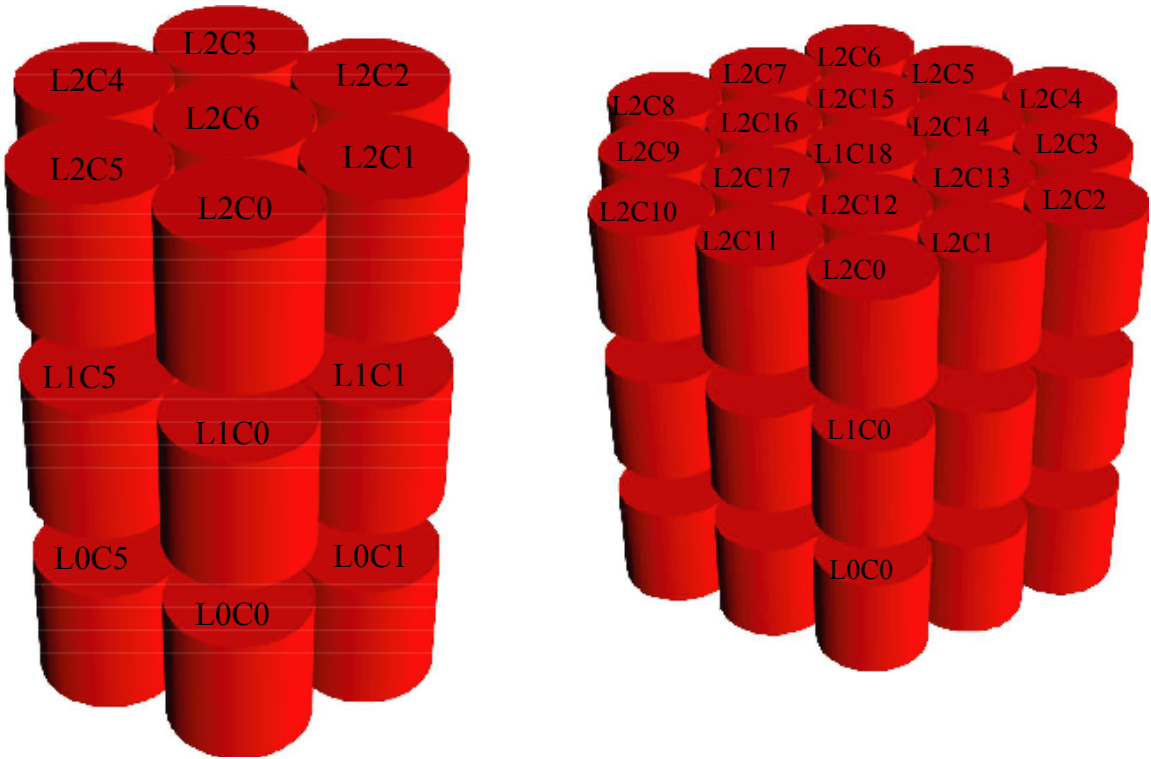

**Figure 3-23 The crystal-layout definition as used in the simulations and quoted in the Tables.**

From this long list of simulated segmentations, we have selected 36 configurations from the 21-detector arrangement and given their FOM in Tables 3-11, 3-12 and 3-13. From this shortened list, one sees that there are a large number of configurations with comparable FOM. Therefore, the Majorana Experiment can reach its design sensitivity with a variety of detector designs that are presently commercially available. Therefore the final choice of configuration is a cost-optimization process.

**Table 3-11 Figures of Merit for various axial and azimuthal segmentations for the various positions of the detectors in the array. Notation: intCo60L0C0 means $^{60}$Co internal to the crystal C0 (outer; C6 is the inner) in a detector in level L0 (top or bottom; L1 is the middle). Tl208L1C6 identifies an external $^{208}$Tl contamination as seen by the C6 (center) crystal of the L1 (middle) plane.**

| Azimuthal Segments | | 1 | 1 | 1 | 1 | 2 | 2 | 2 |
|---|---|---|---|---|---|---|---|---|
| Axial Segments | | 1 | 3 | 5 | 7 | 4 | 6 | 8 |
| intCo60L0C0 | FOM | 1.64 | 2.55 | 3.53 | 4.68 | 3.51 | 4.53 | 5.72 |
| intCo60L0C6 | FOM | 2.43 | 3.73 | 5.23 | 6.02 | 5.18 | 6.26 | 7.73 |
| intCo60L1C0 | FOM | 1.87 | 2.93 | 3.74 | 5.02 | 3.94 | 5.07 | 6.05 |
| intCo60L1C6 | FOM | 3.07 | 4.89 | 6.97 | 9.37 | 6.97 | 8.80 | 12.47 |
| intGe68L0C0 | FOM | 1.36 | 1.84 | 2.24 | 2.65 | 2.30 | 2.64 | 3.04 |
| intGe68L0C6 | FOM | 1.74 | 2.39 | 2.99 | 3.58 | 3.04 | 3.60 | 4.19 |
| intGe68L1C0 | FOM | 1.46 | 1.99 | 2.38 | 2.74 | 2.48 | 2.89 | 3.17 |
| intGe68L1C6 | FOM | 1.97 | 2.74 | 3.58 | 4.09 | 3.64 | 4.21 | 4.95 |
| extTl208L0C0 | FOM | 1.54 | 1.81 | 1.98 | 2.07 | 2.03 | 2.09 | 2.19 |
| extTl208L0C6 | FOM | 2.10 | 2.64 | 2.98 | 3.18 | 3.02 | 3.18 | 3.33 |
| extTl208L1C0 | FOM | 1.74 | 2.02 | 2.18 | 2.36 | 2.29 | 2.32 | 2.50 |
| extTl208L1C6 | FOM | 2.71 | 3.27 | 3.52 | 4.03 | 3.74 | 4.02 | 4.22 |

**Table 3-12 Figures of Merit for various axial and azimuthal segmentations for the various positions of the detectors in the array. The notation is analogous as in Table 3-1.**

| Azimuthal Segments | | 3 | 3 | 3 | 4 | 4 | 4 |
|---|---|---|---|---|---|---|---|
| Axial Segments | | 4 | 6 | 8 | 4 | 6 | 8 |
| intCo60L0C0 | FOM | 3.79 | 4.90 | 6.02 | 4.07 | 5.11 | 6.53 |
| intCo60L0C6 | FOM | 5.67 | 6.95 | 8.10 | 6.32 | 7.35 | 8.95 |
| intCo60L1C0 | FOM | 4.14 | 5.43 | 6.02 | 4.47 | 5.81 | 6.60 |
| intCo60L1C6 | FOM | 7.58 | 9.80 | 12.76 | 8.08 | 9.62 | 13.95 |
| intGe68L0C0 | FOM | 2.40 | 2.75 | 3.14 | 2.58 | 2.94 | 3.35 |
| intGe68L0C6 | FOM | 3.30 | 3.87 | 4.54 | 3.45 | 3.98 | 4.46 |
| intGe68L1C0 | FOM | 2.61 | 3.01 | 3.28 | 2.78 | 3.15 | 3.49 |
| intGe68L1C6 | FOM | 3.80 | 4.32 | 4.99 | 4.21 | 4.66 | 5.54 |
| extTl208L0C0 | FOM | 2.06 | 2.13 | 2.23 | 2.09 | 2.16 | 2.25 |
| extTl208L0C6 | FOM | 3.08 | 3.26 | 3.39 | 3.17 | 3.35 | 3.47 |
| extTl208L1C0 | FOM | 2.35 | 2.36 | 2.54 | 2.41 | 2.43 | 2.57 |
| extTl208L1C6 | FOM | 3.92 | 4.10 | 4.32 | 3.97 | 4.17 | 4.35 |

Because of the flexibility in the final choice of detector size and segmentation, the Collaboration will continue to pursue the investigation of technologies that might reduce costs and simplify the experimental configuration. One possibility, for example, is the segmentation of p-type Ge detectors. At present, detector manufacturers have declined to offer this type of detector because of the cost of reprocessing a sub-standard detector. These detectors have a lithium dead layer approximately 0.5 mm thick and if the detector does not meet specifications, this entire dead layer must be etched off. This removes approximately 10% - 20% of the detector's mass. Because a fairly large fraction of the detectors may likely require reprocessing, this could result in a significant variation in detector size and chemical reprocessing cost. Even if the material can be reclaimed, it requires a larger number of smaller detectors be built potentially affecting the cost. Should this technology become feasible during the period of constructing detectors, it will again be seriously considered. At the time of this writing, it is not the proposed technology, but we are presently doing tests investigating its feasibility.

**Table 3-13 Figures of Merit for various axial and azimuthal segmentations for the various positions of the detectors in the array. The notation is analogous as in Table 3-1.**

| Azimuthal Segments | | 5 | 5 | 5 | 6 | 6 | 6 |
|---|---|---|---|---|---|---|---|
| Axial Segments | | 4 | 6 | 8 | 4 | 6 | 8 |
| intCo60L0C0 | FOM | 4.55 | 5.29 | 6.98 | 4.58 | 5.74 | 7.22 |
| intCo60L0C6 | FOM | 6.92 | 8.10 | 9.53 | 6.86 | 8.05 | 9.46 |
| intCo60L1C0 | FOM | 4.80 | 6.27 | 6.92 | 4.76 | 6.12 | 6.87 |
| intCo60L1C6 | FOM | 8.56 | 10.76 | 16.98 | 9.03 | 10.49 | 14.79 |
| intGe68L0C0 | FOM | 2.66 | 3.01 | 3.32 | 2.74 | 3.04 | 3.48 |
| intGe68L0C6 | FOM | 3.62 | 4.19 | 4.74 | 3.90 | 4.41 | 5.10 |
| intGe68L1C0 | FOM | 2.92 | 3.33 | 3.59 | 2.95 | 3.33 | 3.57 |
| intGe68L1C6 | FOM | 4.36 | 4.86 | 5.55 | 4.45 | 4.99 | 5.58 |
| extTl208L0C0 | FOM | 2.09 | 2.14 | 2.22 | 2.15 | 2.21 | 2.28 |
| extTl208L0C6 | FOM | 3.17 | 3.34 | 3.50 | 3.29 | 3.38 | 3.55 |
| extTl208L1C0 | FOM | 2.44 | 2.45 | 2.56 | 2.46 | 2.45 | 2.59 |
| extTl208L1C6 | FOM | 4.06 | 4.32 | 4.45 | 4.07 | 4.29 | 4.39 |

In our white paper of February 2002, the detector mass of the n-type segmented detectors was arbitrarily chosen to be 2 kg (i.e. 80 mm in diameter by 80 mm in length) as these were the dimensions of the IGEX p-type detectors. Since that time, significant effort has been invested in evaluating feasibility, cost, detector production rates, and the granularity of the array. Although these large detectors optimize the mass of Ge with respect to the number of electronics channels and dead layer losses, they are more difficult to produce than their smaller counterparts. As a result, the added reprocessing of failed detectors increases the cost significantly.

In Table 3-14, we summarize the characteristics of several possible configurations. The estimated price of the various options indicates that unsegmented p-type detectors or a "stock" item like the Clover™ detector sold by Canberra are the most cost effective. Although the "segmented stack" or "SEGA" designs have a preferable background-rejection FOM, these less expensive options meet the required 0νββ sensitivity and hence provide possible feasible designs. Note that the "split p" design is simply a segmentation of the "unsegmented p" detector. Hence its cost would include a price add-on for the labor to segment the detectors. We will continue to investigate the optimization of cost and background rejection, but it is clear that viable options exist.

It is also clear from Table 3-10 that the improved segmentation does not impact the FOM for external radioactivity greatly. However, it is a great aid in rejecting the internal cosmogenic radiations. Increasing the FOM by a factor of 2 is equivalent to doubling the detector size, if the background remained constant. This can be used to argue that doubling the cost of the detectors is justified. However, we can mitigate the internal cosmogenic backgrounds by growing and/or fabricating detectors underground. If this underground operation is technically feasible, it is clearly a cost advantage.

**Table 3-14 A summary of a few of the consider detector configurations, associated characteristics, and their FOMs.**

| Nickname | IGEX P | SEGA | Seg. Stack | Unseg. P | Split P | Clover™ |
|---|---|---|---|---|---|---|
| Type | P-type coax | N-type coax | N-type coax | P-type Coax | P-type Coax | N-type Coax |
| Axial segments | 1 | 6 | 7 | 1 | 2 | 1 |
| Azimuthal segments | 1 | 2 | 1 | 1 | 1 | 2 |
| Diameter | 80 mm | 80 mm | 62 mm | 62 mm | 62 mm | 50 mm |
| Height | 84 mm | 84 mm | 70 mm | 70 mm | 70 mm | 80 mm |
| Mass | 2.2 kg | 2.2 kg | 1.1 kg | 1.1 kg | 1.1 kg | 830 gm |
| est. cost per kg | $45k | $125k | $100k | $23k | $23k plus | $36k |
| Internal Co-60 rejection FOM | 1.9 | 4.0 | 5.6 | 2.5 | 3.3 | 3.5 |
| External Tl-208 rejection FOM | 1.7 | 2.3 | 2.5 | 2.0 | 2.2 | 2.2 |

### *Detector Production and Material Reprocessing*

The detector production and material reprocessing phase will be similar to that used in the IGEX project where special precautions were taken to minimize the loss of the valuable isotopically enriched germanium. The process has been tailored to the higher volume Majorana Project in consultation with several detector-manufacturing companies. The process diagram is shown in Fig. 3-24 using the example of 62-mm diameter by 70-mm long n-type detectors. In the case of p-type detectors of the same dimensions, the productivity could be 3 or 4 detectors from the original 10 kg of metal introduced into the zone refinement from $10^{13}$-$10^{11}$ impurities/cm$^3$. Our Reference Plan design uses 500 kg of $^{enr}$Ge. We estimate that 5% of the $^{enr}$Ge will be lost during processing of the detectors and therefore we will need to purchase approximately 525 kg of $^{enr}$Ge.

In the IGEX experiment, the unrecoverable Ge loss was 5.27% per cycle of purification, zone refinement crystal growth, and detector production. (See Fig. 3-24.) Following a review of the IGEX procedures and data, consultations were held with Eagle Picher Inc., Canberra Inc., and AMETEC. It was concluded that it is feasible to reduce these losses. The largest loss of Ge metal occurred during grinding and machining of the crystals (2.5%). This loss was mainly caused by an inefficient method of collecting the material removed from the crystal. Fortunately, this material can be almost completely recovered by purchasing dedicated grinders and lathes and equipping them with special catch pans and vacuum cleaners. The cost of such machines is small compared to the savings on the cost of enriched germanium. A careful analysis indicates that the unrecoverable losses can be reduced to 3% per cycle from the 5.27% per cycle suffered in the IGEX experiment. In addition, the other losses, such as etch solutions, rinse water, etc., are also reducible if managed by members of the collaboration, so further improvement is possible.

**Table 3-15 A summary of the Ge material accounting during processing of 100 kg of intrinsic Ge into detectors. To convert 100 kg of Ge into 74 crystals requires 11 cycles of the process shown in Fig. 3–24. There is an estimated 3% unrecoverable loss per cycle. The unused and recovered Ge portions in each cycle are combined for the succeeding stage.**

| Intrinsic Ge Input (kg) | Number of Zoned Bars (10 kg ea.) | Number of Detectors (1.1 kg ea.) | Unrecoverable Loss (kg) | Unused Intrinsic Ge (kg) | Recovered Ge (kg) |
|---|---|---|---|---|---|
| 100.0 | 10 | 20 | 3.0 | 0.0 | 75.0 |
| 75.0 | 7 | 14 | 2.1 | 5.0 | 52.5 |
| 57.5 | 5 | 10 | 1.5 | 7.5 | 37.5 |
| 45.0 | 4 | 8 | 1.2 | 5.0 | 30.0 |
| 35.0 | 3 | 6 | 0.9 | 5.0 | 22.5 |
| 27.5 | 2 | 4 | 0.6 | 7.5 | 15.0 |
| 22.5 | 2 | 4 | 0.6 | 2.5 | 15.0 |
| 17.5 | 1 | 2 | 0.3 | 7.5 | 7.5 |
| 15.0 | 1 | 2 | 0.3 | 5.0 | 7.5 |
| 12.5 | 1 | 2 | 0.3 | 2.5 | 7.5 |
| 10 | 1 | 2 | 0.3 | 0 | 7.5 |
| Totals: | n/a | 74 | 11.1 | n/a | 7.5 |

Table 3-15 summarizes the conversion of 100 kg of Ge into 74 detectors demonstrating the effect of this 3% loss. Since each load into the crystal puller requires 10 kg of Ge, 11 cycles are needed to complete the final complement of 1.1-kg detectors. The first cycle grows 10 crystals from which 2 detectors are produced. There is a 3% unrecoverable loss, leaving 75 kg of Ge as input to the second cycle. The second cycle, therefore, can produce seven 10-kg crystals leaving 5 kg of Ge unused. This unused portion from cycle 2 is combined with the residual 52.5 kg of Ge to provide 57.5 kg for cycle 3. After all 11 cycles are complete, 11.1 kg, or 11.1%, of Ge is lost. Accordingly, the Majorana project hopes to reduce this loss by a factor of 2.

The operations represented by the lower left box on the left of Fig. 3-24 entitled "Two 1.1-kg Enriched Ge Detectors" depends very much on the final choice of the type and size detector. In the case of segmented n-type detectors, that phase will involve the masking and segmenting process. In the case of p-type detectors it will involve only the deposition of the Li dead layer and the implantation of the contacts.

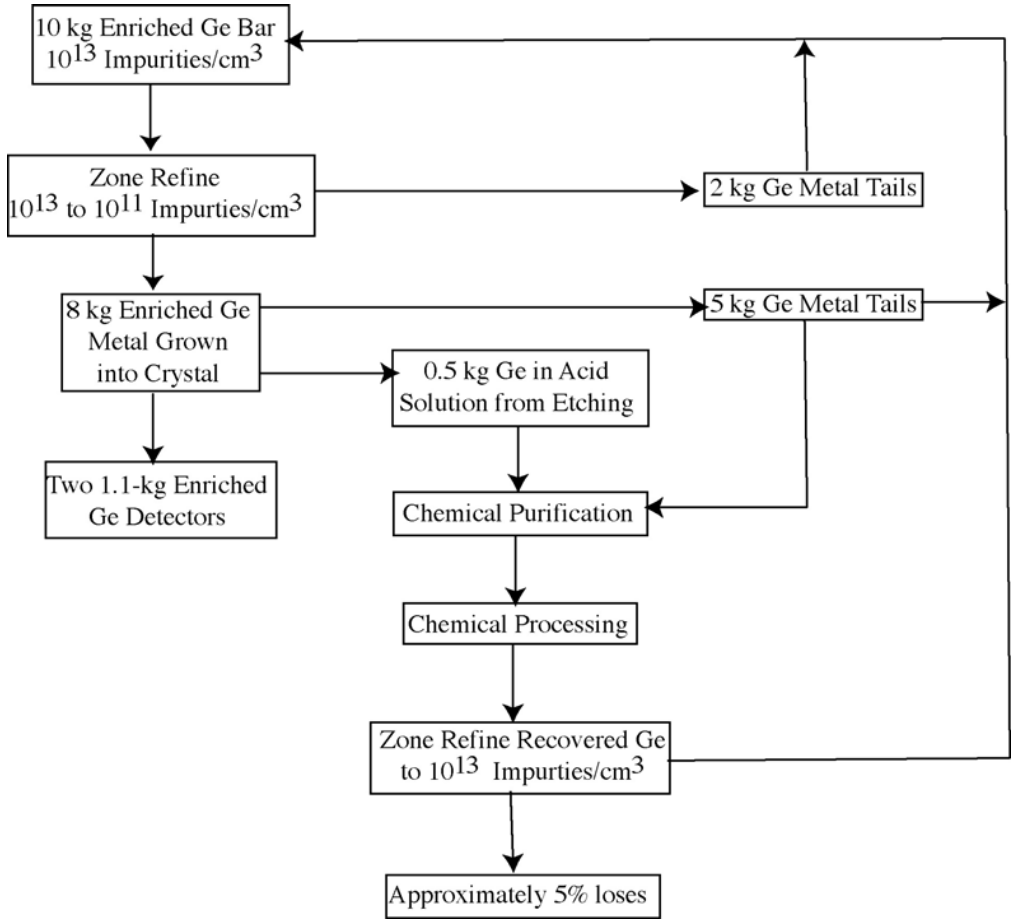

**Figure 3-24. The detector production process diagram.**

### Underground Detector Production

The crystal growing process eliminates all but Ge isotopes from the crystal. Therefore, this step in the detector fabrication process "resets" the [60]Co contamination to zero. If the crystals are grown underground (or grown above ground but rushed underground soon after) and the subsequent detector fabrication is underground, the Co background can effectively be eliminated. Furthermore, it reduces the [68]Ge contamination significantly also. This is an exciting possibility. It effectively increases the sensitivity for a given mass of detector and it reduces the requirement for fine segmentation of the detectors. Both of these features translate into significant cost savings that should more than pay the cost of implementing these activities underground. Although the experiment can be done without this feature, it would clearly be a better experiment if it was included. The possibility is under study by the collaboration with the primary difficulty being the coordination of logistics with commercial detector production companies.

### Detector Receipt Testing

The evaluation of detectors immediately after fabrication is a critical part of effectively mounting the Majorana array. Detector manufacturers typically only deliver a detector after it meets a minimum performance standard. Each delivered detector will then be subjected to a series of measurements designed to establish its initial efficiency, resolution, and the background rejection performance of the segmentation and PSD. The detector characteristics we will measure are:

1. Leakage current versus bias curve
2. Detector efficiency, resolution and peak shape for selected gamma lines
3. Dead layer uniformity and mobility anisotropy
4. Radial pulse-shape discrimination efficacy for selected gamma lines
5. Pulse shape and cross-talk for all segments for selected gamma lines

Our acceptance standard will be the basic parameters for a working detector that have been established through our segmentation R&D program and past experience. In addition, commissioning each detector will include mobility anisotropy measurement of the orientation of the detector crystal to ±1% for solar axion searches.

Regardless of whether the final stage of detector manufacture takes place above ground at a commercial facility or at an underground facility near our laboratory, this testing will require collaboration manpower at a remote site. In either case, our acceptance tests must be done at the production facility, to ensure quality control and assurance of our industrial partners. This effort requires collaboration manpower to perform these measurements as each detector should require roughly 3 days to carry out the basic calibration and electronics measurements. This will also require test apparatus at each production site. Depending on the number of vendors utilized, this could require up to 3 half-time shift persons at vendor sites during detector manufacturer.

### *Crystal Orientation Measurements*

The axion studies require that we know the crystal orientation for all the Ge crystals that comprise Majorana. We have developed a technique for measuring and recording the orientation during the MEGA assembly. We use x-ray diffraction with a Laue camera to orient the crystals. Our germanium crystals are packed in a copper inner can to protect the germanium (the naked crystals are handled at little as possible). There is a roughly one-inch hole in the bottom of the copper can where the inner contact probe is inserted to make the inner electrical connection with the Ge crystal. The serial number of the crystal is scored into the outside of the copper can, making unique identification possible.

The detectors are packed into a copper can inside a clean glove box. Next we wrap this inner can with clear plastic and then pack it inside two nested clear-plastic freezer bags to insure the cleanliness of the copper. The plastic does not affect the x-ray diffraction orientation. An x-ray source illuminates the crystal and a Laue camera determines the orientation of the crystals. The intensity of the x-ray source can be adjusted to provide more intense dots at the camera (increase the current) or alternatively the energy can be adjusted to provide more backscatter dots (increase the voltage)

The crystal is held on a mount that is secured to a goniometer. The goniometer rests on a stage that has more than sufficient range of motion to line up the hole in the end of the can with the x-ray beam. All goniometer angles are set to zero, and the can is rotated to roughly align the backscatter dots with the grid shown in the real-time Laue camera display. The end of the can is lightly scored to mark this starting position. The goniometer has rotation about three axes: roll, pitch, and yaw. These rotations are adjusted until the backscatter dots line up precisely with the gridlines on the Laue camera display. Since Ge has a (001) diamond crystal structure, it is very straightforward to align the dots with the grid. Once the dots are aligned, the angle measurements for roll, pitch, and yaw are recorded for the serial number of the can, and a photographic record is made of the crystal still sitting in the goniometer mount. The crystals are all grown with the axis of cylindrical symmetry being the (100) axis.

A superior, less costly measurement can be made inside the vacuum jacket using anisotropy to determine the orientation. This method allows an assembled device to be measured, thereby reducing the handling of the detector while at atmospheric pressure, and after the crystal is locked in position, reducing mis-alignment in the final assembly.

### 3.7 Cryostat and Crystal Production

The 500-kg, five-year, near-zero-background Majorana Experiment depends on the completion of the assembly of several systems. Completing the assembly requires modeling of the crystals, their environment, and the relevant known physics, to optimize the entire design and to refine sensitivity estimates. Germanium will be enriched and crystals grown, purified, and made into detectors. Cryostats, electronics, and shielding are to be designed, constructed, and made operable. The custom data acquisition system required for handling detector segmentation and pulse digitization will be designed, fabricated, and tested. Data analysis techniques will be developed, and we propose to create a collaboratory infrastructure and social structure to maximize the value extracted from Majorana data. This subsection describes many of these subsystems.

The sensitivity goals of the Majorana instrument can only be realized if a stringent set of requirements is met regarding the physical form and immediate environment of the fiducial mass of germanium. With this in mind, a series of technical goals and requirements critical to the success of Majorana can be exposited for the cryostat and crystal mounting.

To enable background rejection via pulse-shape discrimination, it is critical that the charge-integrating preamplifier of each diode contact have <30 ns rise-time. To achieve this, the detector mounting technique must add little additional capacitance to that intrinsic to the diode contacts. Additionally, cross-talk due to inter-contact capacitive coupling must be minimized. This must be accomplished within the limited palette of materials known to be of sufficient radiopurity to be used close to the germanium crystal. A mounting and contact scheme, illustrated in Fig. 3-26, has been developed to meet these goals.

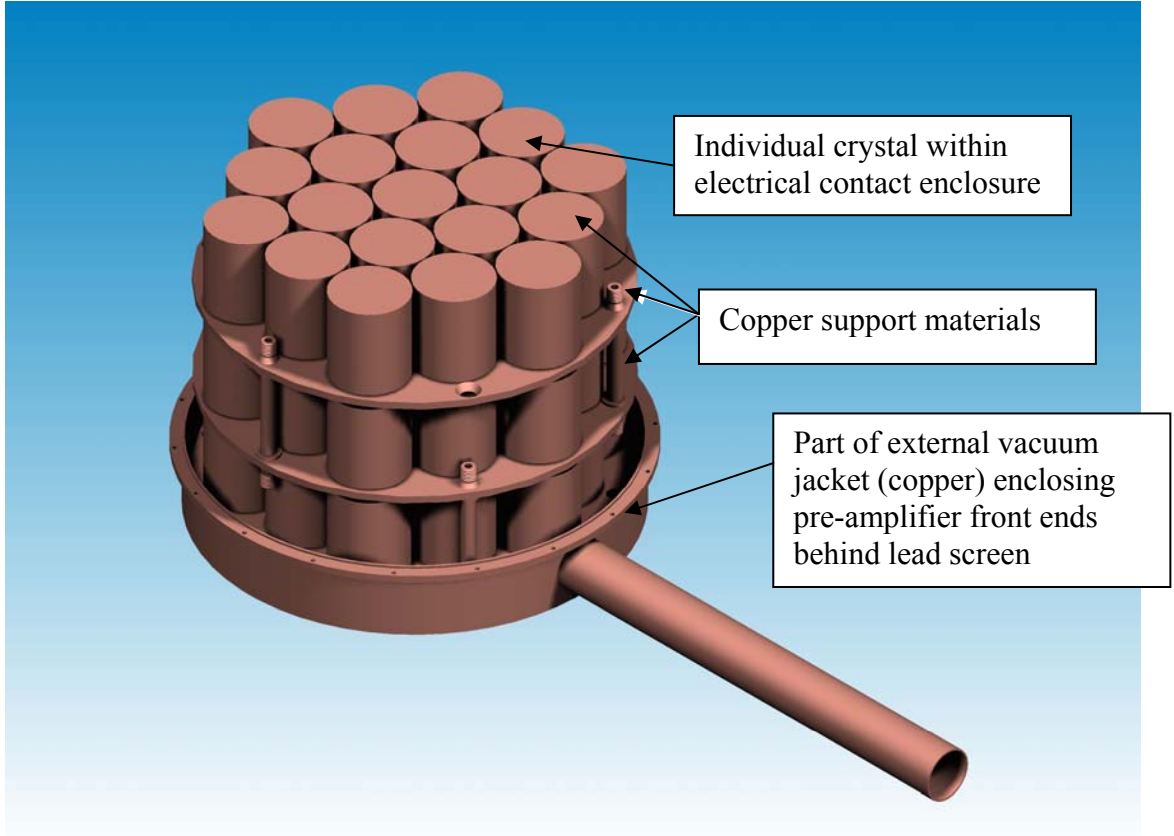

Individual crystal within electrical contact enclosure

Copper support materials

Part of external vacuum jacket (copper) enclosing pre-amplifier front ends behind lead screen

**Figure 3-25. Highly-schematic view of close-packed arrangement of 57 germanium crystals inside a modular Majorana cryostat. Outer vacuum jacket is removed for clarity.**

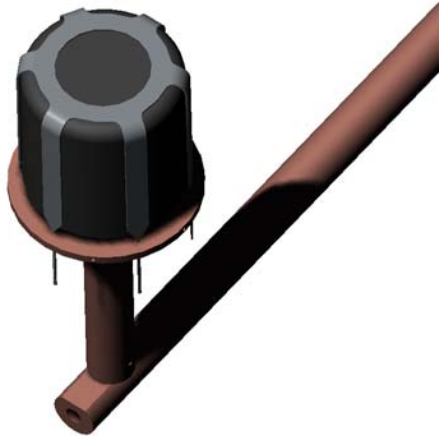

**Figure 3-26. Rendered view of a segmented detector outer support and electrical contact web.**

To maximize self-shielding effects, the crystals must be as closely-packed as possible while still providing the advantages of modular deployment. The background projections for the Majorana instrument depend critically on meeting current state-of-the-art goals as to the radiopurity of the support structures and cryostats housing the crystals. Additionally, the spectroscopic performance of the instrument depends on adequate cooling of the crystals as well as the electrical characteristics of the cryostat, mounting, and charge-sensing components. With these factors in mind, a reference mechanical design has been developed, consisting of ~10 modules with approximate 55 kg of crystals each. A rendering of the internal view of one modular cryostat is shown in Fig. 3-25.

This arrangement allows the germanium crystals to be in close proximity to one another, maximizing their self-shielding potential. This design minimizes the amount of support material per gram of fiducial germanium from about 1- kg Cu : 1- kg Ge in the previous IGEX experiment to about 1- kg Cu : 6.7-kg Ge in the modular cryostat shown here.

A series of detectors, based on the technology of ultra-pure electroformed cryostat components, have been produced by members of the collaboration for the IGEX collaboration and other uses. These deployments have demonstrated the cryogenic performance and long-term stability of this construction technique. Long-term measurements as part of the earlier IGEX effort have established the stringent limits on radiopurity of the electroformed copper support material, as exposited in the sensitivity calculation for Majorana [Bro95].

The crystal geometry, cryostat Reference-Plan design, and crystal mounting technique respond to the unique requirements mandated by the sensitivity and stability goals of Majorana with proven techniques using new designs. Using proven techniques as the reference design instills confidence in the plan.

### 3.8 Shielding

Assuming that all appropriate steps have been taken to eliminate sources of background radiation from the component parts of the detector and shielding materials, the limiting background for the proposed experiment comes from three potential sources: cosmic rays, experiment chamber walls and materials, and airborne radon. Each of these background sources can be mitigated with proper attention to detail. In this section, we describe our Reference Plan for the shielding of Majorana. Past experience indicates that this design will provide the necessary background reduction to permit us to reach the Majorana sensitivity goal.

*Introduction*

First, primordial radionuclides in the walls and construction materials of the experimental area constitute the largest source of background radiation, and the easiest source to mitigate. A sufficiently designed and massive bulk shield composed of radiologically-clean material, typically lead, surrounding the detector systems has been repeatedly demonstrated to effectively eliminate this source of background. Radon daughter products are the most commonly-identifiable background source in low-background experiments. While elimination of radon from the sensitive region of the detectors is non-trivial, it can be done.

Finally, cosmic-ray-related signals may provide the limiting background. Cosmic rays generate background via three separable mechanisms. The most common is via direct interactions in the detector and shield materials. An electronic anticosmic shield can be very effective in eliminating this source of background although it will never be absolutely effective. The second source of cosmic-ray-generated background is from interactions in the walls and other components of the experimental chamber that are not directly protected by the electronic anticosmic shield. This source of background typically manifests itself as a shower of high-energy neutrons, some of which can enter the detector ensemble without generating a blocking pulse in the electronic anticosmic shield, and that subsequently interact with the detector materials and components. This source of high-energy neutron-induced background can be mitigated through a proper choice and arrangement of bulk and electronic shielding materials. However, once again, this source of background cannot be completely eliminated. Last, and least of all, cosmic rays will interact with the detector materials themselves to generate spallation radioisotopes that decay long after the electronic blocking pulse has expired. These background events can only be mitigated by reducing the number of cosmic rays incident on the detector ensemble. The number of incident cosmic rays can only be reduced by increasing the depth of the overburden.

While primordial radioactivities in the chamber walls and radon in the air may vary substantially depending on location and rock composition, both are easily controlled and therefore neither is a significant consideration for site selection. Mitigation methods are required in all cases anyway, and can be just as effective at any site. However, the ultimate limiting source of background, assuming all other sources are properly eliminated, are cosmic rays, and regardless of the efficiency of mitigation effects, they cannot be totally removed. Hence, the deepest site available always offers the potential for the lowest possible background.

**Table 3-16 A summary of the potential backgrounds and their mitigation.**

| Background | Shielding for Mitigation |
|---|---|
| Experimental hall environmental $\gamma$ and n | Outer shield |
| Radon | Gas containment volume |
| Radioactive contaminations in shielding material | Selection of inner shield materials |
| Cosmic rays: prompt, direct signal from $\mu$ | Active veto shield |
| Cosmic rays: prompt, signal from $\mu$-induced secondaries | Active veto shield, Graded outer shield |
| Cosmic rays: $\mu$-induced delayed signals | Depth |

The shielding of the Majorana apparatus is a critical although low-technology component. The various backgrounds that shielding will reduce are summarized in Table 3-16. Shielding reduces background counts from $\gamma$ rays in the experiment hall (from rock, construction materials, and possibly from shielding materials themselves), cosmic-ray $\mu$ penetrating the shielding, and cosmic-μ-induced neutrons. The strategy is to provide extremely low-activity material for the inner layer of γ-ray shielding. Around this will be bulk γ-ray shielding material of lower quality: the outer shield. Finally, outside this bulk shielding will be active μ veto detectors. Cosmic-μ secondary neutrons may potentially require either great depth or an additional shell of μ detection, high-energy neutron moderation (e.g. lead), and hydrogenous moderator incorporating a neutron absorber (e.g. Cd, Ga or B). After a discussion of the available materials for the shield, each layer is discussed, in turn, below.

### Material Selection

To qualify a material for inclusion in the inner area, an exceptional measurement technique is needed. The most elegant and effective method to qualify a material is to have used it in a previous experiment with similar, or even higher sensitivity to contaminants, with no observation of detrimental backgrounds. Unfortunately, this leads to a rather restrictive palette of materials with which to design the apparatus. Although many materials have been used in previous germanium experiments, nearly all were in small quantities and thus are only qualified for use in similar or smaller quantities. For instance, tens of grams of certain plastics have been used inside double-beta decay germanium detectors in the past. We are therefore confident that we can continue their use in such quantities again.

A useful counter-example has been the germanium itself, which would show sharp high-energy alpha peaks (5 MeV) if it contained significant U or Th contamination. Therefore we are confident that we can use significant quantities of Ge. In addition, the production process of the electroformed copper has been developed sufficiently that little radioactive contamination remains in those parts.

Lead is the traditional material of choice for shielding due to its high density. While lead would seem to be subject to the production of all sorts of relatively long-lived γ-emitting radionuclides via μ and fast neutron spallation reactions, experience has failed to identify any of them in γ-ray spectra obtained from ultra-low background, underground experiments shielded with lead. While this is empirical data only, there is a wealth of experimental data to support this position. Lead does produce prompt γ rays that are detectable in these experiments via fast neutron interactions, and mitigating the number of fast neutrons incident on the lead best mitigates these sources of background. A prime example of this type of background is the 1063-keV γ ray from the de-excitation of $^{207m}$Pb formed by an (n,n') reaction on $^{207}$Pb. Another common background engendered by lead is the bremsstrahlung radiation produced during the beta-decay of the $^{210}$Bi daughter of $^{210}$Pb, which is commonly present in contemporary sources of lead, either from smelting operations or from the decay of uranium present in the lead ore. These low-energy γ rays cannot produce counts in the 0νββ region of interest, however, they do represent a potential background for dark matter studies. Several potential methods are available to obtain lead free from $^{210}$Pb, but the most common is to utilize sources of antiquity lead that have had the original $^{210}$Pb contaminations depleted by radioactive decay ($T_{1/2} = 22.3$ y). Competition for these sources of $^{210}$Pb-free lead is extremely high, however, since the semiconductor and microelectronics industries are also in need of this type of lead for solder used in flip-chip circuitry in order to avoid α-particle-induced soft errors. These electronic needs are consuming ~100 tons/y of this so-called low-α lead. Fortunately, our collaboration already possesses 3 tons of antiquity lead with plans to obtain the remaining 2 tons required.

Copper is one of the very few elements that has no relatively long-lived radioisotopes. This is very fortunate, since copper has many wonderful physical, chemical, and electronic properties that make it particularly useful in the fabrication of low-background radiation detectors. The only trick to using copper is to make sure that it is not contaminated with radioactive materials, and that it does not have significant quantities of cosmic-ray generated radioisotopes. The most prolific of these latter contaminants is $^{60}$Co, which is copiously generated by (n,α) reactions on $^{63}$Cu. Electroforming copper in a cosmic-ray free environment has proven to be effective in producing radiologically pure components for detector systems. This, of course, requires use of very high purity copper as the starting material, and appropriate holdback carriers in the plating baths to preclude inclusion of atom quantities of impurities into the electroformed copper [Bro95].

Finally, most plastic materials have been empirically found to be radiologically pure. Since most plastics are produced from oil, and since oil is an old material that has been shielded from cosmic radiation underground, it is reasonable to expect most plastics to be clean. Clearly, there are exceptions to every such general rule, and a good screening program should be employed to verify that any proposed plastic material is actually sufficiently clean. Of major interest in germanium γ-ray spectrometers is the use of Teflon® and plastic scintillator material, which, heretofore, have proven to be sufficiently pure.

### Inner shield

Because of the extreme sensitivity of germanium detectors to γ rays, and the extreme efficiency of the Majorana apparatus, γ rays from the inner region of the shielding are very dangerous. While the special electronics discussed below will offer a degree of protection from γ rays external to the detector, the direct approach is to use only materials with extremely low specific activity from radioactive isotopes. Fortunately, there are few sources of γ rays above 1500 keV. Prime examples of potential high-energy contamination γ rays are:

| | |
|---|---|
| 2200 keV | γ rays from neutron capture on hydrogen in hydrogenous material |
| 2204 keV | $^{214}$Bi, $T_{1/2}$ = 19.9 m, daughter of $^{222}$Rn, $T_{1/2}$ = 3.8 d |
| 2505 keV | Sum energy of $^{60}$Co gamma rays from cosmogenic production in Cu |
| 2614 keV | $^{208}$Tl, $T_{1/2}$ = 3.05 m, daughter of $^{220}$Rn, $T_{1/2}$ = 55 s |

Although the sum energy of the gamma rays from $^{60}$Co exceed 2039 keV (2505 keV), underground electroforming should prevent this spallation isotope from appearing in the Cu support structures. Assuming that there is essentially no hydrogen and/or no thermal neutrons within the inner shield, we can also assume that, as in the previous germanium experiments, we will observe no 2200 keV hydrogen capture. Thus, only the radon daughters [214]Bi and [208]Tl are of concern. If these are indeed only present due to Rn, a few weeks after closing the cave and purging they will no longer represent a concern. On the other hand, if they are present because of contamination by a long-lived solid precursor, they could pose a significant problem, which is addressed below.

Because the chemical nature of the electroforming process tends to suppress elements in differing degrees, it is possible or even likely that the secular equilibrium of the U and Th natural decay chains (Appendix 2) were broken during these measurements. In one particular experiment [Bro95], an inner shield of Cu was produced in a hollow cylinder closed on one end (Marinelli geometry). This part was used as an inner shield for over 90 days and was therefore able to measure or limit concentrations of natural radioactivity in the Cu. Regardless of whether the decay rate of the chain patriarch or the inferred solid precursor isotope is appropriate for the unit conversion to grams per gram, the relevant quantity is the measured specific activity, which translates into the specific activity of the observable $\gamma$ rays: at or less than about $10^{-5}$ Bq/kg.

Another material qualified in previous experiments is lead. Unfortunately, while we can make clean copper in any quantity or shape we require, sources of lead are opportunistic and hence problematic. The sources of lead used include Doe Run mine, Johnson-Matthey/Cominco Inc., 150-year old German ingots, 450-year old Spanish galleon ballast, and 2000-year old Roman anchors. Old lead is preferred because the chemical cleaning that occurs in the smelting process (akin to zone refining) cannot eliminate [210]Pb ($T_{1/2} = 22.3$ y), a lower link in the [238]U chain. Besides controlling particulate deposition (both Rn daughters and ordinary dust), the best solution for the elimination of low-energy activities in common Pb is to use either screened Pb or electroformed Cu in the innermost regions of the shield.

### Outer shield

Primordial radioactivities in the walls and construction materials of the experimental area constitute the largest source of background radiation, but the easiest source to mitigate. A sufficiently designed and massive bulk shield composed of radiologically clean material, surrounding the detector systems, has been repeatedly demonstrated to effectively eliminate this source of external $\gamma$-ray background. The outer shield protects the detectors from gross environmental $\gamma$ rays from the rock, the construction materials, and other shielding materials. Typical rock at an underground location (e.g. SNOLab) contains a few ppm U and Th. Calculations show that a lead shield ~500 g/cm$^2$ thick (~50 cm) is sufficient to reduce the natural $\gamma$-ray radiation from the environment to a negligible level. That is, the count rate in the region of interest due to $\gamma$ rays from the rock walls will be well below that due to $\gamma$ rays from the electroformed copper.

While the U and Th decay chains generate some modest neutron flux from spontaneous fission and ($\alpha$,n) reactions on light isotopes, these sources of neutrons will be relatively low-energy and easily mitigated by any shield design that is capable of attenuating high-energy cosmic-ray engendered neutrons, discussed below. This bulk shield surrounds the innermost layer, which does not contain contemporary lead. This means that the innermost layer of the bulk shield needs to be ~100 g/cm$^2$ of low-α lead (antiquity lead) or the equivalent of some other suitable shielding material, such as plastic or copper.

### Radon Control

Elimination of radon from the sensitive region of the detectors is most easily accomplished by enclosing the entire detector and shield ensemble in an airtight container and pressurizing the internal volume of the container with a radiologically pure gas, typically nitrogen boiled off from liquid nitrogen. The two important parameters are that the entire enclosure consists of metal construction and that there is only one exhaust port for the pressurizing gas. Radon can permeate through plastics and rubber compounds, and if there is a leak anywhere in the enclosure, radon will migrate in through that leak regardless of the positive internal pressure.

### Active Veto Shield

Cosmic rays, of consequence at any reasonable depth, are composed almost exclusively of muons. These μ produce copious quantities of electron-hole pairs in passing through materials, and will produce primary pulses in the Ge detectors whenever they pass directly through the Ge. Furthermore, these μ are capable of undergoing direct interactions with nuclei, resulting in remnant spallation and fragmentation products as well as copious hadrons. While secondary protons are no more debilitating than the primary μ itself, the secondary neutrons can be very high-energy and can travel through significant quantities of material before being thermalized and absorbed. These secondary neutrons undergo further nuclear interactions, resulting in additional new isotope production, some of which will have half-lives and decay energies sufficient to generate background events when these interactions take place in proximity to the detectors. An electronic anticosmic shield can be very effective in tagging those μ, which pass through the primary shield and thus eliminate the primary energy deposition events in the Ge detectors and much of the secondary γ and bremsstrahlung radiation generated in the vicinity of the detectors via interactions in the lead or other materials. This electronic anticosmic shield will likely be an active plastic scintillator composing one of the outermost "layers" of a graded bulk shield.

A 10-cm-thick 4π plastic scintillator constructed of plastic instrumented with photomultiplier tubes will suffice as a μ veto shield. It is immaterial that the energy response/energy threshold of the plastic detectors may vary substantially from end to end because the threshold will be well below the μ through-peak of a perpendicular transversing μ. This active shield response can be recorded as an independent signal, allowing an independent measure of the health of the subsystem. Given that the area of the μ veto system is a few m$^2$, the μ rates expected in any of the available underground labs can be effectively cancelled.

**Table 3-17 Muon flux in selected facilities.**

| Facility | Meters Water Equivalent | μ/m$^2$/y |
|---|---|---|
| WIPP | 1840 | 1E+05 |
| Soudan | 2200 | 6E+04 |
| Gran Sasso | 3800 | 3E+3 |
| NUSL-Cl | 4000 | 1E+3 |
| Sudbury | 6010 | 85 |
| NUSL | 6700 | 65 |
| NUSL-Deep | 7100 | 25 |

### Muon-Induced Neutron Control with the Outer Shield

Muons that do not intercept the anticosmic shield are of no consequence unless they interact in the surrounding environment (rock) sufficiently proximate to the detectors that the reaction products can generate erroneous signals. This source of background typically manifests itself as a shower of high-energy neutrons, some of which could enter the detector ensemble without generating a blocking pulse in the electronic anticosmic shield, and which subsequently interact with the detector materials and components. This source of high-energy neutron induced background can be mitigated through a proper choice and arrangement of bulk and live shielding. However, once again, this source of background cannot be completely eliminated. Preliminary computer modeling has shown that a layered shield consisting of four layers of lead "sandwiching" four layers of plastic was capable of reducing a flux of incident 100-MeV neutrons by more than a factor of 1000. Once the experiment has been sited and the high-energy secondary neutron flux has been estimated, this model can be optimized with respect to cost, size, number and thickness of layers, and specific materials in order to provide sufficient reduction of this potential source of background.

We can conclude at this time that a shield capable of eliminating external γ rays is quite achievable, as is a direct μ veto system. However, additional consideration must be given to the need and optimum design of a shielding or veto system for μ-induced neutrons.

### The Depth Requirement

Muons that induce spallation/fragmentation reactions directly in the detector components can produce numerous radioisotopes (300+) that have sufficiently long half-lives to avoid being canceled by the anticosmic shield and high enough Q-values to generate background events in the energy region of interest. These background processes can only be mitigated by reducing the number of muons incident on the detector ensemble. The number of incident muons can only be reduced by increasing the depth of the overburden. (See Table 3-17.) Preliminary calculations indicate depths of 2000 mwe are sufficient to reduce this source of background to the order of 1 event in the energy region of interest in 2500 kg·y of data. This is a negligible rate even at this relatively shallow depth.

One example of such a concern is the neutron transmutation of germanium into long-lived isotopes. To estimate this background, we take the neutron spectra of Gaitskell [Gai01] at 2000 mwe and compare to that of Hess at zero mwe (Fig. 3-9). At 1 MeV, Gaitskell calculates 10 /MeV/$m^2$/y where Hess reports $3 \times 10^7 10^{11}$ /MeV/$m^2$/y, while at 100 MeV Gaitskell calculates 2 /MeV/$m^2$/y and Hess reports $5 \times 10^7 10^7$ /MeV/$m^2$/y. Since the cross sections for neutron spallation become significant between 20 - 100 MeV for the production of isotopes such as $^{60}$Co and $^{68}$Ge, we can conservatively estimate that the spallation rate at 2000 mwe or deeper will be about a factor of $10^7$ less than above ground. That is, the spallation of the germanium during the entire 5 year operation will be one ten-thousandth of a single day above ground. This estimate ignores the possibility μ-induced π interactions contributing to the *in situ* cosmogenic isotope production. Unfortunately, π have a larger cross section for large ΔA nuclear transmutations but are easier to shield than neutrons. We estimate that this process won't increase the *in situ* production of the estimated neutron-induced rate by more than a factor of a few.

Elastic scattering of neutrons is another concern. While the rate of neutrons reaching the Majorana detectors with sufficient energy (approximately 100 MeV) to create a recoil event registered at 2039 keV is computed [Gai01] to be only around 2-3 per year at 2000 mwe, some consideration to the rejection efficiency of the signals needs to be given. Note however, this background can be eliminated by going to greater depth.

Another concern is neutron inelastic scattering on detector, structural, and shielding materials. These can be greatly reduced by use of aggressive shielding or veto techniques, but the magnitude of this contribution was unidentifiable in previous experiments and a limiting value has not yet been determined from computations.

One empirical measurement that indicates a need for depth (~4000 meters water equivalent or deeper) comes from the IGEX data at the 2450-meter-water-equivalent deep Canfranc tunnel [Aal99]. Approximately 40% of the events in at 2 MeV were in coincidence with the muon veto system. Majorana will be a larger experiment having a Ge-detector surface area of about an order of magnitude greater than IGEX. It will also requires a lower background level averaged over the duration of the experiment: about a factor of 20. Therefore it would be prudent to site the experiment at a depth where the inefficiency of the veto system is not a critical parameter. A site (such as SNOLab or NUSEL) where the muon flux is a factor of 300 below that of the Canfranc laboratory would meet this requirement.

### Mechanical Engineering of the Shield

The Reference-Plan shield arrangement has been designed to use the minimum amount of mass needed to shield a large number of detectors. To reduce the need for low background materials, reduce the footprint of the experiment, and allow multi-crystal γ-ray depositions (for effective identification and suppression), the entire detector mass is shielded with one Pb layer 50 cm thick. The inner cavity occupied by the detector is about $100 \times 70 \times 70$ cm$^3$. The outer dimensions are about $200 \times 170 \times 170$ cm$^3$. Thus the total mass is about 60000 kg of Pb. This entire mass must rest on a μ veto, which can directly carry the load. In addition, the mass over the detector area must be supported above the detectors with great confidence. Fortunately, lead bricks are somewhat self-supporting, but the worst case is that about 200 kg must have an ultra-clean support material.

The strategy now under review is the use of electroformed copper to support the lower portion of the shield, with a common OFHC Cu support midway up in the lead. (This approach is being adapted from the plan for the MEGA system.) Supporting the lead over the detector cavity requires a unique approach. The standard approach would be to simply place a sufficiently thick plate of support material, in this case copper, across the cavity. However, the plate would need to be electroformed because the high background levels that exist in commercial-grade copper. To minimize the required quantity of electroform copper, the following design was suggested and analyzed.

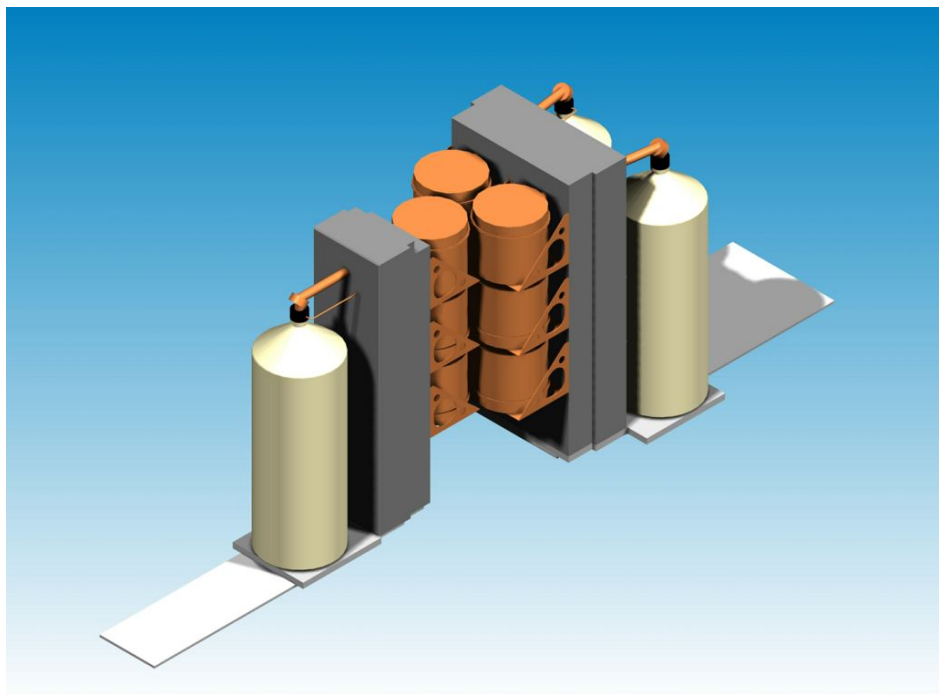

**Figure 3-27. Layout of 9 multi-crystal modules showing monolith slides. The lead shielding is cut short in this drawing to expose the inner structure.**

Support for the lead above the detector cavity is accomplished with two copper plates. The cavity is overlaid with a 0.5-cm-thick electroformed copper plate. This plate achieves two things. First, it provides background shielding for the detector. Second, it provides the structural support for the first layer of lead bricks over the cavity. A 5-cm-thick layer of lead bricks covers the electroformed copper plate. A second plate of commercial type 101 copper, about 1-cm thick supports the additional layers of lead. Design analysis shows that these plates will support up to 40 cm of lead over the cavity with an allowable deflection of <2 mm. This design also keeps the background exposure to the detector minimal. Additional support for the Majorana shield might be provided at low cost by a 'strongback' arrangement of thin copper pieces.

The cave geometry has been chosen to minimize shielding and maximize detector self-shielding and multi-crystal γ-ray detection. However, it will be necessary to access the detector modules for repairs. Furthermore, the Ge detectors for Majorana will be delivered over a few-year time period and hence modules will be brought online in stages requiring minimal impact on already operating modules. A design is needed which facilitates access, maintenance, and the addition of modules.

The Reference Plan is shown in Fig. 3-27. The white rectangles behind the large (50 liter) Dewars are slides, which facilitate the hand- or motor-operated hydraulic removal/insertion of a detector monolith. This design allows the lead cave to be constructed independent of the germanium detector progress. The modules may then be installed individually during the construction and commissioning phase of the experiment. Periodic maintenance of a single detector crystal may be conducted without disruption of the entire apparatus.

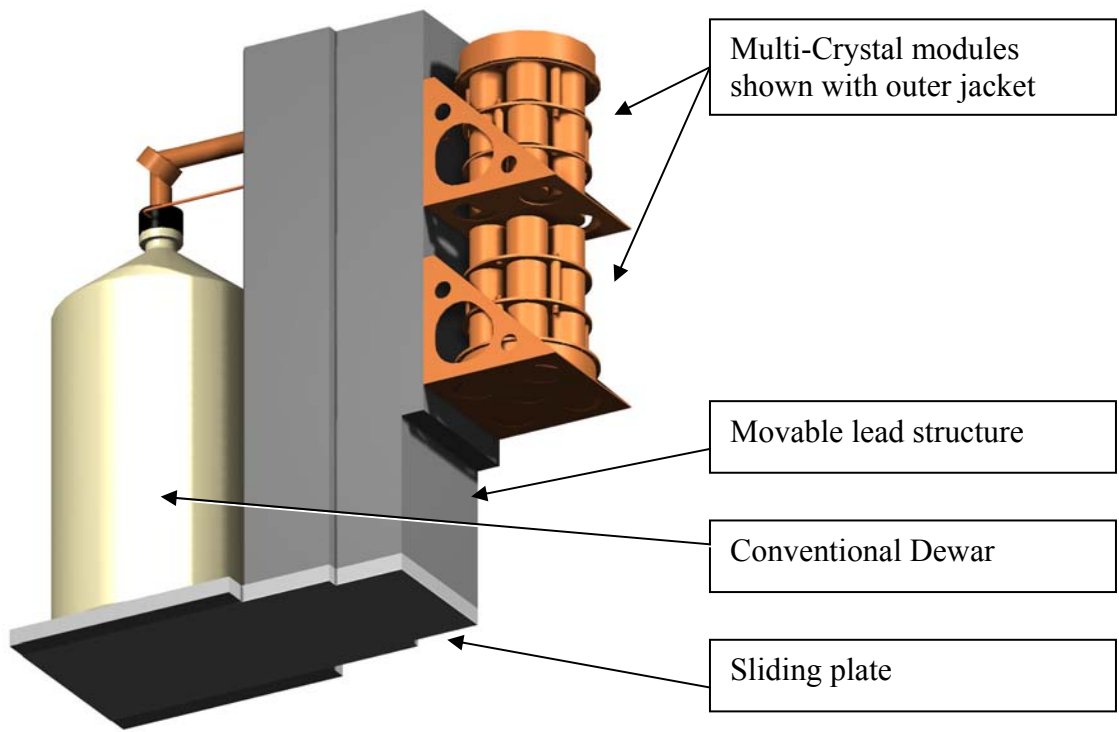

Multi-Crystal modules shown with outer jacket

Movable lead structure

Conventional Dewar

Sliding plate

**Figure 3-28. View of a movable detector monolith. Two multi-crystal modules will be cooled by a single LN Dewar, upper module by dipstick arrangement, the lower by integral connection to the Dewar.**

### *Alternate Cooling/Shielding Investigation*

The standard method for cooling germanium detectors consists liquid nitrogen cryogen providing a heat sink through a copper cold-finger technique protected by a vacuum cryostat. This technique is well developed, well understood, and in the case of the IGEX detectors, executed with cryostats of electroformed copper contributing only a very minor fraction of the background. However it would be overly optimistic to assume that the ultimate background levels will be zero in the Majorana Experiment. Efforts will most likely be invested in attempting to reduce that remaining background, no matter how low it may turn out to be. Notably, opportunities may exist for greatly reducing the amount or types of material close to the detectors. Even if, as expected, these materials are very radiopure, minimizing their use might reduce potential backgrounds or at least simplify the interpretation of the remaining background. It is therefore prudent to set our Reference Plan on the above-mentioned proven technique and investigate alternative configurations in parallel. As a result, we are researching other potential cooling techniques and other potential shielding schemes. This research will continue until the construction schedule is set, at which time the design must be frozen.

There are several alternative cooling/shielding schemes that are being presently discussed within the collaboration and are summarized here.

### Immersed Vacuum Chamber

The traditional alternative cooling technique is the immersed vacuum chamber scheme. The operation of the detectors in vacuum prevents cosmogenically produced isotopes such as atmospheric $^3$H and $^{14}$C from contributing to the low energy background of the detectors by platting out on the sensitive surfaces. In this configuration a tightly packed cluster of germanium detectors would be mounted on a low-mass frame of ultra pure material, inside a vacuum chamber fabricated from ultra pure copper. The chamber must allow high-temperature (100 C) baking under a vacuum to clean all surfaces. When satisfactorily pumped, the entire chamber would be then immersed in a large tank of liquid nitrogen, and the detectors slowly cooled by black-body radiation from the crystals to the cold copper walls maintained at 77 degrees Kelvin. The vacuum would be maintained by a cryo-pump in a configuration that would have the radioactivity of the xeolite or charcoal well shielded from the detector chamber. The power dissipated in the first preamplifier stage could be adjusted to operate the field-effect transistors at an optimum temperature. Unfortunately, radiation makes a poor thermal coupling and therefore it must be demonstrated that the heat load in this multi-crystal array can be effectively removed by this technique. Alternatively, a thermal-coupling gas (e.g. He) could be inserted into the chamber after baking and then removed after cooling and prior to operation. It is also possible to thermally couple the cold copper walls to the crystal array by low mass copper cooling strips or detector holders. These options will be pursued in a research program parallel to the execution to the Reference Plan.

### Immersed Naked Detectors

Another alternative scheme is the direct immersion of naked germanium detectors in liquid nitrogen, as proposed by the GENIUS collaboration [Kla98]. They plan to lower an array of more than 300 detectors of approximately 3 kg each, directly into a large pool of continuously-purified liquid nitrogen. The dimensions of the pool would be large enough to utilize the liquid nitrogen itself as a shield against external radioactivity. Several detector experts agree that liquid nitrogen is an excellent dielectric and in the absence of moisture should not cause surface deterioration or leakage currents. In addition, quick, direct emersion of detectors at approximately 300 degrees Kelvin directly into liquid nitrogen at 77 degrees Kelvin will very likely not damage the detectors due to thermal stresses, because a layer of nitrogen gas will immediately form at the surface, significantly slowing the transfer of heat to the detector. The nitrogen purification system must be extremely efficient in preventing any moisture from freezing out on the surfaces. Prior to direct immersion, the surfaces must be heated in high vacuum to eliminate all moisture. This is in principle possible, but operationally difficult. Minute amounts of moisture frozen on the surface would be devastating to the proper operation of a detector.

A significant research program would be needed to alleviate several potential concerns of this direct immersion configuration. The field-effect transistors (FETs), will not operate well at 77 degrees Kelvin. Therefore, either sufficient power must be dissipated locally, or else the FETs must be removed from the liquid nitrogen, resulting in a long cable length between the first and second stage of the preamplifier. The long cable length required to operate the FET outside of the liquid nitrogen bath will add significant capacitance, resulting in an increase in the energy threshold, as well as considerable degradation of the pulse rise time, and consequently of the quality of pulse-shape discrimination (PSD), necessary to eliminate background. Failure to remove the transistors from the nitrogen might cause nitrogen to boil near the detectors and the transistors, which will raise the energy threshold due to microphonic noise.

The direct immersion technique removes much of the material near the detectors. Unfortunately, it will not eliminate the main sources of background, namely, radioactivity internal to the crystal that is generated by cosmic-ray neutron interactions with the stable isotopes of germanium. Furthermore the PSD effectiveness may be degraded to the point where these backgrounds can't be removed during offline analysis. Due to the rather short neutron interaction length in Ge, PSD is also beneficial in eliminating neutron elastic scattering of nuclei: a significant background for dark matter. In addition, cosmic-ray $\mu$ will create Cherenkov light in the tank. Germanium behaves as an insulating dielectric but only in the absence of light. The Cherenkov light could cause significant surface currents that might harm the sensitive FET, cause pulses, or cause surface damage. Significant research and development will be required to explore and overcome technical challenges of the direct immersion technique.

### Gas-Filled Chamber

A cold gas atmosphere surrounding the detectors, which is circulated from a chiller outside of the detector chamber could cool the detectors. This gas would greatly improve the thermal coupling between the detectors and the chilling liquid. The chamber must be one that can be evacuated and could allow the detectors to be heated to about 100 degrees centigrade and pumped, to clean the surfaces and to completely eliminate moisture. The pressure and temperature of the gas would have to be carefully regulated to prevent arcing when the detectors are at high voltage. The advantage of this configuration is that it can be immersed in a large tank of continuously purified water, ice, or a large volume of scintillator to act as a veto.

### Active Inner Shield

As a nucleus proceeds through a radioactivity chain, it emits a series of radiations. A time coincidence or a delayed coincidence between such radiations can be used to identify and reject events originating from the natural radioactivity chains and many other isotopes. Cosmogenic activities are produced mostly through nuclear reactions producing particle and or $\gamma$ decays from excited states. For short-lived activities (up to ~1 hour), these prompt activities may provide a tag for any potential background. Many activities, $^{60}$Co for example, have coincident $\gamma$ rays. The modularity, segmentation, and pulse shape analysis of the Ge detectors can exploit these various timing signatures to reduce background. In addition however, one could consider an active inner shield contained within the inner passive shield.

The purpose of this shield would be to increase the efficiency for Compton suppression or $\gamma$–$\gamma$ coincidence resulting in a reduced background. We are investigating the potential for such an active inner shield including the possibility of surrounding the Ge detectors with $^{nat}$Ge crystals or a liquid scintillator.

### 3.9   Electronics and Data Acquisition

The instrumentation of the Majorana apparatus is driven by the need for low power dissipation and radiopurity while maintaining fast rise-time performance to support pulse-shape discrimination. Additionally, the electronics chain must provide only a negligible contribution to detector resolution. Field-Effect Transistor front-end modules near the crystal contacts are combined with charge-integrating preamplifiers and digital processing of the preamplifier output pulses to achieve this goal.

To achieve the best current-pulse shape fidelity and lowest-noise operation, it is necessary to amplify the current pulse evolved in a germanium ionization spectrometer without introducing additional capacitance. Any lead-length capacitance added to the irreducible capacitance of the detector electrode under observation reduces the magnitude of the signal by charge-sharing, reducing the signal-to-noise ratio of the induced current pulse relative to thermal fluctuations and other noise sources. The traditional, and very successful, solution to this problem is to locate the first-stage FET of a charge-integrating preamplifier very near the detector electrode under observation. This front-end FET gives considerable voltage gain and provides an output impedance of ~100 ohms, suitable for driving a connection to the rest of the preamplifier circuit. The remaining part of the charge-integrating preamplifier can be located at some distance from the detector electrode, allowing the detector to be shielded from radioactivity in the preamplifier materials. Members of the collaboration, while working on the IGEX experiment, have explored optimizations of the front-end electronics to further reduce background contribution while preserving excellent signal fidelity [Aal99a]. (See Fig. 3-29)

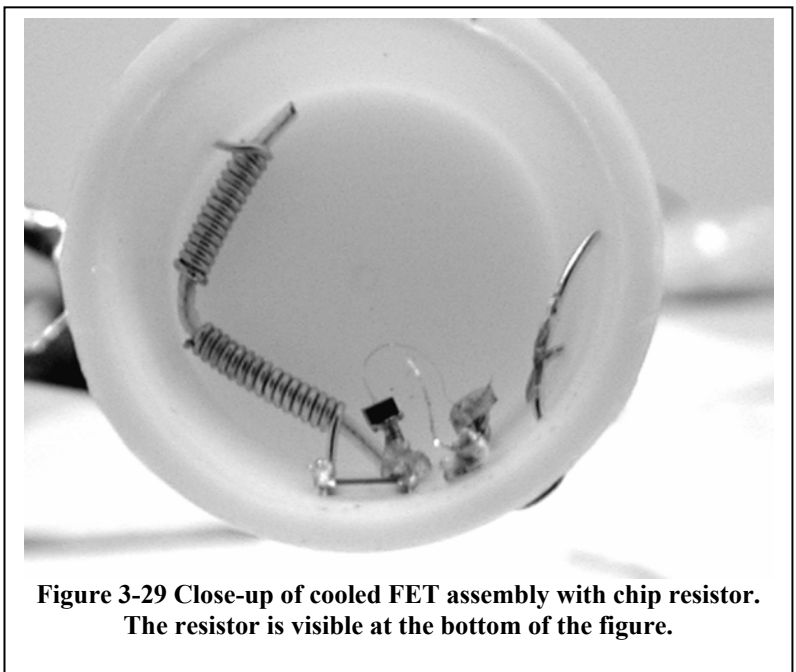

**Figure 3-29 Close-up of cooled FET assembly with chip resistor. The resistor is visible at the bottom of the figure.**

After the recovery of an induced-current signal via charge integration, the signal is ready for further processing. Digital spectroscopy hardware platforms have become commercially available for this task and the data acquisition systems for SEGA and MEGA will use this commercial technology. After digitizing the output of a charge-integrating preamplifier, all subsequent processing is done in the digital domain. This gives remarkable flexibility in matching filter and shaping parameters to the characteristics of the detector. This allows very low energy thresholds (<1 keV) when compared to traditional analog shaping amplifier performance. Additionally, these new hardware platforms make available the digitized information necessary to do pulse-shape discrimination for background rejection based on interaction multiplicity. Other useful analyses, such as pulse-shape discrimination for the localization of single-site interactions and the rejection of microphonic signals are possible.

The number of electronics channels for SEGA and MEGA is rather small. Therefore, the relatively high cost/channel of a purely commercial digitizer/spectroscopy system is offset by its high degree of design refinement and broad built-in capabilities for germanium gamma spectroscopy. In the Majorana detector, the much larger number of channels may necessitate an investigation into lower-cost options. However, the layout and system concept can be well described in terms of the commercial digitizer/spectroscopy hardware. An additional strength of the commercial digitizer is that it is known to meet the technical requirements for spectroscopy and pulse-shape discrimination, and would require much less engineering and technical effort to implement. Therefore that system is described here as it would pertain to Majorana. Other options are discussed at the end of this section.

The basic scheme is shown in Fig. 3-30. Each Ge detector in the Majorana Reference Plan is segmented into a number of disks-shaped active volumes via one inner semi-coaxial contact and a number of outer axially-spaced contacts. For each detector the inner contact is instrumented with one high-bandwidth readout providing data for pulse-shape discrimination and event energy. Each of the outer contacts are instrumented with a low-bandwidth readout providing a digital hit/no-hit indication. Commercial digital spectroscopy modules, the DGF4C series, manufactured by X-ray Instrumentation Associates, will digitize the preamplifier output pulses from the high-bandwidth channel. These CAMAC-based modules capture all energy, timing, coincidence information, and integrated current signals for subsequent pulse-shape discrimination. A binary data stream, defined by the DGF4C firmware, is sent in buffered blocks to the data acquisition and control system. Each DGF4C module has internal counters for accumulating the number of triggers, over/underflow events, real-time, and live-time for each channel.

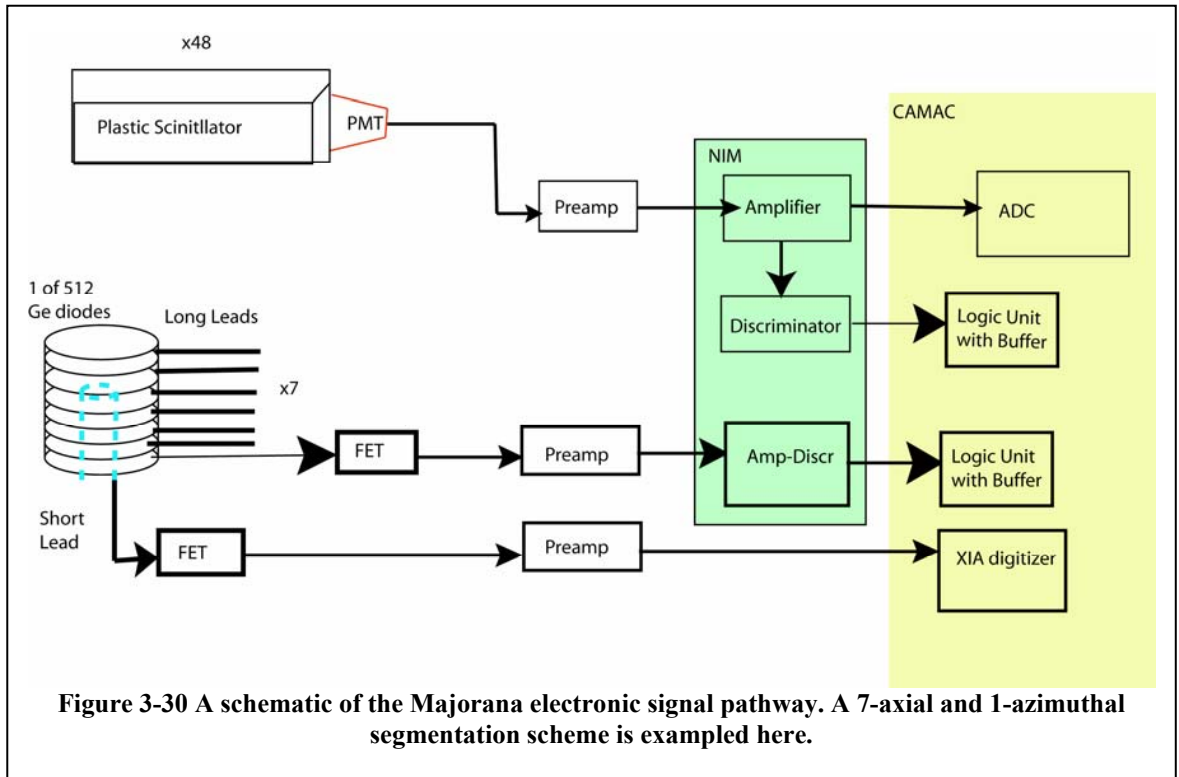

**Figure 3-30 A schematic of the Majorana electronic signal pathway. A 7-axial and 1-azimuthal segmentation scheme is exampled here.**

The signals from the low-bandwidth channels are amplified, discriminated, time-stamped using the DGF4C clock reference, and read into a logic unit. Each DGF4C has 4 channels and therefore a 32-channel logic unit and one DGF4C will acquire data from a group four Ge crystals. Our Reference Plan includes approximately 512 Ge detectors and therefore about 128 such groups. The DGF4C and the logic units have buffers that are continually loaded until one of the buffer of a DGF4C init is full (approximately 10-30 events) or the data acquisition indicates a run-stop. At that point, all the digitizer and logic unit buffers are read out.

A useful feature of the DGF4C is its ability to be interfaced such that triggering and common timing is fairly simple. The ~128 digitizer-logic unit pairs will be distributed among ~12 CAMAC crates. One of these crates will be a Master and the others will be

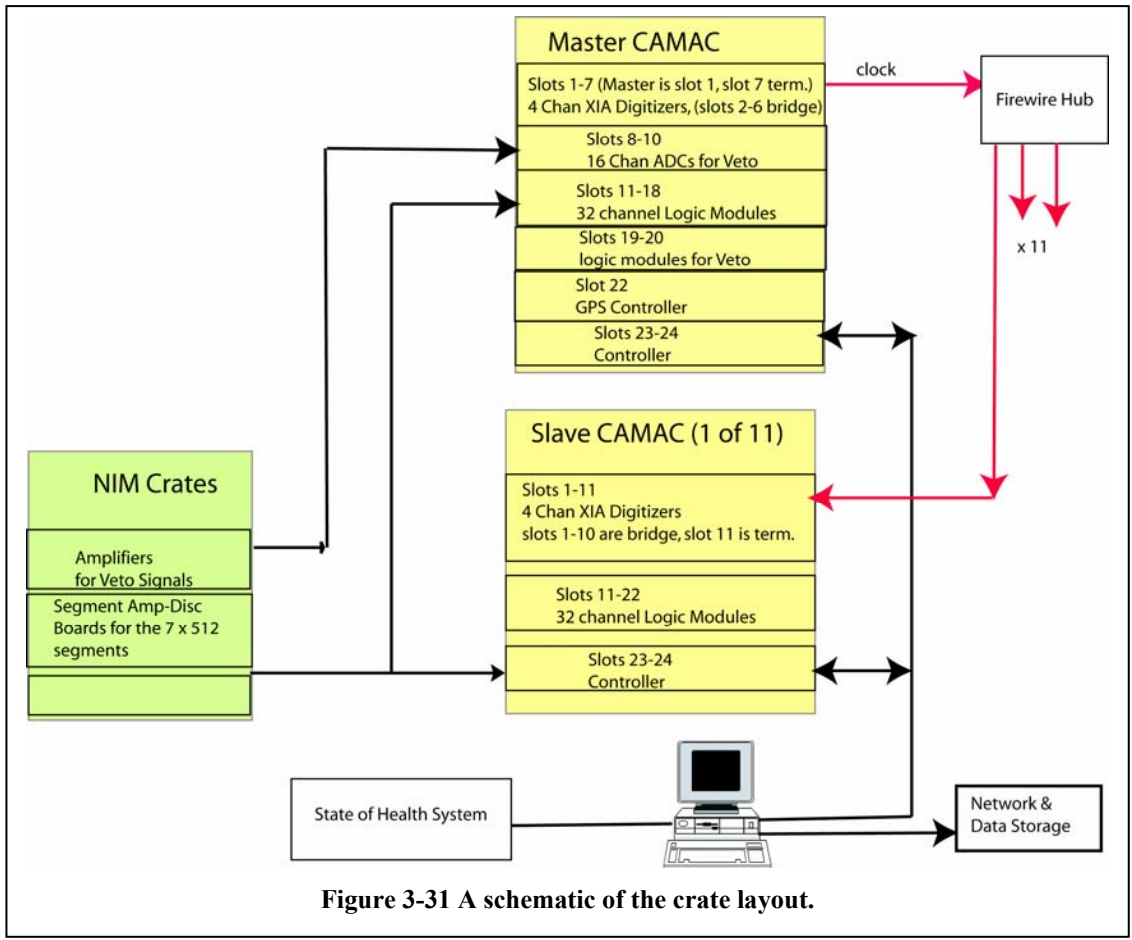

**Figure 3-31 A schematic of the crate layout.**

Slave. (See Fig. 3-31.) One of the DGF4C modules in the Master Crate will be the clock master. The 40 MHz clock from this module will be the input clock for all other modules. This clock increments a 48-bit counter on all the boards. A common sync input for all modules allows a system synch pulse to uniformly reset all DGF4C and logic unit counters. For absolute timing, the clock increments a counter in a GPS module that also resides in CAMAC. The system sync pulse will also reset this counter. The GPS module will then provide an output indicating the absolute time corresponding to a system timer reset and thus allow an absolute time reference for each event.

The communication between a digitizer and its paired logic unit in shown in Fig. 3-32. When the DGF4C triggers, it sets a multiplicity out pulse that latches the logic unit. All 28 channels of the logic unit are then latched into the buffer for latter readout. The logic unit buffers and all the digitizers are reset after a data read. The clock counter, or system time-stamp, for each buffer latch is also recorded.

The electronics system will also include a computer-controlled a high voltage bias supply system for both the HPGe detector array and the phototubes of an anticoincidence shield. Separate, conventional, instrumentation will derive veto signals from an anticoincidence shield. A logical OR of these veto signals is combined into the DGF4C data stream via a Global Second-Level Trigger (GSLT) input, providing a timestamp in the data stream for veto firing.

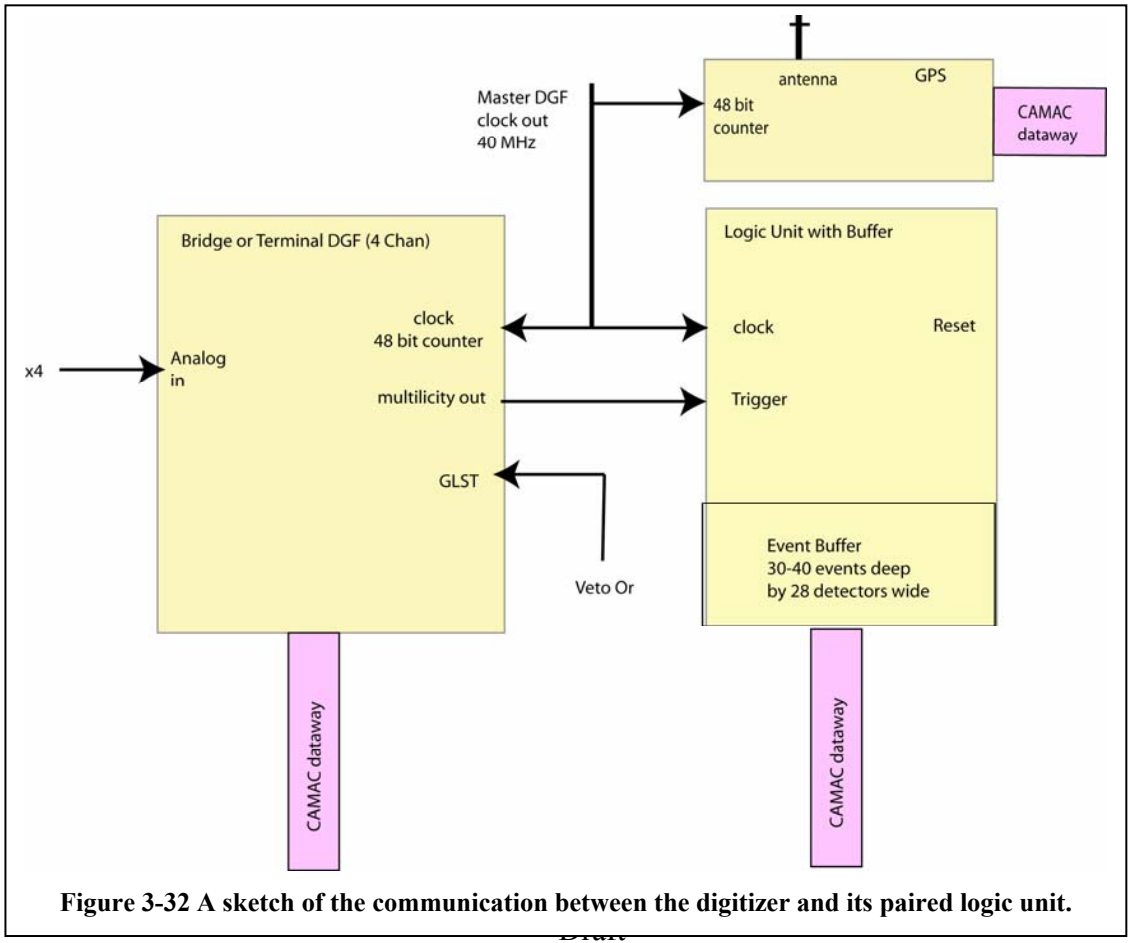

**Figure 3-32 A sketch of the communication between the digitizer and its paired logic unit.**

The time-stamped data stream from each DGF4C module is sent to a control computer, where an event data model utilizing the ROOT framework (http://root.cern.ch) is populated and the ROOT tree subsystem is used for local data storage and serving the data stream to other locations for further analysis. State-of-health data will be continuously monitored for variables such as cryostat temperature, ambient radon level, detector segment leakage current, etc., and will be logged locally and served to other locations for further analysis and real-time monitoring.

The Collaboration also has many years of experience in data acquisition software design, coding, and implementation. This includes, for example, large experiments such as the Sudbury Neutrino Observatory [Ahm01, Ahm02], precision spectrum experiments such as IGEX [Aal99a, Aal99b, Aal99c] and the LANL tritium beta decay experiment [Rob91], and very-distant remote-location operation such as the Russian-American Gallium solar neutrino Experiment collaboration (SAGE) [Abd02] and remote unattended radionuclide sampler/analyzer systems [RadSam] in support of various international treaties. The software control system will implement the various operating modes necessary for the experiment operation. These include normal operation, calibration, and various diagnostic functions. Further discussion of this topic appears in the section, "Computer Science Infrastructure".

By using a combination of highly-evolved low-background electronics near the detector and commercial modules for digitization and energy processing, the technical capabilities of the detector instrumentation are maximized (pulse-shape discrimination, low radioactivity) while minimizing engineering risk and cost. Although a purely commercial option saves engineering costs, the DGF4C digitizers themselves are a significant cost. In the next few paragraphs, we discuss options under investigation, which may lower the overall cost of the electronics and may improve performance.

### *Multi-chip Module Technology*

Further optimizations of the electronics located near the detector electrodes are planned, with multi-chip module technology under consideration. The techniques of multi-chip module construction may allow more sophisticated analog electronics to be implemented near each detector electrode without sacrificing low backgrounds. This has the potential to reduce the number of connections per contact entering the cryostat below the typical four found in most commercial setups (FET source, FET drain, feedback voltage, test pulse injection). With shared power connections and charge-integration completely inside the cryostat, each electrode could have as few as one connection to the outside.

### *Digitizers*

The DGF4C cost about $15k each or roughly $4k/channel. Clearly less expensive digitizers are available. However, the added features of these modules make online processing and triggering easy. Furthermore these are a commercial item and hence are plug-and-play. A less expensive digitizer may require packaging and hence engineering. It will also require triggering electronics and associated engineering. We will assess the relative costs of other possible digitizers. In particular the GRETA collaboration is doing similar research on digitizers for segmented Ge detectors. We may be able to "piggyback" on that effort to reduce any engineering costs. If a more cost-effective solution is found that allows us to meet our pulse-shape discrimination and spectroscopy performance benchmarks, we will alter the Reference Plan.

### VME

CAMAC is not the most modern bus architecture. However, the DGF4C is a CAMAC-based module. If we choose a different digitizer, we will reconsider the use of CAMAC or VME.

### Pulser system

A pulser system to monitor gain stability would be useful. The specifications for such a system are under consideration.

### Computer Science Infrastructure

The classical system QA functions of nuclear data acquisition have changed in the last twenty years: it is now routinely possible to acquire many types of system state-of-health (SOH) data (temperatures, pressures, currents, and so forth) and to produce many products besides raw data, ready for post-processing. The structure of a system for sharing, processing, reporting, and archiving the data and data products can provide a new level of quality and ultimately confidence in results at a minimal cost. The raw data, results and all experiment logs will be electronic and available freely within the collaboration.

Raw data of many kinds will originate from the several underground facilities housing the Majorana Experiment hall, the electroforming laboratory, and other instrumented spaces. The types of raw data may include:

- Time-stamped germanium pulse waveforms and segment-hit patterns
- Time-stamped veto signals
- Periodic SOH data readings (frequency ~1/min)
    - Experiment hall
        - Oxygen levels
        - Detector temperature/leakage current
        - Room temperature and humidity
        - Radon level
        - LN levels
        - LN purge gas flow
        - Electronics temperature
        - Barometric pressure
        - Airborne particle concentration
        - HV status
        - Motion sensors
        - Shielding location sensors
        - Supply Power: current and voltage
    - Electroforming lab

- Radon level
- Airborne particle concentration
- Hydrogen concentration
- Oxygen concentration
- Corrosive fume detection

Data products originating within the underground location could include email alerts when state-of-health parameters exceed preset bounds. In addition, it will be possible to log and share human-generated log entries.

This stream of raw data can be forwarded to all appropriate parties and logged locally in case of communications interruption, but to insure uniform data quality and provide storage/archival, the data will be sent to a central data hosting facility. Each data message will be logged in a chronological way and described in a cost-effective data base.

A web interface to this data base is an effective way to quickly look for alerts and correlate them with anomalies in SOH data, using helper applications for viewing each type of file. This allows human SOH monitoring to occur at all collaborating institutions at all times of day and takes great advantage of the global distribution of collaboration members. A prototype of this system is currently running at PNNL.

Of course, this monitoring of periodically-transmitted files is not instantaneous: if data is sent in, say, 1-day increments, SOH monitoring will be after the fact. For this reason, viewer applications are under development that will allow effectively instantaneous inspection of a live stream of SOH data. This will be quite useful for setup, calibration, and maintenance of the apparatus and related systems.

The receipt of the raw data in a central data hosting facility also implies that unattended auto analysis can take place with the resulting data product distributed. While detailed, parallel analyses will take place in various locations, it is important to have a reasonably high-level and totally uniform method of gauging the basic status of the acquired data. Automated analysis of both germanium spectral data and SOH parameters is planned to take advantage of statistical failure prediction models.

There is a mature body of mathematical/statistical techniques to characterize system reliability and predict system failure. These techniques effectively model the probabilistic dependence structure of SOH measurements and use these models to formulate mean-time-to-failure estimates with associated uncertainties, component failure predictions and preemptive failure detection, and they provide defensible guidance on where to focus engineering efforts to improve system reliability. These techniques include general stochastic processes (e.g., Markov chains and processes) and time-series analysis, reliability models, experimental design and accelerated-life testing methods. Such techniques have been successfully applied to remote unattended radionuclide collection and analysis systems currently in use around the world for treaty verification applications.

### 3.10 Calibration

If the $0\nu\beta\beta$ half-life is $10^{26}$ years, that is an order of magnitude longer than present limits, the Majorana Experiment will detect about 72 signal counts in 5 years of run time. In this case, the minimum statistical error would be approximately 12%. This sets the scale for the size of acceptable systematic uncertainties and therefore how well we need to characterize the detector through calibration. The experimentally determined half-life depends on the live time, amount of source, and the efficiency of the detector. In this section we consider how each of these factors contribute to the overall uncertainty in the half-life determination. Furthermore, we describe how the calibration of the detector will address these uncertainties.

The primary characteristics of the detector array, and hence each individual detector, that we will want to calibrate are:

- The energy scale
- The energy resolution
- The time dependence of energy scale, resolution, and linearity
- The stability of the threshold energy
- The absolute efficiency for double beta decay
- The pulse shape parameters cut efficacy
- The efficiency of the active shield
- The dead time of the active shield and segment coincidence time cuts
- The energy response non-linearity
- The peak shape

#### 3.10.1 Live time

The veto dead time can be estimated as: $f=Rw=FAw$, where $f$ is the fraction of time dead, $R$ is the rate through the detector, $F$ is the flux and $w$ is the time window width. Taking $A$ as a 2-m square (i.e. 4 $m^2$) and the width as a very wide 1 sec, one can create a depth dependent $f$.

| depth | F | R=FA | f |
|---|---|---|---|
| 2000 mwe | $10^5/m^2$ y | 0.013 Hz | 1% |
| 4000 mwe | $3 \times 10^3/m^2$ y | $4 \times 10^{-4}$ Hz | $4 \times 10^{-4}$ |
| 6000 mwe | $2 \times 10^2/m^2$ y | $2 \times 10^{-5}$ Hz | $2 \times 10^{-5}$ |

Even with a 1% dead time, the uncertainty associated with veto dead time should be negligible.

Similarly, the dead time due to the segmentation anti-coincidence requirement will also be small. The coincidence rate (C) can be written: $C= R_{0\nu} R_{tot} w$, where $R_{0\nu}$ is the rate of events in one segment in the region of interest, $R_{tot}$ is the total rate of all segments and $w$ is the time window. Using Reference [Avi91], $R_{tot}$ for energies over 200 keV, is estimated to be $\sim 3 \times 10^{-4}$/kg s. For 500 kg, this becomes ~0.2 Hz and might be a factor of 10 higher for a threshold of 10 keV. Since the timing of Ge detectors is on the order of a μs or less, the timing window can be very small. Thus the fractional loss of live time would be small and therefore its uncertainty will be negligible. ($R_{0\nu}$ should be better than $(0.2/\text{keV kg y})(4 \text{ keV})(500 \text{ kg}) = 1 \times 10^{-5}$ Hz.)

### 3.10.2 Number of $^{76}$Ge atoms

The mass of the Ge can be determined better than 1%. The enrichment will be known to better than 1-2%. The fiducial volume is more difficult, however, as it has to be measured with a source. The source activity can be known to 1-5%. The position of the source will probably be determined to 1-2 mm. One might guess that the source-detector separation distance will be on the order of 10-20 cm and therefore the position is known to about 1-4% of the separation. This will translate into a ~2-8% flux uncertainty due to the position uncertainty.

However, one can use the relative intensity of several gamma rays from a lone source to determine the fiducial volume independent of the position of the source. The dead layer is determined by measuring the relative attenuation of several low-energy gamma-ray lines from the same source, so the location and intensity of the source are inconsequential. As a result, the apparent fiducial volume of the crystal can be determined with a negligible uncertainty. The relative dead-layer effect for internal double-beta decay events versus external gamma-ray transmission still needs to be determined, however. In other words, the internal boundary of the dead layer is not precisely defined as is the external boundary (i.e. the surface of the crystal). At best, this internal boundary represents an interface. However, the approximate size of the dead layer is known and a conservative choice for its size can be used to deduce an upper limit on this uncertainty. For an uncertainty of 100 microns for the dead layer of a P-type detector that is 80 mm in diameter, the fiducial volume uncertainty is 0.2%.

### 3.10.3 Efficiency

If the gain is imprecisely known, then the peak location is actually at a different position than assumed by the analysis. If the number of counts in the peak is too small for the location to be determined by a fit to the peak, then an error in efficiency will result. For a typical IGEX resolution of 3.6 keV ($1\sigma=1.3$ keV), one would expect a region-of-interest selection efficiency of 83.8%. If the gain uncertainty is 0.2 keV (as previously achieved in IGEX) at 2039 keV, one would overestimate the efficiency by 0.5%: a small quantity. Note that the energy of the 0nbb transition is well known ($2039.006 \pm 0.050$ keV [Dou01]) and introduces a negligible uncertainty in the efficiency.

If the resolution is uncertain, one makes a similar type error, although it is two-sided. If the resolution is $4.0 \pm 0.4$ keV for example, one would have an acceptance uncertainty of about 3%.

The pulse-shape-discrimination acceptance is also a contributor. How well this cut acceptance will be known depends mostly on counting statistics from the source. Hence, this uncertainty will be determined by source strength and calibration duration. For example, a 1% uncertainty would result from a 3-Hz counting rate over about an hour. That of course assumes that the PSD cut is similar for all segments. If we need 10000 counts in each segment, we will need a much higher counting rate. Thus a reasonable estimate is a few %.

### 3.10.4  Calibration Specifications

This is the list of specifications that determine what the calibrations are and what they measure.

The required dynamic range of the energy measurements could be fully calibrated with a Pb x-ray source plus an external thorium source.

There are a number of physical processes that need to be calibrated including neutron recoils and double escape peak (DEP) events. An external Th source can provide the DEP signal, although a hot $^{26}$Al source has ideal line energies. A Cf source can provide the neutron recoil event sample.  The source activities are only constrained by the data rate of the data acquisition.

In order to determine absolute efficiencies, sources with modest to long half-lives, reproducible source locations, and good statistics. We will need to calibrate about once every week for about an hour.

The gain for each detector channel needs to be determined precisely enough such that the summed resolution is not degraded beyond the ~4 keV specification discussed above. The native number of channels spanning the energy dynamic range of the detectors will be about 65k. So relative spectrum shifts can be performed without re-binning difficulties.

**Table 3-18 A summary of the systematic errors in the Majorana Experiment.**

| Effect | Uncertainty |
|---|---|
| Statistics ($10^{26}$ year half life) | 12% |
| Live Time | |
| Veto anti-coincidence | <1% |
| Segment anti-coincidence | <1% |
| Number $^{76}$Ge atoms | |
| Ge Mass | <1% |
| Enrichment | 1% |
| Fiducial Volume | |
| Dead Layer Thickness Uncertainty | 1% |
| Acceptance | |
| Gain | <1% |
| Resolution | 3% |
| PSD | Few % |
| Segmentation Cut | Few % |

Using a GPS clock and the planned electronics, we ought to be able to get relative timing between signals to 25 ns and absolute timing to ~100 μs or less. This will be much better than required for any coincidence studies.

The anticipated systematic uncertainties are summarized in Table 3-18. It is readily seen that sufficient calibration is not overly challenging.

### 3.11  Analysis

Majorana is not simply a volume expansion of previous experiments, such as IGEX. It must have superior background rejection. Because it has been conclusively shown that the limiting background in at least some previous experiments has been cosmogenic activation of the germanium itself, it is necessary to mitigate those background sources. Cosmogenic activity fortunately has certain factors, which discriminate it from the signal of interest. For example, while $0\nu\beta\beta$-decay would deposit 2 MeV between two electrons in a small, perhaps 1 mm$^3$ volume, internal $^{60}$Co decay deposits about 318 keV (endpoint) in β energy near the decaying atom, while simultaneous 1173-keV and 1332-keV γ rays can deposit energy elsewhere in the crystal, most probably both in more than one location, for a total energy capable of reaching the 2039 keV region-of-interest. A similar situation exists for internal $^{68}$Ge decay. Thus, deposition-location multiplicity distinguishes double-beta decay from the important long lived cosmogenics in germanium. Isotopes such as $^{56}$Co, $^{57}$Co, $^{58}$Co, and $^{68}$Ge are produced at a rate of roughly 1 atom per day per kilogram on the Earth's surface. Only $^{60}$Co and $^{68}$Ge have both the energy and half-life to be of concern.

To pursue the multiplicity parameter, two approaches are possible. First, the detector current pulse shape carries with it the record of energy deposition along the electric field lines in the crystal; crudely speaking, the radial dimension of cylindrical detectors. This information may be exploited through pulse-shape discrimination, as described below. Second, the electrical contacts of the detector may be divided to produce independent regions of charge collection, the detector segmentation scheme described earlier.

By segmenting the inner contact into two (axial) parts and the outer contact into 6 (azimuthal) parts, as was described earlier in Section 3.6, multiplicity data can be obtained. Other segmentation schemes are as efficient according to our Monte Carlo computations as discussed earlier.

The Monte-Carlo simulation data set shown in Fig. 3-33 is based on this configuration and shows that internal highly-multiple backgrounds like [60]Co can be strongly suppressed at 2039 keV. Cosmic-ray neutrons produce the internal [60]Co modeled in the figure during the preparation of the detector, accumulating after the crystal has been grown. Its elimination by segmentation and pulse-shape discrimination is crucial. Beyond this simple segmentation cut, it may be possible to use the induced signals derived from segments seeing no net charge, adjacent to a segment seeing net charge, to locate a single-site deposition in the axial and azimuthal coordinates of the crystal or to distinguish a single-site deposition from a multiple one. The results of the initial Monte-Carlo simulation described above for a 6-azimuthal and 2-axial segmentation configuration predict a discriminator with an acceptance for 0νββ-decay events of 90.7%. The predicted acceptance for internal [60]Co events is only 13.8%. These acceptance and rejection numbers are typical of the various segmentation configurations being considered.

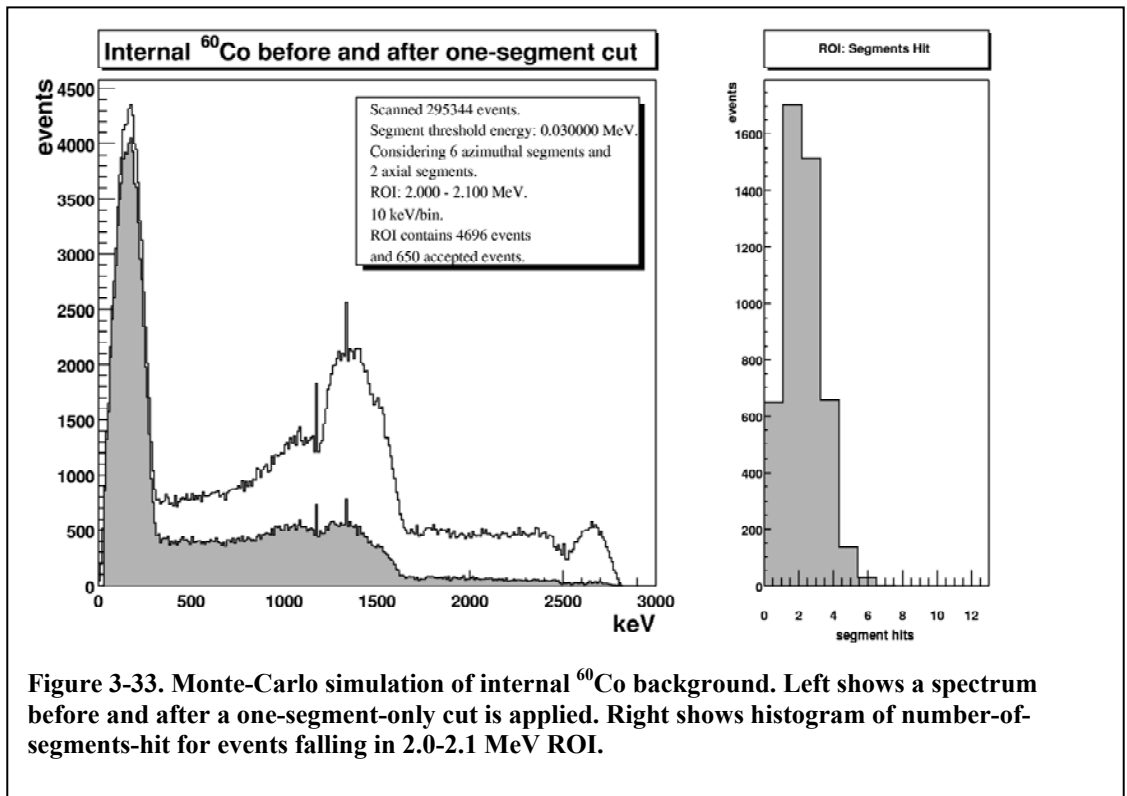

**Figure 3-33. Monte-Carlo simulation of internal [60]Co background. Left shows a spectrum before and after a one-segment-only cut is applied. Right shows histogram of number-of-segments-hit for events falling in 2.0-2.1 MeV ROI.**

Pulse-shape discrimination (PSD) is another way to gain sensitivity to interaction multiplicity, and works by examining the digitized current pulse, as presented by the output of the charge-integrating preamplifiers. Figure 3-34 shows two experimental pulses of approximately equal energy but one is a localized ionization and the other is a multiple site deposition. The two pulses are clearly distinct. Several years of research at PNNL and the University of South Carolina have produced a new PSD technique lacking many of the disadvantages of previous methods.

Early research showed that common methods based on comparison of pulse-shapes to libraries or basis sets of calculated single-site pulses have disadvantages. Comparing each pulse to even a small library of template pulses is computationally intensive. More problematic is the fragility of templates or libraries of calculated pulses in the face of normal variations in experimental conditions. These variations could include changes in operating voltage, differences or inhomogeneities in minority carrier concentration, and variations in the alignment and operating parameters of different preamplifiers.

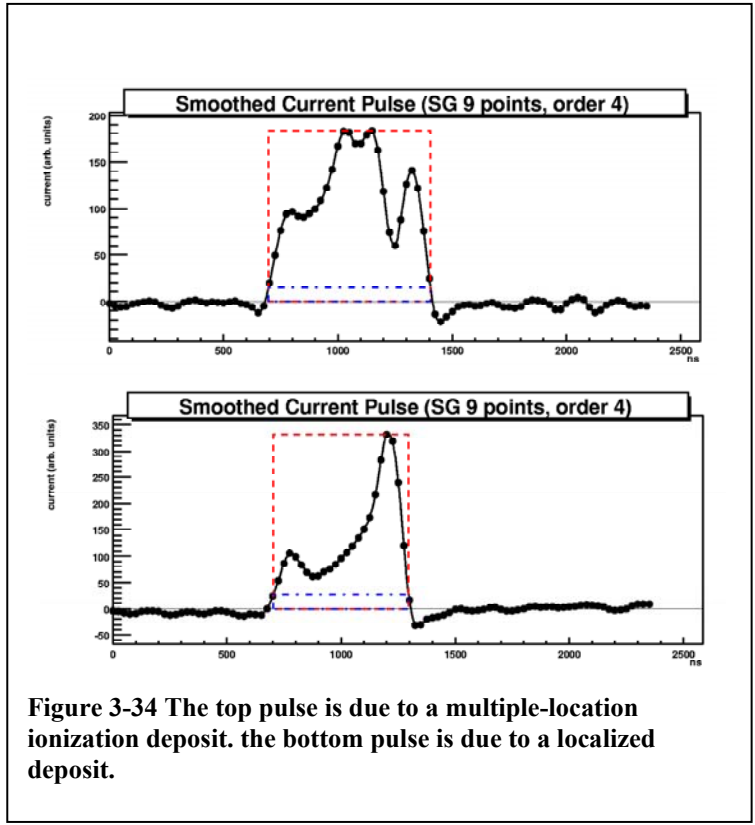

**Figure 3-34 The top pulse is due to a multiple-location ionization deposit. the bottom pulse is due to a localized deposit.**

These problems are avoided with the parametric, self-calibrating, PSD technique recently developed. Using only a short calibration data-set, easily acquired with conventional low-level calibration sources, the discriminator is optimized for each detector/electronics setup. This calibration is fast, allowing the discriminator to be re-optimized frequently to account for any changes in the operating conditions. Using only a small number of parameters extracted from each pulse, this technique has modest computational requirements, allowing analysis to be rapid.

Monte-Carlo data suggest two types of experimental data are useful in quantifying the efficacy of such a discriminator. Events in the full-energy peak of the 1620.6-keV line from [212]Bi were calculated to have an event multiplicity slightly lower than that of the expected internal cosmogenic backgrounds near 2 MeV. Thus these events are a conservative source of background-like events. Events in the 1592.5-keV double-escape peak (DEP) of the 2614.5-keV line from [208]Tl are calculated to have an event multiplicity nearly identical to those from 0ν ββ-decay. Thus these events are a good source of signal-like events.

Applying the new PSD discriminator to an experimental data-set gives the result seen in Fig. 3-35. The white spectrum is from the original data, while the gray spectrum is the result of applying the discriminator. The features of interest are the initial and final peak areas for the two types of events. For the DEP events, the discriminator yielded an acceptance fraction of 80%, and the gray spectrum is normalized by dividing out this fraction. This shows the DEP intensity as the same for the two data sets and facilitates visual analysis of the relative reduction in the intensity of the gamma peak. For an experiment governed by Poisson statistics, the sensitivity scales as the square root of the background, while also scaling linearly with detection efficiency. This suggests defining a figure-of-merit (FOM) as

$$FOM = \frac{\varepsilon_{\beta\beta}}{\sqrt{\varepsilon_{bkg}}}$$

where $\varepsilon_{\beta\beta}$ is the efficiency of a discriminator for 0ν ββ-decay events and $\varepsilon_{bkg}$ is the efficiency for background events. This FOM is then the multiplicative factor increasing or decreasing the half-life limit of the experiment to which the discriminator is applied. For the current implementation of the PSD discriminator, the FOM is 1.56, or a 56% increase in the half-life bound. It is interesting to note that achieving an equal increase in sensitivity by scaling the mass of the proposed experiment would imply increasing the mass from 500 kg to over 1200 kg. Clearly background rejection is, in general, a very cost-effective technology to pursue.

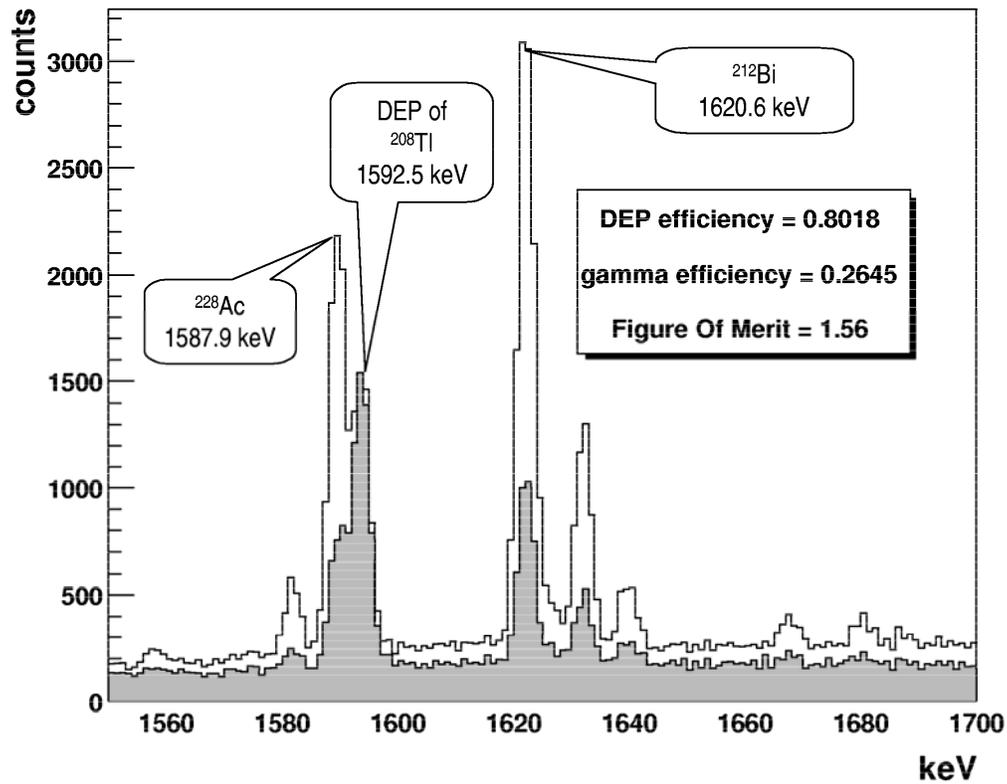

**Figure 3-35 Portion of double escape peak (DEP) spectrum before (white) and after (gray) application of PSD discriminator. Residue (gray)_spectrum has been scaled for equal DEP intensity before and after discrimination.**

Analysis of the data stream from the Majorana instrument will rely on both the traditional techniques of low-background spectroscopy and new segmentation and pulse-shape discrimination methods. Timing information and signals from the active veto shielding are integrated parts of the event stream stored during operation.

An event data model utilizing the ROOT framework ([http://root.cern.ch](http://root.cern.ch)) has been developed for digitized pulses from HPGe detector segments. Subsequent analysis is facilitated using the ability of ROOT to allow flexible access to the part of each event data object required for the particular analysis step. Analysis of the data stream from the Majorana Experiment will proceed in steps, with each step reducing the size and multiplicity of the data stream. The steps are described below.

- *Step 1 – Anticoincidence Time Correlation*

The time of each event is analyzed for correlations with events seen in the anticoincidence shield and/or other events seen in the detector array. Events can be rejected based on observed correlations. Previous experience in the IGEX effort has shown the utility of storing timing information for each germanium detector event relative to the last active veto signal. This allows optimal veto timing to be developed and various veto time windows to be explored in an off-line analysis.

- *Step 2 – Segment Multiplicity Cut*

Events showing energy deposition in more than one detector segment in the array are marked as rejected. Monte-Carlo models of detector segmentation have driven the calculated efficacy of this technique, as described in the Majorana sensitivity calculation. More complete Monte-Carlo models are underway, and will further refine the conservative estimates of the efficacy of this cut.

- *Step 3 – Pulse-Shape Discrimination: Noise Rejection*

Experimental rejection of noise spikes and microphonics has been demonstrated using sophisticated post-processing of each event reaching this phase. Leakage current spikes and high voltage system leakage are two examples of the types of low-rate noise that can be identified. Additionally, analysis of the baseline noise between pulses is a useful state-of-health diagnostic for the detectors of the Majorana instrument. Electromagnetic interference, increased leakage current, or thermal fluctuations will become apparent before they have a deleterious effect on detector resolution.

- *Step 4 – Pulse-Shape Discrimination: Multi-Site Rejection*

Over the past few years, researchers at PNNL and USC have developed and tested an improved class of pulse-shape discrimination techniques. These optimal, self-calibrating, parametric discrimination techniques differ greatly from earlier methods in that they are easily calibrated to individual detector characteristics. These pulse-shape discrimination techniques can be applied to reject multi-site background events, as discussed earlier. This analysis is facilitated using existing codes built with the ROOT object-oriented C++ framework. Additionally, detector segmentation results can be improved by analyzing induced current pulses on all relevant detector segments for consistency with the signature of a single-site interaction.

- *Step 5 – Pulse-Shape Discrimination: Event Localization*

Experimental data shows that, for single-site interactions, information about the spatial location of the interaction can be extracted from pulse-shape discrimination result, as well as from an analysis of induced currents on all relevant detector segments. Each remaining valid event data object is updated with the best possible information as to the location of the interaction.

- *Step 6 – Cuts based on inhomogeneities in activity map*

Single-site events associated with areas of abnormally high activity, for example, due to surface contamination or an isolated "hot spot" in support material, can be rejected via an appropriate small reduction in fiducial volume of the overall detector array. An ongoing map of detector activity is generated as data collection progresses, allowing the identification of any problematic areas. This data can also guide the replacement of any components having higher-than-expected levels of radioactivity.

- *Step 7 – Spectrum Analysis*

The final set of selected events represent single-site energy depositions. It is this set that will form the energy spectrum that will be analyzed for evidence of double-beta decay.

The rich, multi-parametric nature of the data generated by the Majorana apparatus allows an array of analysis techniques to be applied. This data set will become a resource to which new, more optimal, analysis techniques can be applied as the experiment continues to operate and as new techniques are developed. Additionally, as described in the next section, these data will be available for alternative analysis and the possible extraction of physics results not envisioned at the outset.

### 3.12 Underground Facilities

The underground facility ultimately selected is pivotal in the design of the Majorana apparatus. However, common features of the key Majorana underground spaces can be identified based on the function of the space. Minor variations on the basic outline presented here will be required to adjust to the specific conditions found in the potential underground location.

The spaces identified for the Majorana Experiment include an environment-controlled experiment hall, an electroforming laboratory, and potentially an underground detector manufacturing laboratory. In the following subsections we discuss these requirements.

### Basic Requirements

Since the Majorana apparatus will consist of several multi-germanium-crystal modules, periodic installation activity will be required during construction. The periods between these incremental additions will see operation identical to that during the running period after construction is complete, in order to detect problems with the newly installed components, make repairs, and provide feedback into the construction process. The Majorana apparatus will therefore require ease of access and a flexible system providing (1) pre-install staging and work area, (2) expanded space for installation, and (3) a small, isolated space for running.

The Majorana hall will require an air-locked entry, a control room space, the space for the actual apparatus, and a multi-function anteroom. The requirements for the Majorana Experiment hall can be listed as:

Apparatus
- 5 x 4 m footprint
- Cleanable surfaces
- Scrubbed air
- Air-conditioning to ~ 20 C with great stability
- Humidity control

Staging/Installation/Anteroom
- 5 x 4 m footprint
- Cleanable surfaces
- Scrubbed air
- Removable barrier to apparatus

Control Room
- 4 x 4 m footprint
- Monitoring station
- Cabling runs for 24 crates in 4 racks
- Controlled temperature for electronics
- Broadband connectivity to the Internet
- Power: <20kW conditioned
- Some uninterruptible power supply capability

Several approaches are possible in organizing the space needed for Majorana: an organic concept, with all the non-proprietary spaces connected, either in a 'square' or 'linear' arrangement depending on the nature of the underground space, or separate functional facilities, each with airlocks, dressing rooms, and so forth.

### Required Infrastructure

There are several infrastructure features that the Majorana Project will require. Some provided by the laboratory and some not. The typical facility-provided infrastructure items include for example power (filtered and unfiltered), provision for exhausting nitrogen gas and filtered hood effluent. Also, broadband network connectivity is required and expected from the facility.

Air conditioning is a typical requirement: it improves stability and extends the lifetime of electronic and electrical devices and improves the productivity of human workers. In the case of ultra-low level experiments, the potassium in sweat contains enough $^{40}$K to spoil a run with a single drop inside the hand-built shield. In addition to normal air conditioning, the temperature of the electronics and the apparatus itself must be maintained stably to prevent gain shifts during the period between calibrations. Typical temperature dependence of gain in an ORTEC 572 amplifier, for example, is around 10 ppm/degree C. Thus a 10-degree shift lasting for a substantial time would shift the 2039 keV region by 0.2 keV. This is not a large effect, but can be easily prevented.

Other examples of atypical infrastructure include ultra-clean air. In past experiments, the requirement of excluding radon from the spaces around the detectors was achieved by venting nitrogen boil-off gas into the lead cave. This greatly inhibits the inflow of radon into a well-sealed lead cave, but does nothing to prevent the deposition of radon daughters in the inner spaces during construction and maintenance. A supply of air, scrubbed of radon and subsequently filtered of particulates, could eliminate this source of background in the low energy region.

Another experiment-specific type of infrastructure is the provision of liquid nitrogen into the apparatus for cooling. If an adequate supply is provided at the surface, a simple manifold system external to the Majorana Experiment hall would be cheap and beneficial. This manifold would allow the introduction of liquid from 160-liter Dewars with an absolute minimum of human attention and no entry into the apparatus chamber. This would require careful design to prevent the inadvertent inclusion of radon and moisture-laden air. In the event that a local supply of liquid nitrogen is unavailable, a set of nitrogen distillation systems could provide the supply, requiring only power. The nitrogen could then be introduced through the same manifold system. Adequate space near the Majorana Experiment hall would be required for the distillation station.

*Detector Production Underground*

The details of the detector manufacturing are considered proprietary by commercial suppliers, but certain parts are well known. Very clean air will be needed in a space the equivalent of 4 meters by 20 meters. Zone refining, crystal pulling, crystal cutting, surface treatment, and testing will be housed in separate rooms within this space. Portions of the process require Class 10 air. Although the detectors could be manufactured at the companies' present locations, the possibility of fabricating them underground is exciting. For this purpose, one must consider the processes of crystal pulling and the subsequent detector fabrication separately as they would require independent facilities.

After the Ge is enriched, it will be free of cosmogenic isotopes. However, during the period for delivery to a crystal-pulling site and then a detector production site, the Ge will reside on the surface and the cosmogenic isotopes will begin to grow in. When the crystals are zone refined, the problem isotope $^{60}$Co will be once again removed. Thus if the detector fabrication and possibly its associated zone refinement was done underground, this dangerous long-lived background would be greatly reduced. There is a very clear advantage for this process to take place underground.

$^{68}$Ge will also be produced during the time after enrichment and prior to delivery underground. Although the zone refinement that takes place during crystal growth and detector production will not remove this isotope, there is an advantage to limiting the time the material resides above ground. The half-life of $^{68}$Ge is 278 days sets the time scale for delivery underground. If the crystal growing process also took place underground, it would greatly reduce the amount of this troublesome isotope.

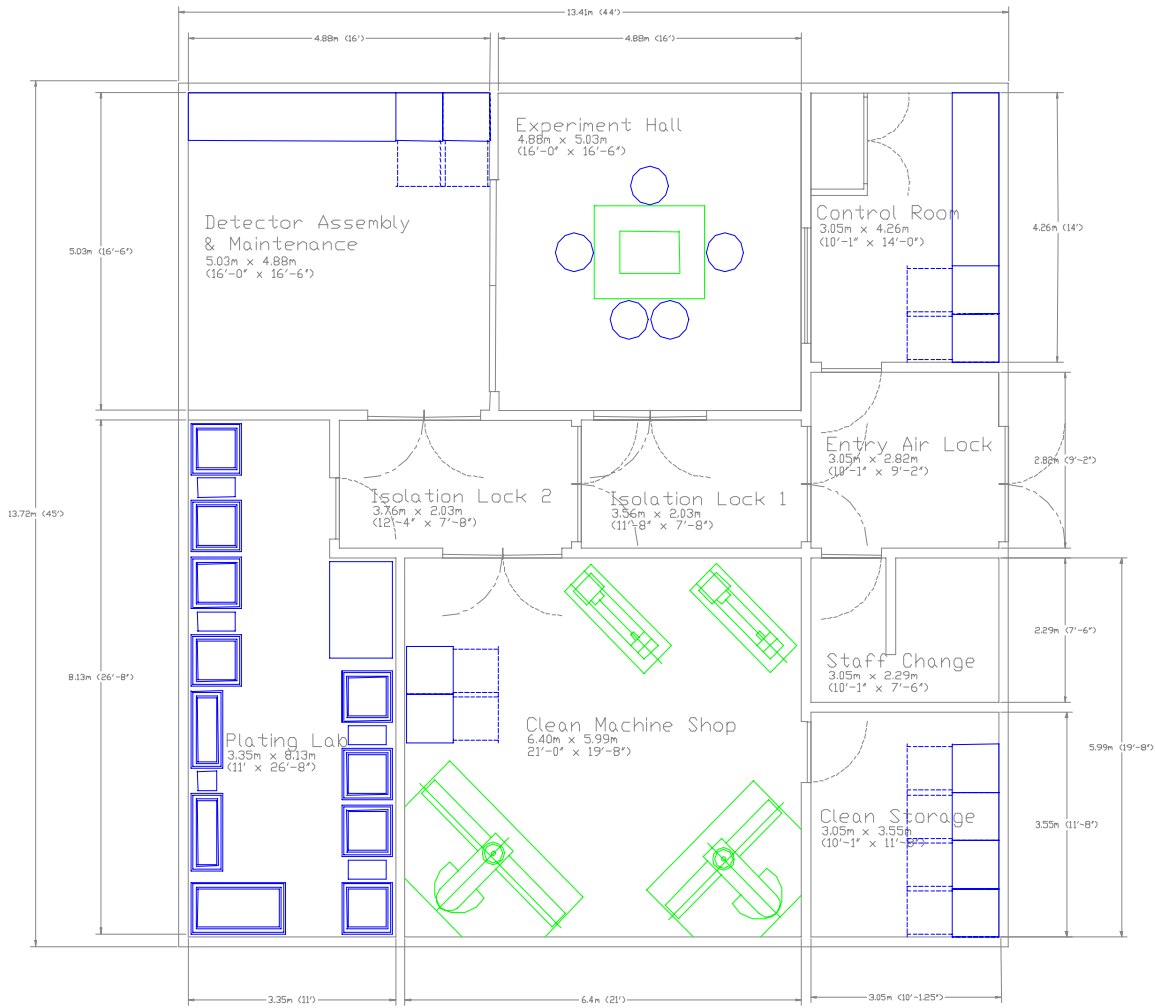

**Figure 3-36. Notional layout of Majorana production and experiment spaces.**

We propose that both of these activities take place underground by contracting with a detector production company to provide these services.

*Electroforming Underground*

The electroforming process is critical to the project because of the need for low-background materials for support structures, vacuum jackets, and so forth. Several plating baths will be required to prepare all the required parts during the production campaign. Fortunately, during production most baths require only periodic monitoring and attention.

Electroforming takes place in a copper sulphate bath in which a current is passed from a large copper electrode to a mandrel, of appropriate shape. Approximately 100 A at 1-3 V are required to form a considerable thickness of copper on the mandrel each day. After a few days of copper application, the mandrel and the forming part must be removed from the bath and have a thin layer removed to break grain boundaries in the copper and to equalize the growth rate independent of the varying electric fields in the bath. Thus a clean machine shop must be part of the electroforming facility.

The preparation of the baths requires that the $CuSO_4$ must be recrystallized multiple times to establish purity. A fume hood for this purpose is required. Alternatively, we may use self-contained nitrogen boxes to prevent Rn daughters from contaminating the baths. Storage space for raw copper, materials, and parts in progress will be required, as well. The electroforming laboratory requirements can be listed as:

Plating Area:
- Plating area requires 4 x 8 x 3 m room
- Requires spill containment lining
- Shared $10^{-6}$ torr dry vacuum system
- Fume extractor for etching
- Flammable and hazardous gas sensors
- Receives HEPA-filtered air supply
- Radon-scrubbed air for lowest-level work
- Air-lock entry, washable walls
- Power required ~ 12 kW 120/240 VAC
- Air-conditioning to ~ 20 C

Machining Area
- Clean shop area requires 4 x 8 x 3 m room
- Receives HEPA-filtered air supply
- Air-lock entry, washable walls
- Power required ~ 24 kW 120/240 VAC
- Air-conditioning to ~ 20 C

Storage Area
- Materials storage area requires 3 x 4 x 3 m room
- Radon-proof storage lockers with purge gas and vacuum capability
- Shared $10^{-6}$ torr dry vacuum system
- Receives HEPA-filtered air supply
- Air-lock entry, washable walls
- Power required ~ 2.4 kW 120 VAC
- Air-conditioning to ~ 20 C

Because the cosmogenic isotope $^{60}$Co is readily produced in Cu, there is a great advantage to electroforming the Cu underground. Its important to note that electroforming should remove any Co from the Cu, so the process is very effective in eliminating this potential source of background.

### Potential Sites for Majorana

We have considered three possible underground sites for Majorana in North America. Here we briefly list and describe these sites. Our clear preference would be to site at the National Underground Science and Engineering Laboratory (NUSEL), should it be built. Our strategy for making a site selection is based on the current lack of a definite plan to build NUSEL. We plan to pursue the option of sitting Majorana in SNOLab. Toward this goal, we will respond to the anticipated call for Letters of Intent (LOI) by the SNOLab management expressing our potential desire to site there. We would also solicit SNOLab-member Canadian institutions to join our collaboration to buttress our local activities associated with the lab. Unfortunately, competition for the limited space at SNOLab is a significant concern and several other collaborations are also expected to seek occupancy. Hence it is not assured that our LOI will be selected. In that event we would plan to site at WIPP and begin to re-design the shield as necessary to maximize the detector's capabilities at that relatively shallow site.

### NUSEL

In May 2001, based on the recommendation of the Bahcall Committee and the endorsement of the Nuclear Physics Long Range plan, a proposal was submitted to National Science Foundation to convert the Homestake Mine into a National Underground Science and Engineering Laboratory (NUSEL). The proposal calls for 5-year construction plan to provide an underground laboratory at the 7400-foot level along with "campus like" support facilities located on the surface.

During the 5 year NUSEL construction period it is planned to support and maintain an active scientific program, including the establishment of an ultra low background counting facility and a cosmogenic decay storage facility for materials.
This results in an excellent synergy with the Majorana Project. The NUSEL plan for developing the underground laboratory calls for customized "built to order" halls. The NUSEL and Majorana timescales are such that Majorana would be in a position to specify custom chamber requirements and NUSEL would be in a position to respond.

The local support infrastructure needs of Majorana are expected to be compatible with those anticipated to be available during the 5 year NUSEL construction program. The depth of this laboratory and its excellent proposed surface and underground infrastructure make it extremely attractive. The small footprint of the Majorana experiment is consistent with being accommodated in space that already exists at the Homestake mine.

### SNOLab

In mid 2002, a proposal to expand the underground laboratory space near the Sudbury Neutrino Observatory (SNO) was approved and funded. This facility located in Sudbury,

Ontario, Canada is at a deep site (6800 feet) and will provide a modest amount of lab space (about 900 m$^2$). The scope of this lab will not be near that of the proposed NUSEL, but would provide space for two small-footprint experiments such as Majorana. SNOLab will be contained within the working INCO nickel mine and therefore will have limited access through a rather small lift. The SNO experience indicates that working in these conditions is possible but not ideal.

### WIPP

The U.S. Department of Energy currently operates the Waste Isolation Pilot Plant (WIPP) near Carlsbad, New Mexico, as a disposal site for transuranic waste. The constructed underground facilities include four shafts, an experimental area, an equipment area, a maintenance area, and connecting tunnels. These underground facilities were excavated in the Salado Formation, 655 meters (2,150 feet) beneath the land surface. DOE now has proposed to expand the availability of WIPP facilities and infrastructure to scientists who wish to conduct experiments there. The relatively-low background radiation of the salt walls in the WIPP underground facility is one of the factors that makes the site an attractive environment for experiments. However the shallowness of the site would create operating difficulties for Majorana. The neutron shield would have to be extreme and still the background would not be as low as reasonably achievable. Nevertheless, some Majorana associated auxiliary experiments (SEGA and MEGA) are being prepared for installation in WIPP at the time of this writing.

# 4    The Ongoing Majorana R&D Program

A short research and development program is underway to optimize and streamline the design of Majorana. These R&D activities are centered around the SEGA and MEGA projects, which themselves have physics goals. The SEGA project will study how segmentation and pulse shape methods are optimized with a specific geometry of segmentation, and MEGA will study how many crystals (18) are operated together to establish and demonstrate electronic and mechanical methods for Majorana. We want to emphasize that these are optimization of engineering details and are not studies to demonstrate technological feasibility. The physics and engineering goals for SEGA, MEGA and Majorana are summarized in Fig. 4-1. The Reference Plan design of Majorana is capable of reaching our sensitivity goal. However, the design may be subject to modest modifications for optimization and possible cost reduction.

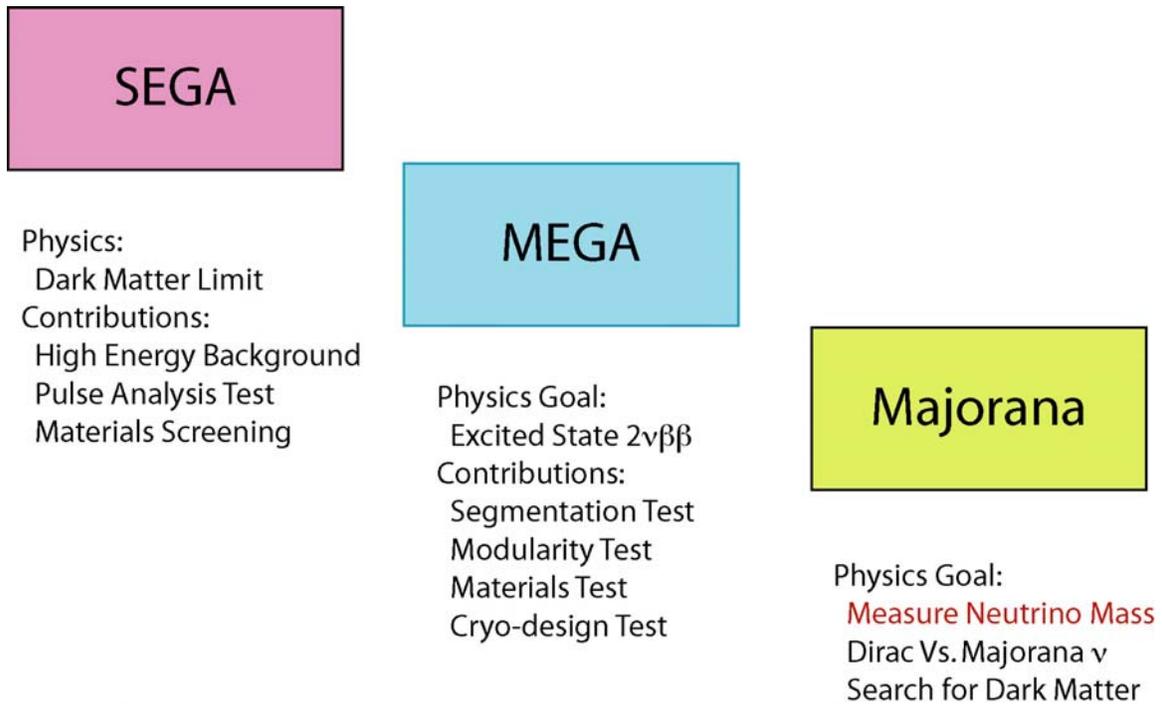

**Figure 4-1 Approach showing engineering prototypes (SEGA and MEGA) leading to the final experiment, Majorana.**

### 4.1 SEGA, MEGA and Majorana

#### *SEGA Description*
The SEGA experiment consists of an isotopically enriched 2 x 6 segmented detector. We describe segmentation in terms of k × j segments where there are k segments along the Z axis of the right circular cylinder (i.e. disks) and j segments about the ϕ direction (i.e. pie slices). The approach for segmentation currently in use mandates that the germanium detector must be n-type due to the ease of lithography of the boron implanted contacts on the outside of the crystal.

The main R&D goal of SEGA is the determination of the most effective combination of signal processing and segmentation. Our signal processing and segmentation effort is aimed at separating single-site energy deposition events, such as double-beta decay, from multi-site events, such as [68]Ge and [60]Co decays. For a single-site energy deposition in a segmented detector, the only segment with a net charge is that containing the deposition site. Furthermore, by comparing induced signals in neighboring segments, one can determine if the signal was actually a multiple site event within a single segment [Vet00].

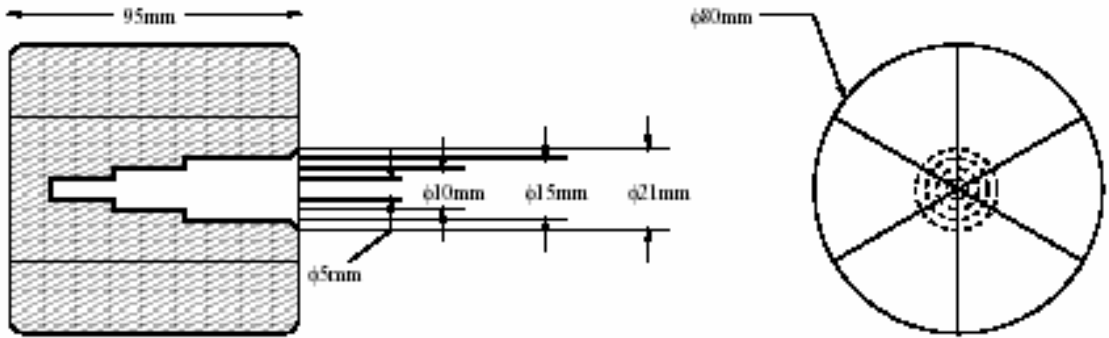

**Figure 4-2 Segmentation layout of the SEGA detector. The dimensions are those originally designed and not those of the delivered part.**

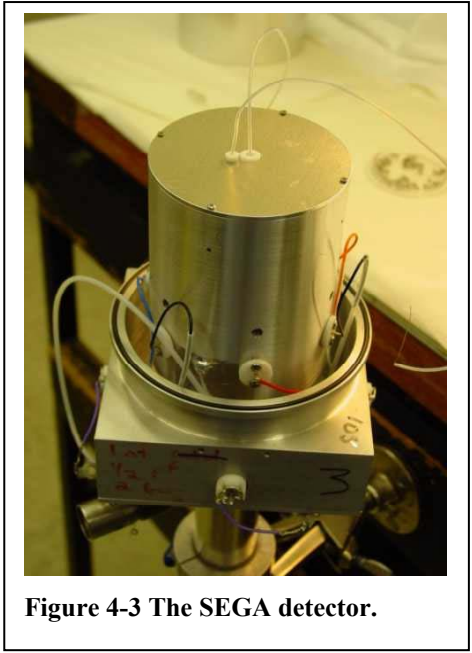

**Figure 4-3 The SEGA detector.**

Figure 4-3 shows a photograph of the SEGA detector during testing at TUNL.

*MEGA Description*
MEGA will consist of a pair of segmented germanium detectors surrounded by a toroid of 16 fairly large (70% relative efficiency) p-type germanium detectors. The toroidal apparatus will approximate the cryogenic and electronic challenges in the construction of the multi-crystal modules of Majorana.

The first challenge will be the cooling of the detectors as the crystals must operate below about 125K, optimally around 90K. In comparison, the noise level for the preamp front-end located near the crystal (for low capacitance and high bandwidth) typically reaches a minimum at 145K. These temperatures can be simultaneously optimized by engineering the thermal conductivity of the FET mount and clever use of the few milliwatts of power dissipated by each front-end FET.

Another seemingly pedestrian but fundamental challenge is the provision for electronic feedthroughs. Typical germanium detectors have the luxury of a single feedthrough with four contacts for a single detector segment. With several segments per crystal, the situation in the Majorana detectors is not as simple. We propose to try new pre-amp schemes with many of the electrical contacts common. We are considering the use of Multi-Chip Module (MCM) technology that would place the entire pre-amp within the cryostat. While this invites some risk of radiological contamination, the masses are small and modest shielding together with material assay should eliminate this concern.

The shielding configuration of the MEGA apparatus provides a challenge similar in kind

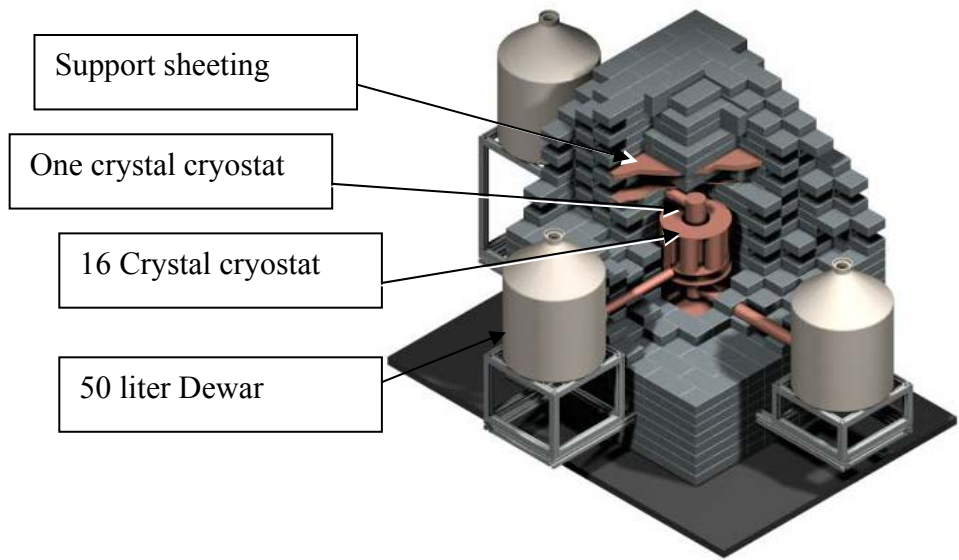

**Support sheeting**

**One crystal cryostat**

**16 Crystal cryostat**

**50 liter Dewar**

**Figure 4-4 Cutaway view of preliminary design of the MEGA apparatus including shielding.**

but smaller in magnitude than Majorana (as described in section 3.7). Because the toroidal detector ring will occupy an area in the shield of about 30 cm by 30 cm, clean support for the lead above will have to be arranged. This shielding design will be tested with the construction of MEGA.

The Monte Carlo simulation of this multi-crystal array poses an interesting task, similar to the analysis of the multi-segment data. The effect of various contaminants within multiple materials and locations must be known to guide Majorana construction. However it is possible to validate Monte Carlo code by the measurement of signals from known sources in the MEGA geometry. Shortly after introduction underground, the detection of photons escaping the inner detectors, for instance from $^{58}$Co ($T_{1/2}$ =71 d), will allow testing of suppression of multi-gamma isotope backgrounds using multiple crystals.

Table 4-1 R&D Engineering Issues for Majorana Implementation

**Primarily SEGA**

- Pulse-shape discrimination performance
- Segmentation performance vs. granularity
- Advanced uses of segmentation signals

**MEGA and SEGA**

- Background models
- "No-Hit" segment signal analysis methods
- Multiple-scatter event tracking and reconstruction for background identification
- Front-end electronics
  - Radiopurity
  - Rise-time performance
  - Ease of assembly and testing
- Measurement of fast neutron background

**MEGA**

- Detector mounting scheme
- Detector support material radiopurity
- Cryostat mechanical and thermal design

## 4.2   SEGA and MEGA Science Goals

SEGA and MEGA will address the engineering and testing issues listed in Table 4-1. However, this R&D program is not without exciting, degree-producing physics goals as well. These include:

- Double Beta Decay
  - $2\nu\beta\beta$: MEGA measurement of $2\nu\beta\beta$ to $0^+$ excited-state in a number of isotopes
  - $2\nu\beta\beta$: precision re-measurement of $^{76}$Ge decay to the ground state. (Note: the similarity of the $2\nu$ signal to the $0\nu$ signal will be an important test of the electronic signal processing efficacy of the Majorana approach)
- Dark  Matter:
  - Rapid (<1 year) exclusion of the DAMA CDM result from SEGA or MEGA
- Solar Axion:
  - MEGA will be the largest-active-mass Ge solar axion experiment to date and will be the first of any single crystal experiment with known orientation of the crystalline axes.

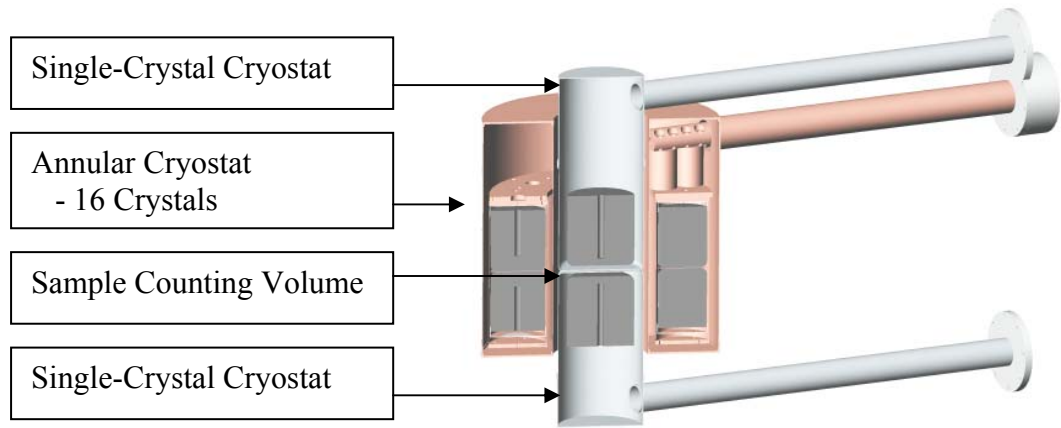

Single-Crystal Cryostat

Annular Cryostat
 - 16 Crystals

Sample Counting Volume

Single-Crystal Cryostat

**Figure 0-5 Cutaway view of the MEGA apparatus with 3 separate cryostats. The inner detectors may be removed and a sample introduced between them. The outer 16 crystals are cooled together much like the multi-crystal Majorana apparatus. All the support materials are of electroformed copper.**

# 5 Project Schedule

The Majorana Project will be implemented over approximately 5-6 years, but physics milestones will be achieved much earlier. In this section we summarize our anticipated schedule and milestones. We consider the fast acquisition plan of enriching 200 kg/y as the most economical and the fastest to the physics goals. We also anticipate that we can have detectors fabricated at a rate that will roughly coincide with the $^{enr}$Ge production. Hence we use that as our example here. Figure 6-1 shows our proposed straw man schedule including the critical decision milestones.

After construction start is approved (CD-3), the contracts for the enriched isotope, site preparation and detector fabrication would be let. It is anticipated that $^{76}$Ge will begin to appear approximately 6-12 months after the contract is signed and that detector fabrication would begin at that point. After and additional 6-12 months, those detectors would be ready to assemble into cryostats, the shield and then begin operation. Because, even 50-60 kg of $^{enr}$Ge represents a significant increase in sensitivity over previous

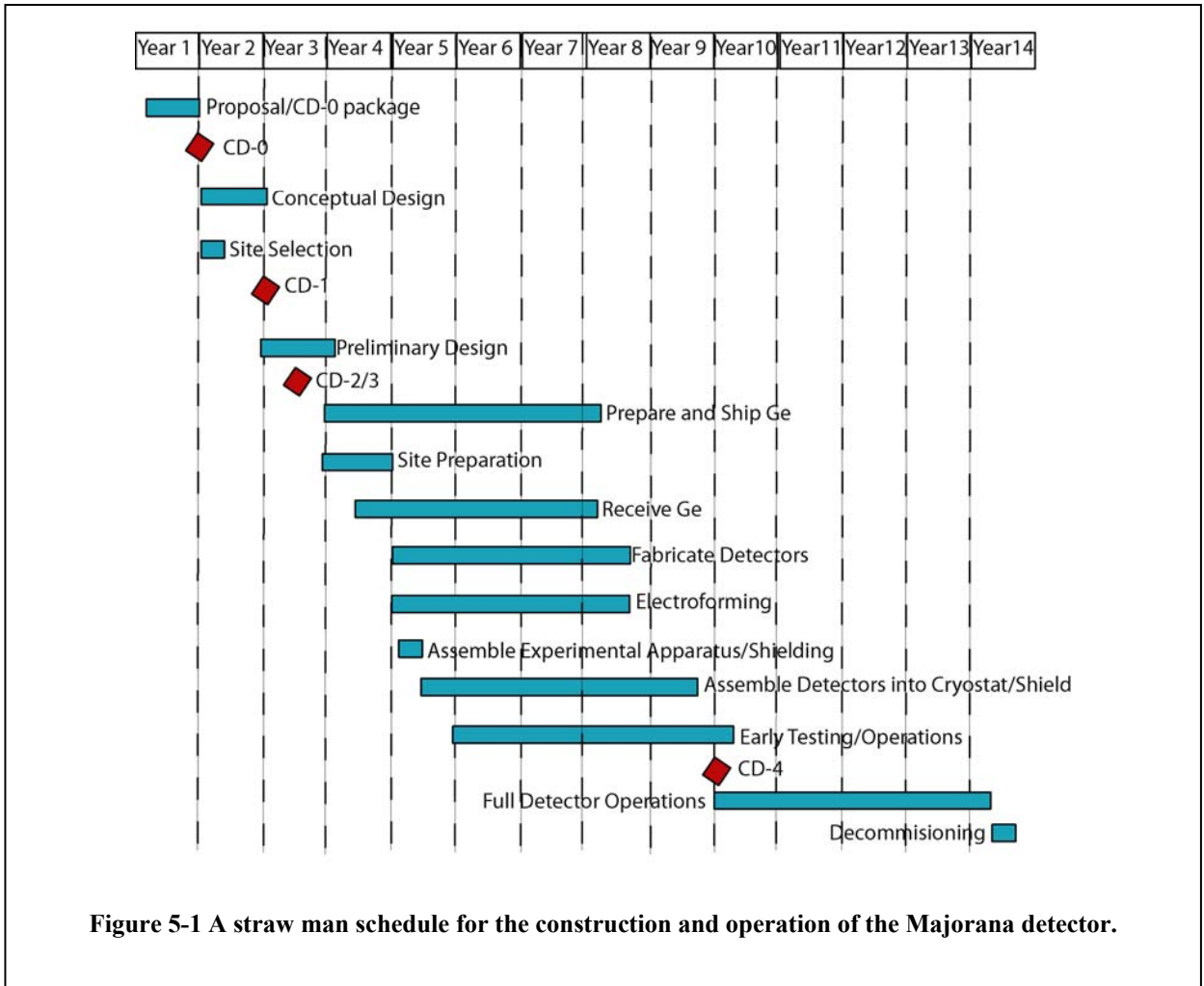

**Figure 5-1 A straw man schedule for the construction and operation of the Majorana detector.**

experiments, operating a partial detector is very fruitful.

The straw man schedule assumes that Critical Decision 0 (CD-0) is obtained soon (Beginning of Year 2 as defined in Fig. 6-1) as the preparation process is nearing completion. During the following year, the conceptual design and site selection will be completed. Towards the end of the subsequent year (Year 3) the design will be completed. Because the required engineering to prepare for the construction of Majorana is minimal, we propose to obtain CD-2 and CD-3 simultaneously in the last half of Year 3.

Possible design alternatives must either be implemented into the Reference Plan or eliminated from consideration prior to acquisition. The research considering such alternatives must therefore be complete early in the project's development so that designs can be modified without causing delay. The possible major alternatives are: 1) the choice of detector type (p or n), size and segmentation. 2) alternative cooling options. 3) the use of an active shield. 4) underground crystal growing. 5) and the underground site choice. Table 6-1 summarizes the timing for the decision on each of these options. The decision dates chosen are the latest permissible to prevent any delay in the Preliminary Baseline/Proposed Work Plan milestone (i.e. CD-1). Many of these options, such as the detector type, have little impact on other aspects of the overall apparatus. Therefore the decision can be left until a late date. Other questions, such as whether to use an active shield do impact other aspects of the design and must be answered earlier. The infrastructure required for growing crystals underground is significant enough, that the decision is required much earlier.

**Table 5-1 A summary of the decision dates for possible alternatives to the Reference Plan.**

| Possible Alternative | Decision Date |
|---|---|
| Choose p- or n-type crystal | CD-1 milestone date |
| Choose detector size | CD-1 milestone date |
| Choose segmentation | CD-1 milestone date |
| Choose cooling configuration | During CD-0 time period |
| Decide whether to use an active shield | Prior to CD-0 |
| Decide whether to grow crystals underground | Prior to CD-0 |
| Choose underground lab site | CD-1 milestone date |

During the first year of construction (Year 4), we anticipate that ECP would deliver only 50 kg of $^{enr}$Ge as the plant is being readied for higher production rates. We would expect this delivery to occur near year's end and that the detector manufacture of the initial detectors would be ready to receive it. During this year, the underground site would have to be prepared. The data acquisition electronics must be purchased and assembled during this year.

During the second year of construction (Year 5) the first 50 kg of Ge will have been built into detectors and we anticipate receiving the next 200 kg. The experiment will begin to operate as soon as the detectors are ready so the shield, cooling and data acquisition need to be ready. In the next year (Year 6), we receive the next 200 kg of Ge and the detectors manufactured from the Year-5 Ge will be built into detectors. The physics results from the initial 50 kg should be ready during this year. Even with this limited mass of Ge, the experiment will provide $0\nu\beta\beta$ limits that exceed previous values.

We receive the final 75 kg Ge in the 4[th] year of construction (Year 7) and the previous year's Ge is built into detectors. Results from the first 250 kg will begin to be available. The final detectors will be built in the 5[th] year of construction (Year 8) and by year's end, the full array should be assembled. Allowing for 1 year to finish any checkout and final startup, the full array should be ready for operation and hence a CD-4 milestone in Year 9. At this point the full array is operational, the construction project is complete and we enter the operating stage of the experiment.

The experiment, composed of the full 500-kg array, will operate for a number years until the statistical precision reaches the systematic limit. How many years this will take depends on the background levels, but we estimate about 3-5 years of full-array operation. The straw man schedule is drawn for 5 years of operation (Year 10-14). During the final months of the project, we would decommission the experiment (last half Year 14). This could mean the complete dismantling of the detector and distribution of the usable items or perhaps a reconfiguration for other experimental goals. In fact, its unprecedented sensitivity to low levels of radioactivity may very well make the Majorana apparatus useful to future environmental and national security programs.

Figure 6-2 shows the sensitivity of the experiment as a function of time with respect to the straw man schedule. The improvement in the sensitivity is dramatic even during the construction period (Year 4-9) because data can be taken with a partial array.

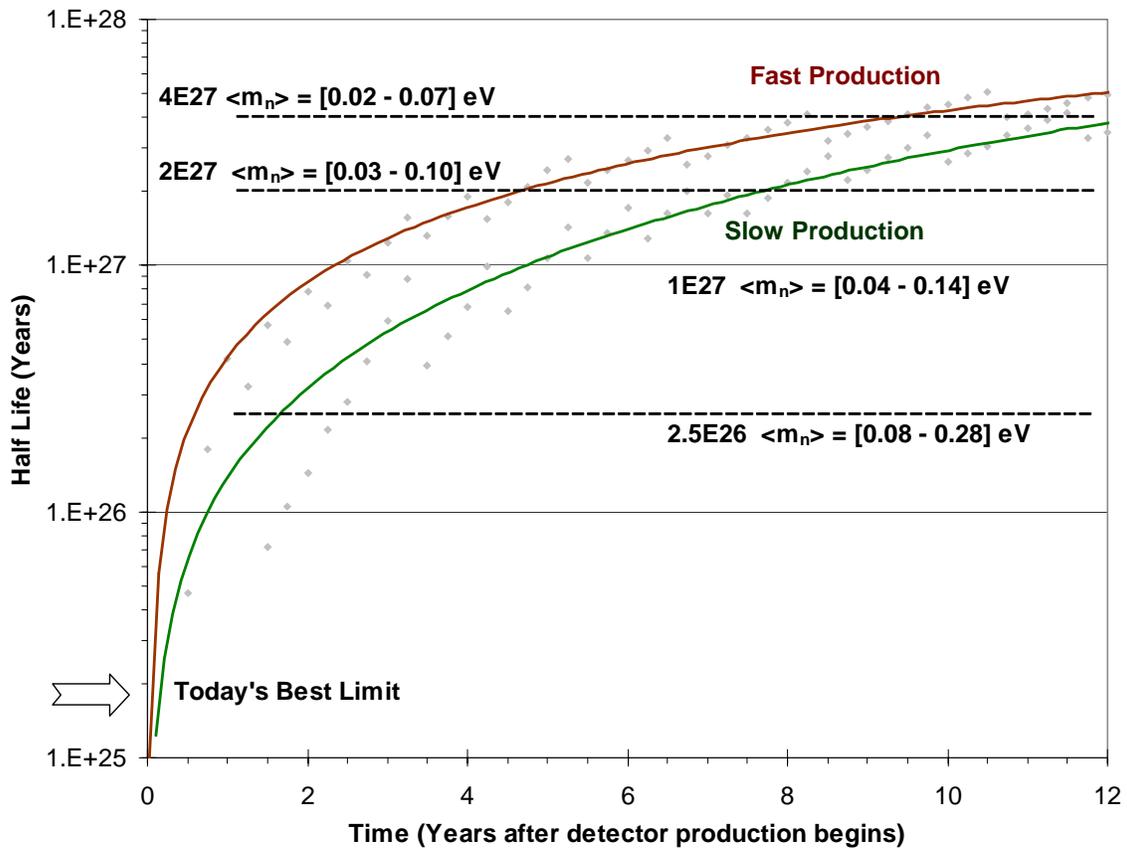

**Figure 6-2 A plot of sensitivity growth assuming data acquisition during construction.**

# 8.0   Appendices

## Appendix 2. Simplified Decay Chain Data

| Nuclide | Half-life | Major γ energy (keV) and intensity (%) | |
|---|---|---|---|
| $^{238}U_{92}$ | 4.51 x 10$^9$ Y | | |
| \| α | | | |
| $^{234}Th_{90}$ | 24.1 d | 63.3 | 4.49 |
| \| β$^-$ | | 92.6 | 5.16 |
| $^{234}Pa^m_{91}$ | 1.17 m | 766.6 | 0.21 |
| | | 1001.4 | 0.59 |
| (99.87%) β$^-$   IT(0.13%) | | | |
| $^{234}Pa_{91}$ | 6.75 h | 131.2 | 20.0 |
| | | 226.8 | 11.4 |
| | | 569.3 | 13.5 |
| | | 882.0 | 28.0 |
| β$^-$ | | 926.4 | 24.9 |
| | | 946.0 | 12.0 |
| $^{234}U_{92}$ | 2.47 x 10$^5$ Y | 53.2 | 0.12 |
| \| α | | | |
| $^{230}Th_{90}$ | 8.0 x 10$^4$ Y | 67.7 | 0.38 |
| \| α | | 143.9 | 0.05 |
| $^{226}Ra_{88}$ | 1602 Y | 186.1 | 3.5 |
| \| α | | | |
| $^{222}Rn_{86}$ | 3.823 d | | |
| \| α | | | |
| $^{218}Po_{84}$ | 3.05 m | | |
| (99.98%) α   β$^-$ (0.02%) | | | |
| $^{214}Pb_{82}$ | 26.8 m | 241.9 | 7.46 |
| | | 295.2 | 19.20 |
| $^{218}At_{85}$ | ~2 s | 351.9 | 37.10 |
| β$^-$   α | | | |
| $^{214}Bi_{83}$ | 19.9 m | 609.3 | 46.10 |
| | | 768.4 | 4.88 |
| (99.98%) β$^-$   α (0.02%) | | 934.0 | 3.16 |
| | | 1120.3 | 15.00 |
| | | 1238.1 | 5.92 |
| | | 1377.6 | 4.02 |
| | | 1408.0 | 2.48 |
| | | 1729.6 | 3.05 |
| | | 1764.5 | 15.90 |
| | | 2204.1 | 4.99 |
| $^{214}Po_{84}$ | 164 µs | 799.7 | 0.01 |
| $^{210}Tl_{81}$ | 1.3 m | 296.0 | 79.16 |
| | | 795.0 | 98.95 |
| | | 1060.0 | 12.37 |

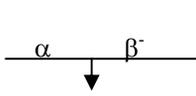

|        |        |
|--------|--------|
| 1210.0 | 16.82  |
| 1310.0 | 20.78  |

## Simplified Decay Scheme for $^{238}$U (continued)

| Nuclide | Half-life | Major γ energy (keV) and intensity (%) | |
|---------|-----------|---------|---------|
| $^{210}$Pb $_{82}$ | 22.3 Y | 46.5 | 4.05 |
| $^{210}$Bi $_{83}$ | 5.01 d | | |
| $^{210}$Po $_{84}$ | 138.4 d | 803.0 | 0.0011 |
| $^{206}$Tl $_{81}$ | 4.19 m | | |
| $^{206}$Pb $_{82}$ | Stable | | |

(~100%) β⁻    α (0.00013%)

α    β⁻

## Simplified Decay Scheme of $^{232}$Th

| Nuclide | Half-life | Major γ energy (keV) and intensity (%) | |
|---------|-----------|-----------|-----------|
| $^{232}$Th $_{90}$ | 1.41 x 10$^{10}$ Y | | |
| $\quad$ α | | | |
| $^{228}$Ra $_{88}$ | 5.75 Y | | |
| $\quad$ β$^-$ | | | |
| $^{228}$Ac $_{89}$ | 6.15 h | 099.6 (D) | 1.37 |
| | | 129.1 | 2.45 |
| | | 209.3 | 3.88 |
| | | 270.2 | 3.43 |
| | | 328.0 | 2.95 |
| | | 338.3 | 11.25 |
| | | 409.5 | 1.94 |
| | | 463.0 | 4.44 |
| | | 772.4 (D) | 1.58 |
| $\quad$ β$^-$ | | 794.9 | 4.34 |
| | | 835.7 | 1.68 |
| | | 911.2 | 26.60 |
| | | 964.8 | 5.11 |
| | | 969.0 | 16.17 |
| | | 1588.2 | 3.27 |
| | | 1630.6 | 1.60 |
| $^{228}$Th $_{90}$ | 1.910 Y | 84.37 | 1.6 |
| $\quad$ α | | 216.0 | 0.3 |
| $^{224}$Ra $_{88}$ | 3.64 d | 241.0 | 3.97 |
| $\quad$ α | | | |
| $^{220}$Rn $_{86}$ | 55 s | 549.7 | 0.1 |
| $\quad$ α | | | |
| $^{216}$Po $_{84}$ | 0.15 s | | |
| $\quad$ α | | | |
| $^{212}$Pb $_{82}$ | 10.64 h | 238.6 | 43.6 |
| $\quad$ β$^-$ | | 300.1 | 3.34 |
| $^{212}$Bi $_{83}$ | 60.6 m | 39.86 | 1.10 |
| | | 288.1 | 0.34 |
| | | 452.8 | 0.36 |
| | | 727.3 | 6.65 |
| (64.0%) β$^-$ $\quad$ α (36.0%) | | 785.4 | 1.11 |
| | | 1620.6 | 1.51 |
| $^{212}$Po $_{84}$ $\quad\quad$ $^{208}$Tl $_{81}$ | 304 ns | | |
| | 3.05 m | 277.4 | 6.31 |
| | | 510.8 | 22.60 |
| | | 583.2 | 84.50 |
| $\quad$ α $\quad$ β$^-$ | | 860.6 | 12.42 |
| | | 2614.5 | 99.20 |
| $^{208}$Pb $_{82}$ | Stable | | |

## Simplified Decay Scheme for $^{235}$U

| Nuclide | Half-life | Major γ energies (keV) and intensity (%) | |
|---|---|---|---|
| $^{235}$U $_{92}$ | 7.1 x 10$^8$ Y | 143.8 | 10.9 |
| | | 163.3 | 5.00 |
| α | | 185.7 | 57.50 |
| | | 205.3 | 5.00 |
| $^{231}$Th $_{90}$ | 25.5 h | 81.5 (D) | 1.29 |
| β$^-$ | | 84.2 | 6.60 |
| $^{231}$Pa $_{91}$ | 3.276 x 10$^4$ Y | 27.4 | 9.3 |
| | | 283.7 | 1.60 |
| | | 300.0 | 2.39 |
| α | | 302.7 | 2.24 |
| | | 330.1 | 1.31 |
| $^{227}$Ac $_{89}$ | 21.6 Y | | |
| (98.6%)β$^-$    α (1.4%) | | | |
| $^{227}$Th $_{90}$ | 18.718 d | 49.9 | 0.52 |
| | | 50.1 | 7.28 |
| | | 236.0 (D) | 11.65 |
| | | 256.0 (D) | 7.6 |
| $^{223}$Fr $_{87}$ | 22 m | 50.8 | 34.0 |
| | | 80.0 | 8.16 |
| α    β$^-$ | | 234.6 | 3.4 |
| $^{223}$Ra $_{88}$ | 11.43 d | 122.3 | 1.19 |
| | | 144.2 | 3.26 |
| | | 154.2 | 5.59 |
| | | 269.4 | 13.6 |
| α | | 323.9 | 3.9 |
| | | 338.3 | 2.78 |
| | | 444.9 | 1.27 |
| $^{219}$Rn $_{86}$ | 4.0 s | 271.2 | 9.9 |
| α | | 401.7 | 6.64 |
| $^{215}$Po $_{84}$ | 1.78 ms | | |
| (~100%) α    β$^-$ (0.00023%) | | | |
| $^{211}$Pb $_{82}$ | 36.1 m | 404.8 | 3.83 |
| | | 427.0 | 1.72 |
| | | 831.8 | 3.8 |
| $^{215}$At $_{85}$ | ~0.1 ms | | |
| β$^-$    α | | | |
| $^{211}$Bi $_{83}$ | 2.14 m | 351.0 | 12.76 |
| (0.28%)β$^-$    α (99.7%) | | | |
| $^{211}$Po $_{84}$ | 0.52 s | 569.65 | 0.53 |
| | | 897.8 | 0.52 |
| $^{207}$Tl $_{81}$ | 4.79 m | 897.8 | 0.24 |
| α    β$^-$ | | | |
| $^{207}$Pb $_{82}$ | Stable | | |

## Appendix 3. Representation of the table of isotopes from $^{53}$Mn to $^{77}$As.

### 3.11 **Appendix 4. Alpha signals in primordial decay chains.**

| E (keV) | Chain % | Isotope | Chain | E (keV) | Chain % | Isotope | Chain |
|---------|---------|---------|-------|---------|---------|---------|-------|
| 3830 | 2.000E-01 | Th-232 | Th-232 | 5181 | 1.098E-03 | Po-218 | U-238 |
| 3953 | 2.300E+01 | Th-232 | Th-232 | 5212 | 3.600E-01 | Th-228 | Th-232 |
| 4010 | 7.700E+01 | Th-232 | Th-232 | 5304.5 | 1.000E+02 | Po-210 | U-238 |
| 4039 | 2.300E-01 | U-238 | U-238 | 5340.5 | 2.670E+01 | Th-228 | Th-232 |
| 4147 | 2.300E+01 | U-238 | U-238 | 5423.3 | 7.270E+01 | Th-228 | Th-232 |
| 4196 | 7.700E+01 | U-238 | U-238 | 5449 | 4.900E+00 | Ra-224 | Th-232 |
| 4216.2 | 1.000E-04 | U-234 | U-238 | 5489.7 | 9.992E+01 | Rn-222 | U-238 |
| 4314.6 | 7.800E-03 | Ra-226 | U-238 | 5607.1 | 1.445E-02 | Bi-212 | Th-232 |
| 4367.8 | 3.100E-01 | Th-230 | U-238 | 5612.7 | 2.337E-02 | Bi-212 | Th-232 |
| 4476 | 1.200E-01 | Th-230 | U-238 | 5685.6 | 9.510E+01 | Ra-224 | Th-232 |
| 4524 | 1.100E-03 | Po-210 | U-238 | 5747 | 9.700E-02 | Rn-220 | Th-232 |
| 4601.9 | 5.550E+00 | Ra-226 | U-238 | 5768.1 | 2.154E-01 | Bi-212 | Th-232 |
| 4604.7 | 2.400E-01 | U-234 | U-238 | 5985 | 1.800E-03 | Po-216 | Th-232 |
| 4621 | 2.340E+01 | Th-230 | U-238 | 6002.5 | 9.978E+01 | Po-218 | U-238 |
| 4687.5 | 7.630E+01 | Th-230 | U-238 | 6050.8 | 9.055E+00 | Bi-212 | Th-232 |
| 4723.7 | 2.740E+01 | U-234 | U-238 | 6090.1 | 3.457E+00 | Bi-212 | Th-232 |
| 4784.5 | 9.455E+01 | Ra-226 | U-238 | 6288.3 | 9.990E+01 | Rn-220 | Th-232 |
| 4986 | 7.850E-02 | Rn-222 | U-238 | 6778.5 | 1.000E+02 | Po-216 | Th-232 |
| 5093.6 | 1.760E-02 | Ra-224 | Th-232 | 6892.4 | 1.060E-02 | Po-214 | U-238 |
| 5138.7 | 5.000E-02 | Th-228 | Th-232 | 7687.1 | 9.999E+01 | Po-214 | U-238 |
| 5175 | 1.800E-01 | Th-228 | Th-232 | | | | |

Table A4.1 Assuming equal amounts of $^{238}$U and $^{232}$Th; with both chains in equilibrium, these are the relative intensities of the alpha lines sorted by energy.